\documentclass[a4paper,onecolumn,11pt, accepted=2024-10-08]{quantumarticle}
\pdfoutput=1
\usepackage[utf8]{inputenc}
\usepackage[english]{babel}
\usepackage[T1]{fontenc}
\usepackage{amsmath}
\usepackage{amsthm}
\usepackage{amssymb}
\usepackage{algorithm}
\usepackage{subfigure}
\usepackage{float}

\usepackage[linkcolor=mymagenta,
            citecolor=mymagenta]{hyperref}
\usepackage{qcircuit}
\usepackage{cleveref}
\usepackage{tikz}
\usepackage{lipsum}

\usepackage{makecell}
\usepackage{pdflscape}

\usepackage{adjustbox}
\usepackage{yhmath}
\usepackage{graphics}

\usepackage{makecell}

\usepackage[numbers]{natbib}

\usepackage{mathtools}

\definecolor{mymagenta}{RGB}{200, 0, 100}
\newcommand{\lrb}[1]{\left ( #1 \right )}
\newcommand{\lrsb}[1]{\left [ #1 \right ]}
\newcommand{\lrcb}[1]{\left \{ #1 \right \}}

\newcommand{\abs}[1]{\left | #1 \right |}
\newcommand{\norm}[1]{\left \| #1 \right \|}

\newcommand{\ket}[1]{\left| #1 \right\rangle}
\newcommand{\bra}[1]{\left\langle #1 \right|}

\newcommand{\zon}{\{0,1\}^n}
\newcommand{\zo}{\{0,1\}}

\renewcommand{\exp}[1]{\operatorname{exp} \lrb{#1}}
\newcommand{\poly}[1]{\operatorname{poly} \lrb{#1}}
\renewcommand{\log}[1]{\operatorname{log} \lrb{#1}}
\newcommand{\logb}[1]{\operatorname{log}_2 \lrb{#1}}
\renewcommand{\ln}[1]{\operatorname{ln} \lrb{#1}}

\newcommand{\R}{\mathbb{R}}
\newcommand{\C}{\mathbb{C}}
\newcommand{\Z}{\mathbb{Z}}

\renewcommand{\P}[1]{\mathbb{P}\lrb{#1}}
\newcommand{\E}[1]{\mathbb{E}\lrsb{#1}}

\newcommand{\sgn}[1]{\operatorname{sgn}\lrb{#1}}

\renewcommand{\arcsin}[1]{\operatorname{arcsin}\lrb{#1}}

\newcommand{\mysin}[1]{\operatorname{sin}\lrb{#1}}
\newcommand{\mycos}[1]{\operatorname{cos}\lrb{#1}}

\newcommand{\tcount}[1]{{\rm T}_{\rm count}\lrb{#1}}
\newcommand{\tdepth}[1]{{\rm T}_{\rm depth}\lrb{#1}}
\newcommand{\numancilla}[1]{{\rm \# ancilla}\lrb{#1}}

\newcommand{\A}{\mathcal{A}}

\newcommand{\relu}[1]{\left(#1 \right)^+}

\begin{document}

\title{Option pricing under stochastic volatility on a quantum computer}

\author{Guoming Wang}
\email{guoming.wang.cs@gmail.com}
\orcid{0000-0002-0768-2644}
\author{Angus Kan}
\email{ckan@wesleyan.edu}
\orcid{0000-0002-5699-365X}
\affiliation{ORCA Computing}

\maketitle

\begin{abstract}
We develop quantum algorithms for pricing Asian and barrier options under the Heston model, a popular stochastic volatility model, and estimate their costs, in terms of T-count, T-depth and number of logical qubits, on instances under typical market conditions. These algorithms are based on combining well-established numerical methods for stochastic differential equations and quantum amplitude estimation technique. In particular, we empirically show that, despite its simplicity, weak Euler method achieves the same level of accuracy as the better-known strong Euler method in this task. Furthermore, by eliminating the expensive procedure of preparing Gaussian states, the quantum algorithm based on weak Euler scheme achieves drastically better efficiency than the one based on strong Euler scheme.
Our resource analysis suggests that option pricing under stochastic volatility is a promising application of quantum computers, and that our algorithms render the hardware requirement for reaching practical quantum advantage in financial applications less stringent than prior art.
\end{abstract}

\section{Introduction}
\label{sec:introduction}

Recently, there has been rising interest in leveraging quantum technologies to tackle the computational problems in finance better than what is possible in the classical world (see~\cite{herman2023quantum} for a survey). In particular, multiple quantum algorithms~\cite{Rebentrost2018quantum,Stamatopoulos2020option,Stamatopoulos2022towardsquantum,An2021quantumaccelerated,Chakrabarti2021thresholdquantum, alcazar2022quantum, bouland2023quantum,Stamatopoulos2024derivativepricing}
have been proposed to price financial derivatives and assess financial risks faster than their classical counterparts. In this work, we focus on the pricing of options, i.e., derivative contracts that grant buyers the right, but not the obligation, to buy or sell an underlying asset at an agreed-upon price and date or within a specific time frame. Determining the fair market value of an option is of paramount importance in quantitative finance. Yet it could be difficult to solve precisely due to the stochastic nature of financial markets. Classical Monte Carlo methods are often employed to draw random paths from which one can estimate the discounted average payoff of the option at expiration~\cite{shreve2004stochastic,shreve2005stochastic,Kloeden_Platen_1992}. By utilizing quantum techniques, especially amplitude estimation~\cite{Brassard_2002}, one can achieve quadratic speedup in this process and thus obtain more accurate estimate of the target quantity in shorter time, provided we have sufficiently powerful quantum hardware.

Nevertheless, previous works on quantum algorithms for derivative pricing have been restricted to the basic Black-Scholes model ~\cite{black1973pricing,merton1973theory} which assumes that the asset price follows a geometric Brownian motion (GBM) with constant drift and volatility. Even though this model is highly successful and easy to use, it has several limitations, including the assumption that the volatility of asset price remains constant over time. This assumption is unrealistic, because market volatility not only fluctuates over time but also appears to be random. To better reflect this reality, various stochastic volatility models have been proposed, and the \emph{Heston model}~\cite{heston1993closed} is one of the most popular among them. This model is described by two coupled stochastic differential equations (SDEs), one for the asset price and one for the asset volatility. Derivative pricing under the Heston model is more difficult than the one under the Black-Scholes model, as it often has no closed-form solutions and thus often involves expensive numerical simulation. In fact, finding efficient (classical) algorithms for this problem remains an active topic of research in (classical) computational finance until today~\cite{ beliaeva2010simple,  alos2012decomposition, CHIARELLA20122034,  he2018closed}.

In this work, we initiate the study of option pricing under the Heston model on a quantum computer. Specifically, we develop quantum algorithms for pricing two common exotic options -- Asian and barrier options -- under this stochastic volatility model by combining classical numerical methods for SDEs~\cite{Kloeden_Platen_1992} and quantum amplitude estimation. We first apply three numerical schemes, including strong Euler, weak Euler and order 2.0 Taylor methods, to this problem and empirically evaluate their performance. Here weak Euler method is a variant of the better-known strong Euler method (also known as the Euler-Maruyama method) in which Gaussian random variables are replaced with simpler discrete random variables. Surprisingly, despite its simplicity, weak Euler method achieves the same level of accuracy as the other two in our experiments. Furthermore, although
order 2.0 Taylor method is significantly more complicated than strong and weak Euler methods, we do not witness any advantage of the former over the latter in our experiments. 

In light of these findings, we then devise quantum algorithms for option pricing under the Heston model by combining strong and weak Euler schemes and iterative quantum amplitude estimation~\cite{grinko2021iterative}. This enables us to achieve quadratic speedup in the estimation of the discounted average payoff over traditional Monte Carlo methods. Furthermore, to better understand the practicality of our algorithms, we explicitly construct the circuits for the unitary operations in these algorithms, and estimate the costs and errors of these algorithms on four example instances under typical market conditions. Here we assume a Clifford + T gate set, and use T-count, T-depth and the number of logical qubits in the circuit as our cost metrics, since T gates typically dominate the computational costs~\cite{Nielsen_Chuang_2000,bravyi2005universal}. In developing the algorithm based on strong Euler scheme, we also optimize the circuits of~\cite{mcardle2022quantum} for preparing quantum states encoding Gaussian distributions and significantly reduce their T-counts and T-depths, which might be of independent interest. 

Under the settings delineated in Tables \ref{tab:case_study_heston_specs}, \ref{tab:case_study_option_specs} and \ref{tab:case_study_paramter_setting}, we obtain the resource estimates summarized in Table \ref{tab:case_study_resource_costs}. They indicate that the quantum algorithm based on weak Euler scheme is far more efficient than the one based on strong Euler scheme, mainly because it avoids the expensive procedure of preparing Gaussian states. Furthermore, our resource analysis suggests that option pricing under stochastic volatility could be a promising application of quantum computers, and the hardware requirement for reaching quantum advantage using our algorithms is less stringent than previous ones for similar tasks.

The remainder of the paper is organized as follows. In Section \ref{sec:preliminary}, we review the basics of mathematical finance and numerical methods for SDEs which are necessary for understanding this work. Then in Section \ref{sec:numerical_methods_for_Heston_model}, we apply three numerical methods to option pricing under the Heston model and empirically evaluate their performance. Next, in Section \ref{sec:option_pricing_heston_model_quantum_algorithms}, we develop our quantum algorithms for the same problem based on strong and weak Euler schemes and estimate their costs and errors on four example instances. Finally, Section \ref{sec:conclusion} concludes this paper and points out directions for future work.

\section{Preliminaries on mathematical finance and numerical methods for stochastic differential equations} 
\label{sec:preliminary}

In this section, we review the basics of mathematical finance (including stochastic volatility, exotic options and no-arbitrage pricing) as well as numerical methods for stochastic differential equations (SDEs). Readers who are familiar with these contents can skip to the next section.

\subsection{Mathematical finance}

\subsubsection{Stochastic volatility}
The seminal Black-Scholes model~\cite{black1973pricing,merton1973theory} assumes that the price $S_t$ of the asset follows a geometric Brownian motion with constant drift and volatility:
\begin{align}
    dS_t = r S_t dt + \sigma S_t dW_t,
\end{align}
under the risk-neutral measure, where $W_t$ is a Brownian motion under this measure \footnote{We can use Girsanov's theorem to convert a process $dS_t = \mu S_t dt + \sigma S_t dW_t$ under the real-world measure to an equivalent process $dS_t = r S_t dt + \sigma S_t d\tilde{W}_t$ under the risk-neutral measure. In this paper, we always work under the risk-neutral measure and hence drop the tilde notation for better readability. }, $r$ is the risk-free interest rate, and $\sigma$ is the volatility of the process. Then by Ito's lemma, one can prove that the asset price at any time $t$ is given by
\begin{align}
S_t= S_0 \exp{\lrb{r-\frac{\sigma^2}{2}}t + \sigma W_t}    
\end{align}
and obeys a log-normal distribution~\cite{Kloeden_Platen_1992}. This knowledge enables us to derive analytical formulas for the prices of certain options under the Black-Scholes model. For example, the value of an European call option with current underlying price $S$, strike price $K$ and time to expiration $\tau$ is given by:
\begin{align}
    C(S, \tau) = S N(d_+) - K e^{-r \tau} N(d_-),
\end{align}
where 
\begin{align}
d_{\pm} &= \frac{1}{\sigma \sqrt{\tau}}\lrsb{\ln{\frac{S}{K}} + \tau\lrb{r \pm \frac{\sigma^2}{2}}},
\end{align}
and $N(y)=\frac{1}{\sqrt{2\pi}}\int_{-\infty}^y e^{-z^2/2} dz$ is the cumulative distribution function (cdf) of a standard normal distribution. There also exist analytical formulas for the prices of barrier, digital and geometric Asian options under the Black-Scholes model. However, the pricing of other exotic options, e.g., arithmetic Asian options, and American options is more difficult and often requires numerical methods, e.g., Monte Carlo simulations, finite difference methods, and the binomial tree method~\cite{shreve2004stochastic,shreve2005stochastic}.

\begin{figure}[htbp!]
 \centering
 \subfigure[Asset price process]{
    \includegraphics[width=0.45\linewidth]{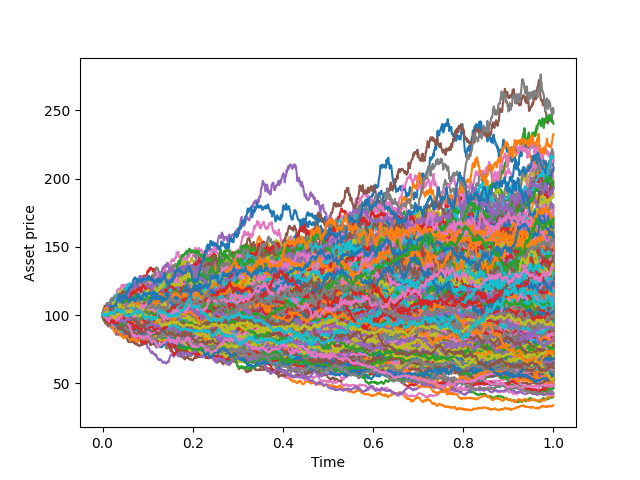}}
 \subfigure[Asset variance process]{
    \includegraphics[width=0.45\linewidth]{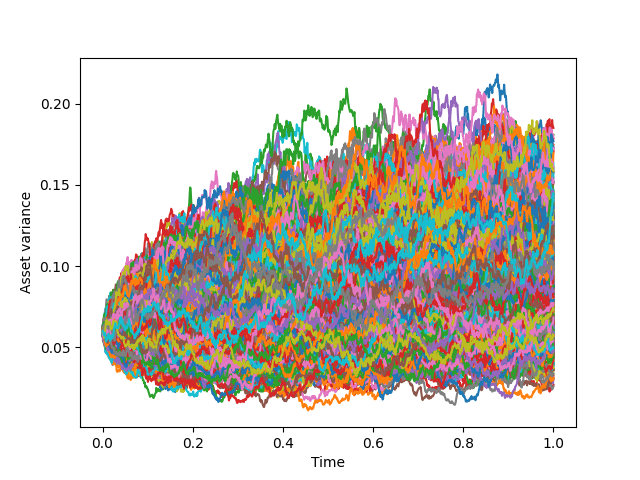}}         
    \caption{The evolution of asset price and its variance under the  Heston model, in the case $S_0=100$, $\nu_0=0.06$, $r=0.05$,  $\rho=-0.1$, $\kappa=2$, $\theta=0.09$ and $\xi=0.2$, from time $0$ to time $1$. Here this stochastic process is simulated by strong Euler method with $1024$ time steps, and $1000$ random paths are shown in the plots.}
    \label{fig:heston_model_example}
\end{figure}

Although the Black-Scholes model is easy to use, it relies on the unrealistic assumption that the volatility remains a constant throughout the process. In reality, the volatility of asset prices not only fluctuates over time but also exhibits random behaviors. To better capture this phenomenon, various stochastic volatility models were proposed, and the \textbf{Heston model}~\cite{heston1993closed} is one of the most popular among them. In this model, the instantaneous variance of the asset price follows a Cox–Ingersoll–Ross (CIR) process and exhibits mean reversion towards a long-term value. Formally, under the risk-neutral measure, the dynamics of the asset price $S_t$ is governed by a system of stochastic differential equations:
\begin{align}
    dS_t &= r S_t dt + \sqrt{\nu_t} S_t dW_t^1, \label{eq:st}\\
    d\nu_t &= \kappa (\theta - \nu_t) dt + \xi \sqrt{\nu_t} dW_t^2. \label{eq:nut}
\end{align}
where $r$ is the interest rate, $\theta$ is the long-term variance, $\kappa$ is the rate of mean reversion, and $\xi$ is the volatility of the price volatility $\sqrt{\nu_t}$. It is known that if the parameters obey the Feller condition 
\begin{align}
2\kappa \theta > \xi^2,    
\end{align}
then the process $\nu_t$ is strictly positive. Henceforth, we will assume that this condition holds. Moreover, $W^1_t$ and $W_t^2$ are two Brownian motions with correlation coefficient $\rho \in [-1, 1]$, i.e., $\E{W_t^1 W_t^2} = \rho t$ (or $dW_t^1 dW_t^2=\rho dt$) \footnote{Empirical evidence suggests that for stocks, $\rho$ is often negative, indicating a tendency for the asset price and its volatility to be negatively correlated. This reflects the common observation that volatility tends to increase as the asset price decreases, and vice versa, a phenomenon often referred to as the \emph{leverage effect} in financial markets.}. Figure \ref{fig:heston_model_example} illustrates an example of the evolution of asset price and its variance under the Heston model.

Let $R_t=\ln{S_t/S_0}$ be the log return at time $t$. We are often more interested in $R_t$ than $S_t$ itself. Using Ito's lemma, we can show that $R_t$ follows the stochastic process:
\begin{align}
dR_t = \lrb{r - \frac{\nu_t}{2}} dt + \sqrt{\nu_t} dW_t^1.
\label{eq:rt}
\end{align}

For technical reasons, it is more convenient to work with two independent Brownian motions instead of two correlated ones. Fortunately, we can find another Brownian motion $V_t$ such that $W^2_t$ and $V_t$ are independent and $W^1_t$ can be treated as a mixture of $W_t^2$ and $V_t$:
\begin{align}
dW^1_t = \rho dW^2_t + \sqrt{1-\rho^2} dV_t.
\end{align}
As a consequence, we can re-write $dR_t$ as
\begin{align}
dR_t = \lrb{r - \frac{\nu_t}{2}} dt  + \sqrt{\nu_t} \lrsb{\rho dW^2_t + \sqrt{1-\rho^2} dV_t}.
\label{eq:rt2}
\end{align}

In general, we do not have analytical expressions for the joint distribution of $R_t$ and $\nu_t$ (or even just the distribution of  $R_t$), although we know that the distribution of $\nu_t$ is related to a non-central chi-squared distribution and approaches a Gamma distribution as time becomes large. As a consequence, we need to resort to numerical methods, especially Monte Carlo simulations, to price many options under the Heston model~\footnote{\cite{heston1993closed} gives a simple way to price European vanilla options under the Heston model. It is, to our knowledge, unknown whether the method can be generalized to price more sophisticated options.}.

\subsubsection{Exotic options}
An option is called \emph{vanilla} or \emph{path-independent} if its payoff depends only on the price of the underlying asset at the expiration date; otherwise, it is called \emph{exotic} or \emph{path-dependent}, as its payoff depends on the path of the asset price between now and the expiration date. Moreover, an option can be American-style or European-style. The former allows holders to exercise their rights at any time before or on the expiration date, while the latter only allows holders to do so on the expiration date. We will consider European-style exotic options in this work.

Specifically, we will focus on the pricing of two exotic options - (arithmetic) Asian option and barrier option - under the Heston model. These options play a major role in quantitative finance, not only as intensively traded contracts on their own, but also as the building blocks of a large variety of structured products. Meanwhile, there are no known analytic formulas for their prices under the Heston model. Classically, one often needs to utilize Monte Carlo simulation to estimate their prices. Quantum computing offers the potential to solve these tasks faster.

\textbf{Asian options} are a type of exotic options where the payoff depends on the average price of the underlying asset over a certain period of time instead of its price at expiration.

Formally, let $A_T = \frac{1}{T} \int_0^T S_t dt$ be the average price between time $0$ and $T$. Then the payoff of an Asian call option with strike $K$ and expiration time $T$ is given by
\begin{align}
f(S_{0:T}) = \relu{A_T-K},
\end{align}
where we use the notation $S_{0:T}$ to denotes the path of asset price from time $0$ to $T$, and $x^+=\max(x, 0)$ is conventional. Similarly, the payoff of an Asian put option with strike $K$ and expiration time $T$ is given by
\begin{align}
f(S_{0:T}) = \relu{K-A_T}.
\end{align}

\textbf{Barrier options} are a type of exotic option in which payout depends on whether or not the underlying asset reaches or exceeds a predetermined price, i.e., a barrier. A barrier option can be \emph{knock-out} or \emph{knock-in}. While the former expires worthless if the underlying asset exceeds the barrier, the opposite is true for the latter, i.e. it has nonzero value only if the underlying asset reaches the same level. 

Formally, let $M_T = \max\limits_{0 \le t \le T} S_t$ and $N_T = \min\limits_{0 \le t \le T} S_t$ be the maximum and minimum prices of the asset between time $0$ and $T$, respectively. Then the payoff of an up-and-in, up-and-out, down-and-in, or down-and-out call option with strike $K$, barrier $B$ and expiration time $T$ is given by
\begin{align}
f(S_{0:T}) = 1_{M_T \ge B} \cdot \relu{S_T-K}, \\
f(S_{0:T}) = 1_{M_T \le B} \cdot \relu{S_T-K},\\
f(S_{0:T}) = 1_{N_T \le B} \cdot \relu{S_T-K}, \\
f(S_{0:T}) = 1_{N_T \ge B} \cdot \relu{S_T-K}, 
\end{align}
respectively. The payoff of a knock-in or knock-out put option can be defined similarly. We simply replace $\relu{S_T-K}$ with $\relu{K-S_T}$ in the above equations.

\subsubsection{No-arbitrage pricing}
Financial derivatives are priced based on the principle that no trader should make risk-free profit by buying the derivative and selling a replicating portfolio or vice versa. That is, the market should be free of arbitrage. Thus, the price of the derivative should be set at the same level as the value of the replicating portfolio. It can be shown that the value of the replicating portfolio equals the expected discounted payoff of the derivative under the risk-neutral measure.

Formally, assuming that the risk-free interest rate is a constant $r$, the no-arbitrage price of a derivative with payoff function $f$ and expiration time $T$ is given by
\begin{align}
   C(x, v) = e^{-rT} \E{f(S_{0:T}) | S_0=x, \nu_0=v},
\label{eq:no_arbitrage_price}   
\end{align}
where $\E{\cdot}$ is the expectation under the risk-neutral measure, $S_0=x$ and $\nu_0=v$ are the spot price and variance at time $0$. 

There are normally no closed-form expressions for the value $C(x, v)$ of the derivative, and one needs to compute it numerically. There are two common ways to achieve so~\cite{Kloeden_Platen_1992,shreve2004stochastic}: (1) use Monte Carlo simulation to generate paths of the underlying asset under the risk-neutral measure and use these paths to estimate the risk-neutral expected discounted payoff; or (2) reduce the problem to a partial differential equation and solve this PDE by numerical techniques, e.g., finite difference. Here we will follow the first approach and investigate how much advantage we can gain by quantizing it. We leave it as future work to explore the quantum version of the second approach.

\subsection{Numerical methods for SDEs}
\label{sec:methods_for_sdes}

In this subsection, we briefly review the common numerical techniques for solving stochastic differential equations. For more details, please refer to~\cite{Kloeden_Platen_1992}.

Consider a system of SDEs:
\begin{align}
    dX_t^k = a^k(X_t) dt + \sum_{j=1}^m b^{k, j}(X_t) dW^j_t,
    \label{eq:sde_standard_form}
\end{align}
for $k=1,2,\dots,d$, where each $a^k$ and $b^{k,j}$ are functions from $\R^d$ to $\R$, and $W^1_t, W^2_t, \dots, W^m_t$ are independent Brownian motions. \emph{Note that here and henceforth the superscripts $k$ and $j$ are not exponents, but labels for the dimensions of the system and the Brownian motions, respectively}.

This system of SDEs describes a $d$-dimensional Ito process driven by $m$ Brownian motions:
\begin{align}
    X_t = X_0 + \int_0^t a(X_s) ds + \sum_{j=1}^m \int_0^t b^j(X_s) dW^j_s,
    \label{eq:ito_process}
\end{align}
where $X_t=(X_t^1, X_t^2, \dots, X_t^d)$, $a=(a^1, a^2, \dots, a^d)$, and $b^j=(b^{1,j}, b^{2,j}, \dots, b^{d, j})$ for $j=1,2,\dots,m$. Note that the first integral in Eq.~\eqref{eq:ito_process} is a Riemann integral, while the others are Ito integrals. Our goal is to approximate the probability distribution of $X_t$ for given initial condition $X_0$ \footnote{In general, $X_0$ can be random. But in this work, it is sufficient to assume that $X_0$ is deterministic.}.

\subsubsection{Strong and weak convergence criteria}
To simulate the Ito process Eq.~\eqref{eq:ito_process}, we partition the time interval $[0, T]$ into $N$ subintervals $0=\tau_0<\tau_1<\dots<\tau_N=T$ and iteratively construct random variables $Y_0$, $Y_1$, $\dots$, $Y_N$ such that $Y_n$ is an approximation of $X_{\tau_n}$ for $n=0,1,\dots,N$. Here we only consider the simplest equidistant case, i.e., $\tau_n=nh$, where $h=T/N$, for each $n$.

For example, the Euler-Maruyama scheme recursively defines the $Y_n$'s as follows:
\begin{align}
Y_{n+1} = Y_n + a(Y_n) h + \sum_{j=1}^m b^j(Y_n) \Delta W^j_n,
\end{align}
for $n=0, 1, \dots, N-1$ with initial value $Y_0=X_0$ and $\Delta W^j_n = W^j_{\tau_{n+1}}-W^j_{\tau_n}$. Note that the $\Delta W^j_n$'s are independent and identically distributed (i.i.d.) Gaussian random variables with expected value zero and variance $h$.

In the stochastic setting, a time discrete approximation $Y_n$ of an Ito process $X_t$ can converge to the true process in two different senses, one strong and one weak.

\textbf{Strong convergence.} We say that $Y_N$ converges to $X_T$ in the strong sense with order $\gamma \in (0, \infty)$ if there exists a finite constant $K$  such that
\begin{align}
\E{\norm{X_T-Y_N}} \le K h^{\gamma}.
\end{align}
In this case, the trajectories, i.e., the sample paths, of the approximation are close to those of the Ito process. This implies that the probability distributions of $f(X_T)$ and $f(Y_N)$ are also close as long as the function $f$ satisfies some mild conditions. In fact, if $f$ is Lipschitz continuous, i.e., there exists a constant $L>0$ such that $|f(x)-f(y)|\le L\norm{x-y}$, for all $x,y \in \R^d$, then the strong convergence of $Y_N$ to $X_T$ immediately implies the convergence of $f(Y_N)$ to $f(X_T)$:
\begin{align}
    |\E{f(X_T)-f(Y_N)}|
    \le \E{|f(X_T)-f(Y_N)|}
    \le L \E{\norm{X_T-Y_N}}
    \le LK h^{\gamma}.
\end{align}

\textbf{Weak convergence.} While strong methods give faithful pathwise approximations of Ito processes, they are often expensive to implement. In many practical situations, one is only interested in the expectation of some function $f$ of the final value $X_T$ of the Ito process, and it suffices to just have a good approximation of the probability distribution of $f(X_T)$. This can be done more efficiently by weak methods.

We say that $Y_N$ converges to $X_T$ in the weak sense with order $\beta \in (0, \infty)$ if for every $g \in \mathcal{C}_P^{2(\beta+1)}$, there exists a finite constant $K$ such that
\begin{align}
    |\E{g(X_T)} - \E{g(Y_N)}| \le K h^{\beta}.
\end{align}
Here $\mathcal{C}_P^l$ is the set of $l$-times continuously differentiable functions from $\R^d$ to $\R$ whose partial derivatives up to order $l$ have polynomial growth. In particular, all polynomials belong to this space for every $l$.

Note that as $h \to 0$, $Y_N$ and $X_T$ will have similar moment properties, which implies that the distribution of $Y_N$ will converge to the distribution of $X_T$. However, it is difficult to bound the rate of this convergence in general.

\subsubsection{Strong approximations of Ito processes}
Next, we introduce some well-known first- and second-order strong and weak methods for SDEs. These methods rely on the following differential operators:
\begin{align}
L^0=\frac{\partial}{\partial t} + \sum_{k=1}^d a^k \frac{\partial}{\partial x^k} + \frac{1}{2} \sum_{k,l=1}^d \sum_{j=1}^m b^{k,j}b^{l,j} \frac{\partial^2}{\partial x^k \partial x^l},   
\label{eq:l0}
\end{align}
and
\begin{align}
L^j=\sum_{k=1}^d b^{k,j} \frac{\partial}{\partial x^k}, ~&\forall 1\le j\le m.
\label{eq:lj}
\end{align}

The Euler-Maruyama scheme is also known as the \textbf{strong Euler scheme} and has order $\gamma=0.5$. Here we repeat its update rules:
\begin{align}
Y^k_{n+1} = Y^k_n + a^k h + \sum_{j=1}^m b^{k, j} \Delta W^j_n,    
\end{align}
for $k=1,2,\dots,d$ and $n=0, 1, \dots, N-1$, where $\Delta W_n^j \sim {\cal N}(0, h)$ for each $j$, and the $\Delta W_n^j$'s are independent from each other.

The \textbf{Milstein scheme} has strong order $\gamma=1.0$ and has the following update rules:
\begin{align}
    Y_{n+1}^k = Y_n^k + a^k h + \sum_{j=1}^m b^{k,j}\Delta W_n^j 
    + \sum_{j_1, j_2=1}^m L^{j_1}b^{k,j_2} I^{j_1, j_2}_n,    
\end{align}
for $k=1,2,\dots, d$ and $n=0, 1, \dots, N-1$, where
\begin{align}
I^{j_1, j_2}_n = \int_{\tau_n}^{\tau_{n+1}} \int_{\tau_n}^{s_2} dW_{s_1}^{j_1} dW_{s_2}^{j_2}
\end{align}
for $j_1, j_2 \in \{1,2,\dots,m\}$. When $j_1=j_2=j$, we have
\begin{align}
I^{j, j}_n = \frac{1}{2}\lrsb{(\Delta W_n^j)^2 - h}.
\end{align}
But when $j_1 \neq j_2$, the double Ito integral $I^{j_1, j_2}_n$ cannot be easily expressed in terms of the Brownian increments $\Delta W_n^{j_1}$ and $\Delta W_n^{j_2}$. This term is avoided only when $m=1$, in which case the update rules become:
\begin{align}
    Y_{n+1}^k = Y_n^k + a^k h + b^k \Delta W_n + \frac{1}{2}\lrb{\sum_{l=1}^d b^l \frac{\partial b^k}{\partial x^l}}\lrsb{(\Delta W_n)^2-h}.
\end{align}
In this work, we need to simulate the Heston model which is a $2$-dimensional SDE system, and it will be quite expensive to implement the Milstein scheme for this system. So we will not develop a quantum algorithm based on this scheme here. But we remark that one could utilize this method to simulate other stochastic volatility models such as the constant elasticity of variance (CEV) model.

\subsubsection{Weak approximations of Ito processes}
In weak methods, we have more degrees of freedom than with strong methods and can replace the increments $\Delta W^j_n$'s and multiple Ito integrals (if any) in strong methods by more convenient random variables with similar moment properties.

Specifically, the (simplified) \textbf{weak Euler scheme} has order $\beta=1.0$ and has the following update rules:
\begin{align}
Y^k_{n+1} = Y^k_n + a^k h + \sum_{j=1}^m b^{k, j} \Delta \hat{W}^j_n,    
\end{align}
for $k=1,2,\dots,d$ and $n=0, 1, \dots, N-1$, where 
\begin{align}
\P{\Delta \hat{W}^j_n = \pm \sqrt{h}}=\frac{1}{2}    
\end{align}
for each $j$, and the $\Delta \hat{W}_n^j$'s are independent from each other. Note that $\Delta \hat{W}^j_n$ has mean $0$ and variance $h$ which are the same as those of $\Delta {W}^j_n$ in strong Euler scheme. More importantly, it is much easier to generate the discrete random variables $\Delta \hat{W}^j_n$'s than the continuous random variables $\Delta {W}^j_n$'s both classically and quantumly.

The (simplified) \textbf{order $2.0$ weak Taylor scheme} \cite{Kloeden_Platen_1992} has more complicated update rules:
\begin{align}
    Y_{n+1}^k =& Y_n^k + a^k h + \frac{1}{2} L^0 a^k h^2 \nonumber\\
    &+ \sum_{j=1}^m \lrcb{b^{k,j} + \frac{1}{2}h \lrb{L^0b^{k,j}+L^ja^k}}\Delta \hat{W}^j_n \nonumber\\
    &+\frac{1}{2} \sum_{j_1, j_2=1}^m L^{j_1}b^{k, j_2} \lrb{\Delta \hat{W}^{j_1}_n \Delta \hat{W}^{j_2}_n + V^{j_1, j_2}_n},
\end{align}
for $k=1,2,\dots,d$ and $n=0, 1, \dots, N-1$, where 
\begin{align}
\P{\Delta \hat{W}^j_n=\pm \sqrt{3h}}=\frac{1}{6},~&~
\P{\Delta \hat{W}^j_n=0}=\frac{2}{3},    
\end{align}
and
\begin{align}
\P{V^{j_1, j_2}_n = \pm h} &= \frac{1}{2}, ~\mathrm{for}~1 \le j_2<j_1,\\
V^{j_1, j_1}_n &= -h,\\
V^{j_1, j_2}_n &=-V^{j_2, j_1}_n, ~\mathrm{for}~j_1 < j_2 \le m,
\end{align}
for $j_1=1,2,\dots, m$. The $\Delta \hat{W}_n^j$'s and $V^{j_1, j_2}_n$'s for $1\le j_1<j_2\le m$ are independent from each other. Note again that it is fairly easy to generate the discrete random variables $\Delta \hat{W}_n^j$'s and $V^{j_1, j_2}_n$'s on both classical and quantum computers.

\section{Numerical methods for option pricing under the Heston model}
\label{sec:numerical_methods_for_Heston_model}
In this section, we develop three methods for simulating the Heston model which are based on the strong Euler, weak Euler and order 2.0 weak Taylor schemes for SDEs, respectively. Then we apply them to price Asian and barrier options under the same model, and empirically evaluate the performance of these methods in these tasks. Our simulation results suggest that the weak schemes perform as well as (or even better than)  strong Euler scheme. Furthermore, we do not see any advantage of order 2.0 weak Taylor scheme over the other schemes in our experiments \footnote{It is possible, however, that high-order methods do have advantages over first-order ones when one wants to estimate the option's value within very high accuracy.}. It turns out that the simplest weak Euler scheme is the most cost-effective choice among the three.

\subsection{Numerical simulation of the Heston model} 
\label{sec:simulation_for_heston_model}
To employ the numerical methods of Section \ref{sec:methods_for_sdes} to simulate the Heston model, we need to first convert the model into the standard form Eq.~\eqref{eq:sde_standard_form}. Let $X_t=(X_t^1, X_t^2)=(R_t, \nu_t)$ and $B_t=(B_t^1, B_t^2)=(W_t^2, V_t)$. Then we can rewrite the Heston model as the following 2-dimensional SDE system:
\begin{align}
dX_t^1 &= a^1(X_t) dt + b^{1,1}(X_t) dB_t^1 + b^{1,2}(X_t) dB_t^2, \\
dX_t^2 &= a^2(X_t) dt + b^{2,1}(X_t) dB_t^1 + b^{2,2}(X_t) dB_t^2,   
\end{align}
where 
\begin{align}
&a^1(X_t) = r - \frac{1}{2}\nu_t,~~~b^{1,1}(X_t) = \rho \sqrt{\nu_t}, ~~~b^{1,2}(X_t) = \sqrt{1-\rho^2} \sqrt{\nu_t},\\
&a^2(X_t) = \kappa (\theta - \nu_t),~~~b^{2,1}(X_t) = \xi \sqrt{\nu_t}, ~~~b^{2,2}(X_t) = 0.    
\end{align}
It follows that
\begin{align}
&\frac{\partial a^1}{\partial \nu_t}=-\frac{1}{2},\\
&\frac{\partial b^{1,1}}{\partial \nu_t}= \frac{1}{2}\rho \nu_t^{-1/2}, ~~~\frac{\partial^2 b^{1,1}}{\partial \nu_t^2}= -\frac{1}{4}\rho \nu_t^{-3/2},\\
&\frac{\partial b^{1,2}}{\partial \nu_t}= \frac{1}{2} \sqrt{1-\rho^2} \nu_t^{-1/2},~~~
\frac{\partial^2 b^{1,2}}{\partial \nu_t^2}= -\frac{1}{4} \sqrt{1-\rho^2} \nu_t^{-3/2},\\
&\frac{\partial a^2}{\partial \nu_t}=-\kappa,\\
&\frac{\partial b^{2,1}}{\partial \nu_t}= \frac{1}{2} \xi \nu_t^{-1/2},~~~
\frac{\partial^2 b^{2,1}}{\partial \nu_t^2}= -\frac{1}{4} \xi \nu_t^{-3/2}. 
\end{align}
All the other up-to-second-order partial derivatives of $a^k$ or $b^{k,j}$ with respect to $R_t$ or $\nu_t$ are zero.

Next, we define the differential operators $L^0$, $L^1$ and $L^2$ as in Eqs.~\eqref{eq:l0} and \eqref{eq:lj} (here we have two variables $x^1=R_t$ and $x^2=\nu_t$). Using the facts:
\begin{enumerate}
\item Applying $\frac{\partial}{\partial x^1}$ to $a^k$ or $b^{k,j}$ gives $0$;
\item Applying $\frac{\partial^2}{\partial x^i \partial x^j}$ to $a^k$ or $b^{k,l}$ yields $0$ unless $i=j=2$;
\item $b^{2,2}=0$,
\end{enumerate}
we can simplify $L^0$, $L^1$ and $L^2$ into:
\begin{align}
&L^0 = \dfrac{\partial}{\partial t} + a^2 \frac{\partial }{\partial x^2} + \frac{1}{2} (b^{2,1})^2  \frac{\partial^2 }{\partial (x^2)^2}, \\
&L^1 = b^{2,1} \frac{\partial}{\partial x^2}, \\
&L^2 = 0.    
\end{align}

Now we apply three numerical methods to the Heston model. In all of these methods, we start with $Y^1_0=0$ and $Y^2_0=\nu_0$. Then they will be updated differently:
\begin{enumerate}
\item Strong Euler method: The update rules are
\begin{align}
    Y^k_{n+1}= Y^k_n + a^k h + \sum_{j=1}^2 b^{k,j} \Delta W^j_n,
    \label{eq:strong_euler_for_heston}
\end{align} 
for $k=1,2$ and $n=0,1,\dots,N-1$, where $\Delta W^j_n \sim {\mathcal N}(0, h)$, and the $\Delta W^j_n$'s are independent from each other;
\item Weak Euler method: The update rules are
\begin{align}
    Y^k_{n+1}= Y^k_n + a^k h + \sum_{j=1}^2 b^{k,j} \Delta \hat{W}^j_n,
    \label{eq:weak_euler_for_heston}    
\end{align} 
for $k=1,2$ and $n=0,1,\dots,N-1$, where 
\begin{align}
\P{\Delta \hat{W}^j_n=\pm\sqrt{h}}=\frac{1}{2},    
\end{align}
and the $\Delta \hat{W}^j_n$'s are independent from each other;
\item Order 2.0 weak Taylor method: The update rules are
\begin{align}
Y^k_{n+1}= &Y^k_n + a^k h + \frac{1}{2} L^0a^k h^2 
+\sum_{j=1}^2 \lrsb{b^{k,j} + \frac{1}{2}h(L^0b^{k,j}+L^ja^k)}\Delta \hat{W}^j_n \nonumber\\
&+ \frac{1}{2} \sum_{j_1,j_2=1}^2 L^{j_1} b^{k, j_2}\lrb{\Delta \hat{W}^{j_1}_n\Delta \hat{W}^{j_2}_n+V^{j_1,j_2}_n}\\
=&Y^k_n + a^k h + \sum_{j=1}^2 b^{k,j} \Delta \hat{W}^j_n
+ \frac{1}{2} L^0a^k h^2 
+\frac{1}{2}\sum_{j=1}^2 h(L^0b^{k,j}+L^j a^k) \Delta \hat{W}^j_n \nonumber \\
&+ \frac{1}{2} \sum_{j_2=1}^2 L^1 b^{k, j_2}\lrb{\Delta \hat{W}^{1}_n \Delta \hat{W}^{j_2}_n+V^{1,j_2}_n},    
\end{align} 
for $k=1,2$ and $n=0,1,\dots,N-1$, where the second step follows from $L^2=0$, and 
\begin{align}
\P{\Delta \hat{W}^j_n=\pm\sqrt{3h}}=\frac{1}{6}, ~&~   
\P{\Delta \hat{W}^j_n=0}=\frac{1}{3},
\end{align}
and $V^{1,1}_n=V^{2,2}_n=-h$, $V^{1,2}_n=-V^{2,1}_n$,
$\P{V^{1,2}_n=\pm h}=1/2$. The $\Delta \hat{W}^j_n$'s and $V^{1, 2}_n$'s are independent from each other. We can further simplify the update rules into:
\begin{align}
Y^1_{n+1}=
&Y^1_n + a^1 h + b^{1,1} \Delta \hat{W}^1_n + b^{1,2} \Delta \hat{W}^2_n
+ \frac{1}{2} L^0a^1 h^2  \nonumber \\
&+ \frac{1}{2} h \lrsb{(L^0b^{1,1}+L^1 a^1) \Delta \hat{W}^1_n +L^0b^{1,2} \Delta \hat{W}^2_n} \nonumber \\
&+ \frac{1}{2} L^1 b^{1, 1}\lrsb{(\Delta \hat{W}^{1}_n)^2-h}
+ \frac{1}{2} L^1 b^{1, 2} \lrb{\Delta \hat{W}^{1}_n \Delta \hat{W}^{2}_n+V^{1,2}_n},    
\end{align}
\begin{align}
Y^2_{n+1}= &Y^2_n + a^2 h + b^{2,1} \Delta \hat{W}^1_n
+ \frac{1}{2} L^0a^2 h^2 \nonumber \\
&+\frac{1}{2} h(L^0b^{2,1}+L^1 a^2) \Delta \hat{W}^1_n 
+ \frac{1}{2} L^1 b^{2, 1}\lrsb{(\Delta \hat{W}^{1}_n)^2-h}.    
\end{align}
\end{enumerate}

Given an option with payoff function $f$, we can use each of the above methods to generate a random path $\vec Y = (Y_0, Y_1, Y_2, \dots, Y_N)$, and translate them into the price path $\vec S=(S_0, S_{h}, S_{2h}, \dots, S_T)$, and then compute the payoff $f(\vec S)$ of this path. By repeating this procedure sufficiently many times and computing the mean of the collected samples, we can estimate the value of this option under the Heston model within desired accuracy.

\subsection{Empirical results}
\label{sec:classical_simulation_results}
Next, we apply the methods in Section \ref{sec:simulation_for_heston_model} to several instances of option pricing under the Heston model to test their performance in practice. Specifically, we consider four settings of the Heston model which are listed in Table \ref{tab:classical_simulation_heston_specs}
and estimate the values of an Asian option and a barrier option under each setting. The specifications of these options are given in Table \ref{tab:classical_simulation_option_specs}.

\begin{table}[htbp]
    \centering
\begin{tabular}{ |c|c|c|c|c|c|c|c| } 
 \hline
Setting & $r$ & $\rho$ & $\kappa$ & $\theta$ & $\xi$ & $S_0$ & $\nu_0$ 
\\
 \hline 
No. 1 & $0.03$ &  $-0.1$ & $2$ & $0.12$ 
& $0.3$ & $100$ & $0.1$ \\ 
 \hline
No. 2 & $0.03$ &  $0$ & $2$ & $0.03$ 
& $0.2$ & $100$ & $0.03$ \\ 
 \hline
No. 3 & $0.05$ &  $-0.1$ & $2$ & $0.09$ 
& $0.2$ & $100$ & $0.06$ \\ 
 \hline
No. 4 & $0.05$ &  $-0.1$ & $2$ & $0.04$ 
& $0.2$ & $100$ & $0.05$ \\ 
 \hline 
\end{tabular}
\caption{The settings of the Heston model studied in Section \ref{sec:classical_simulation_results}. Each setting includes the model parameters $r$, $\rho$, $\kappa$, $\theta$, $\xi$ and initial condition $S_0$ and $\nu_0$.}
\label{tab:classical_simulation_heston_specs}
\end{table}

\begin{table}[htbp]
    \centering
\begin{tabular}{ |c|c|c|c|c|c| } 
 \hline
Instance & Setting &  Option type & $K$ & $B$ & $T$ 
\\
 \hline 
No. 1 & No. 1 & Asian~call & $90$ & N/A & $1.0$ \\ 
 \hline
No. 2 & No. 1 & Down-and-out put & $110$ & $70$ & $1.0$ \\ 
 \hline
No. 3 & No. 2 & Asian~put & $110$ & N/A & $1.0$ \\ 
 \hline
No. 4 & No. 2 & Up-and-out call & $90$ & $130$ & $1.0$ \\ 
 \hline 
No. 5 & No. 3 & Asian call & $90$ & N/A & $1.0$ \\ 
 \hline
No. 6 & No. 3 & Down-and-in put & $110$ & $80$ & $1.0$ \\ 
 \hline 
No. 7 & No. 4 & Asian put & $110$ & N/A & $1.0$ \\ 
 \hline 
No. 8 & No. 4 & Up-and-in call & $90$ & $120$ & $1.0$ \\ 
 \hline
\end{tabular}
\caption{The problem instances studied in Section \ref{sec:classical_simulation_results}. The specification of each instance includes the setting of the Heston model and the type, strike $K$, barrier $B$ (if needed) and expiration time $T$ of the option.}
\label{tab:classical_simulation_option_specs}
\end{table}

In each experiment, we utilize the strong Euler method, weak Euler method and weak Taylor method (order 2.0) to generate $5$ million random paths, and compute the mean and standard deviation of the payoffs of the option for the collected paths in each case. We vary the number of time steps $N$ in powers of two starting from $2$, $4$, $8$, $\dots$ up to $1024$, and observe how fast the estimated value converge to the true value as $N$ grows large for each method \footnote{We use the result of the strong Euler method with $N=1024$ time steps as the best estimate of the true value, as this method is known for its robust performance in producing pathwise accurate simulations in a variety of stochastic models and has been widely employed as a benchmark in the numerical solution of SDEs \cite{Kloeden_Platen_1992, higham2001analgorithmic, andersen2001extended, Glasserman2003montecarlo, milstein2004stochastic, giles2008multilevel, lord2010comparison}. Although this estimate contains a bias due to time discretization, the bias is small enough to be negligible for our purposes.}.

Figures \ref{fig:setting1_asian_option}, \ref{fig:setting1_barrier_option}, \ref{fig:setting2_asian_option}, \ref{fig:setting2_barrier_option}, \ref{fig:setting3_asian_option}, \ref{fig:setting3_barrier_option}, \ref{fig:setting4_asian_option} and \ref{fig:setting4_barrier_option} illustrate the results of our experiments. In each plot, there are three curves with error bands for the strong Euler method, weak Euler method and weak Taylor method (order 2.0) respectively. The upper and lower bands are 3 standard deviations +/- from the mean of the 5 million samples. Furthermore, there is also a vertical line with error bands which indicates our best estimate of the true value which is obtained from the $N=1024$ strong Euler method. The upper and lower bands are again 3 standard deviations +/- from the mean of the 5 million samples in that case.

\begin{minipage}{\linewidth}
  \centering
  \begin{minipage}{0.45\linewidth}
      \begin{figure}[H]
          \includegraphics[width=\linewidth]{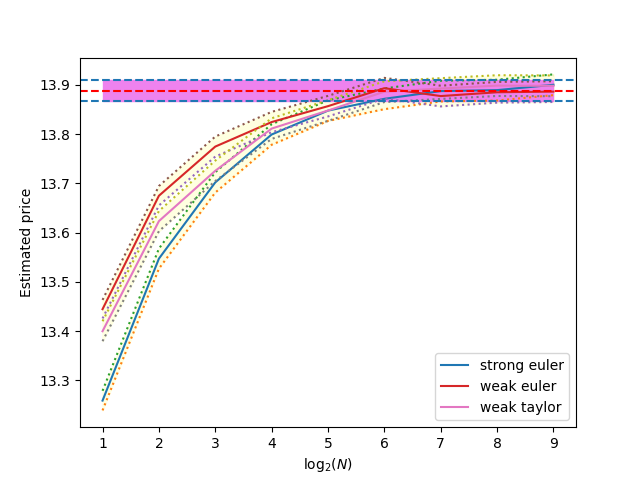}
          \caption{The estimated value of an Asian call with strike $K=90$ and expiration time $T=1$, in the case $S_0=100$, 
          $\nu_0=0.1$, $r=0.03$, $\rho=-0.1$, $\kappa=2$, $\theta=0.12$, $\xi=0.3$, in different methods with varying number of time steps. 
          }
          \label{fig:setting1_asian_option}
      \end{figure}
  \end{minipage}
  \hspace{0.05\linewidth}
  \begin{minipage}{0.45\linewidth}
      \begin{figure}[H]
          \includegraphics[width=\linewidth]{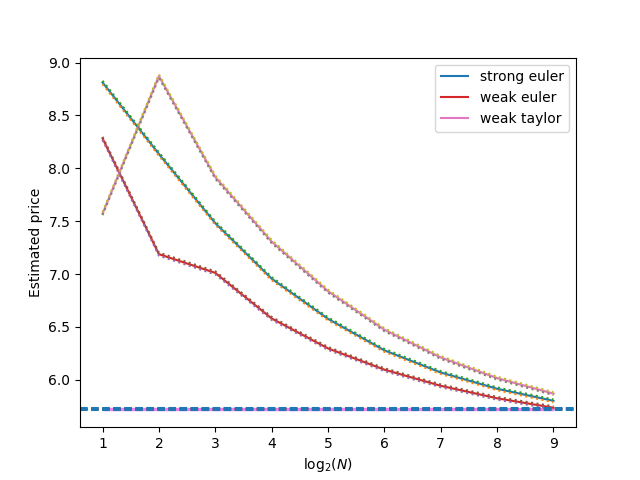}
          \caption{The estimated value of a down-and-out put  with strike $K=110$, barrier $B=70$ and expiration time $T=1$, in the case $S_0=100$, 
          $\nu_0=0.1$, $r=0.03$, $\rho=-0.1$, $\kappa=2$, $\theta=0.12$, $\xi=0.3$, in different methods with varying number of time steps. }
          \label{fig:setting1_barrier_option}
      \end{figure}
  \end{minipage}
\end{minipage}

\begin{minipage}{\linewidth}
  \centering
  \begin{minipage}{0.45\linewidth}
      \begin{figure}[H]
          \includegraphics[width=\linewidth]{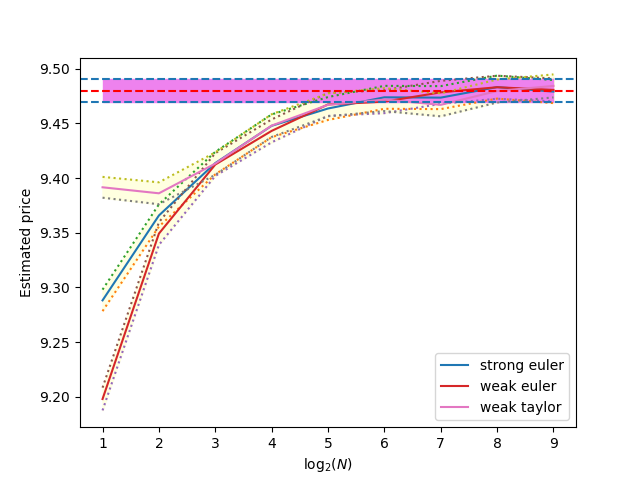}
          \caption{The estimated value of an Asian put with strike $K=110$ and expiration time $T=1$, in the case $S_0=100$, $\nu_0=0.03$, $r=0.03$, $\rho=0$, $\kappa=2$, $\theta=0.03$, $\xi=0.2$, in different methods with varying number of time steps.}
          \label{fig:setting2_asian_option}          
      \end{figure}
  \end{minipage}
  \hspace{0.05\linewidth}
  \begin{minipage}{0.45\linewidth}
      \begin{figure}[H]
          \includegraphics[width=\linewidth]{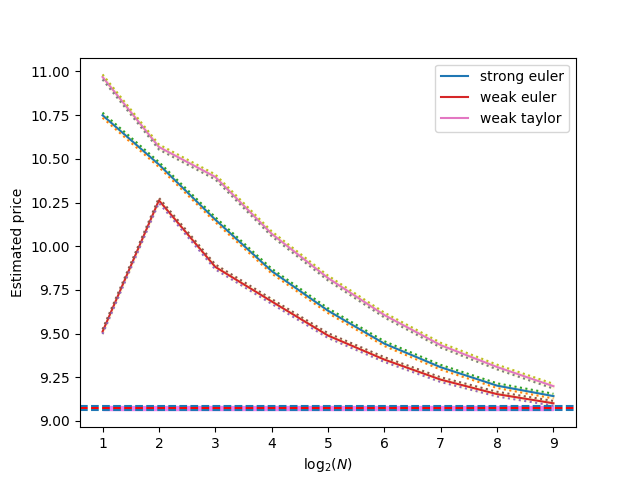}
          \caption{The estimated value of an up-and-out call with strike $K=90$, barrier $B=130$ and expiration time $T=1$, in the case $S_0=100$, $\nu_0=0.03$, $r=0.03$, $\rho=0$, $\kappa=2$, $\theta=0.03$, $\xi=0.2$, in different methods with varying number of time steps.          
          }
          \label{fig:setting2_barrier_option}          
      \end{figure}
  \end{minipage}
\end{minipage}

\begin{minipage}{\linewidth}
  \centering
  \begin{minipage}{0.45\linewidth}
      \begin{figure}[H]
          \includegraphics[width=\linewidth]{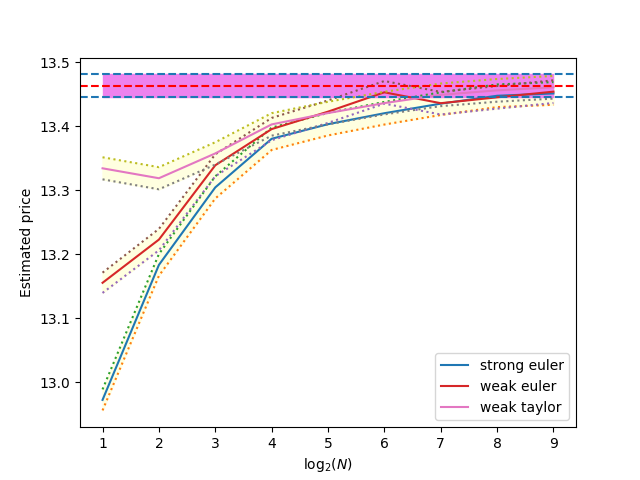}
          \caption{The estimated value of an Asian call with strike $K=90$ and expiration time $T=1$, in the case $S_0=100$, $\nu_0=0.06$, $r=0.05$, $\rho=-0.1$, $\kappa=2$, $\theta=0.09$, $\xi=0.2$, in different methods with varying number of time steps.}
          \label{fig:setting3_asian_option}          
      \end{figure}
  \end{minipage}
  \hspace{0.05\linewidth}
  \begin{minipage}{0.45\linewidth}
      \begin{figure}[H]
          \includegraphics[width=\linewidth]{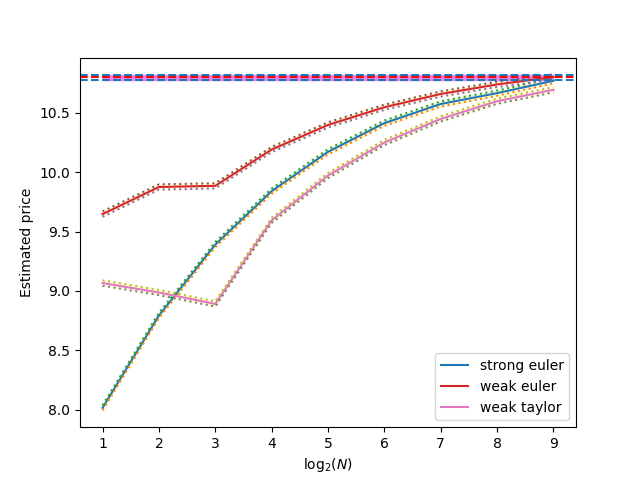}
          \caption{The estimated value of a down-and-in put with strike $K=110$, barrier $B=80$ and expiration time $T=1$, in the case $S_0=100$, $\nu_0=0.06$, $r=0.05$,
          $\rho=-0.1$, $\kappa=2$, $\theta=0.09$, $\xi=0.2$, in different methods with varying number of time steps.}
          \label{fig:setting3_barrier_option}          
      \end{figure}
  \end{minipage}
\end{minipage}

\begin{minipage}{\linewidth}
  \centering
  \begin{minipage}{0.45\linewidth}
      \begin{figure}[H]
          \includegraphics[width=\linewidth]{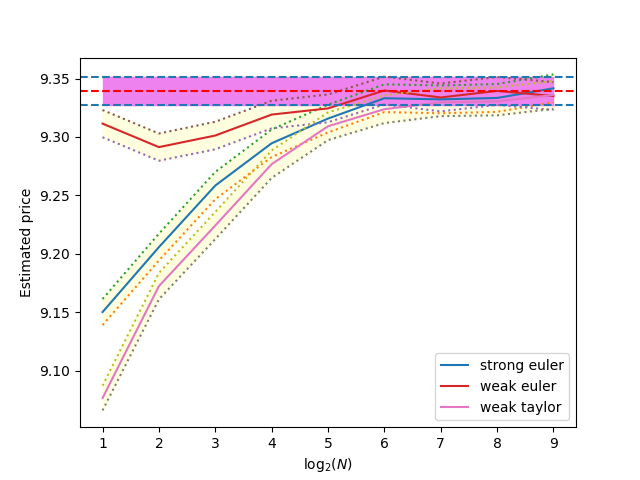}
          \caption{The estimated value of an Asian put with strike $K=110$ and expiration time $T=1$, in the case $S_0=100$, $\nu_0=0.05$, $r=0.05$, $\rho=-0.1$, $\kappa=2$, $\theta=0.04$, $\xi=0.2$, in different methods with varying number of time steps.}
          \label{fig:setting4_asian_option}          
      \end{figure}
  \end{minipage}
  \hspace{0.05\linewidth}
  \begin{minipage}{0.45\linewidth}
      \begin{figure}[H]
          \includegraphics[width=\linewidth]{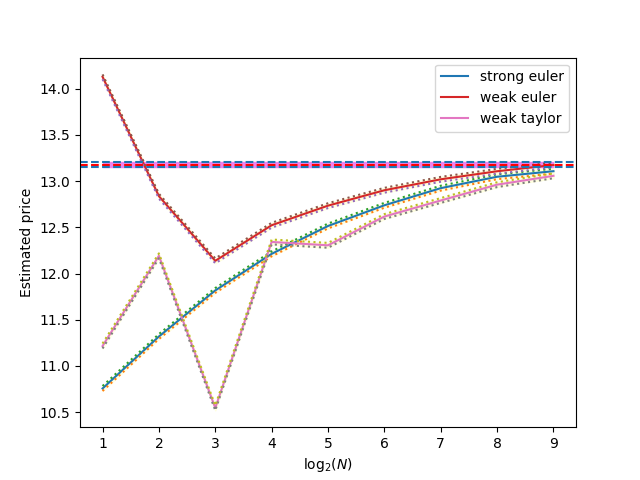}
          \caption{The estimated value of an up-and-in call with strike $K=90$, barrier $B=120$ and expiration time $T=1$, in the case $S_0=100$, $\nu_0=0.05$, $r=0.05$, $\rho=-0.1$, $\kappa=2$, $\theta=0.04$, $\xi=0.2$, in different methods with varying number of time steps.}
          \label{fig:setting4_barrier_option}          
      \end{figure}
  \end{minipage}
\end{minipage}

In all of the above experiments, the strong Euler, weak Euler, and order 2.0 weak Taylor methods exhibit similar performance. Note that for Asian and barrier options, their payoff functions depend on the path of the asset price between time $0$ and time $T$. This differs from the traditional setting in numerical SDEs where the payoff function depends only on the asset price at time $T$. Thus, the strong Euler, weak Euler, and weak Taylor (order 2.0) methods might exhibit different convergence behaviors in this path-dependent setting, and there is currently no theoretical proof of their convergence orders in our specific tasks. Nevertheless, our simulation results indicate that the accuracy of these methods improves rapidly as $N$ increases under typical market conditions.

Interestingly, despite its simplicity, the weak Euler method often performs best among the three methods (in the sense that its output converges to the ideal value faster than the others) in our experiments. This suggests that for pricing Asian and barrier options under the Heston model, it is not crucial to have a faithful pathwise approximation of the Ito process. Instead, it suffices to replace the real probability distribution of the price path with one that has similar moment properties.

Moreover, we do not observe any significant advantage of the order 2.0 weak Taylor method over the strong and weak Euler methods. Note that we do not aim to estimate the option's value with extremely high accuracy, e.g., with an error of less than $0.001$ in the above settings. It is possible that higher-order methods start to show their advantages over lower-order ones as we enter high-precision regimes, as predicted by existing literature, where higher-order Taylor methods tend to outperform others in high-precision scenarios or when very fine time discretization is required \cite{Kloeden_Platen_1992, Glasserman2003montecarlo, milstein2004stochastic}.

Our experiments also suggest that, in general, Asian options are easier to price than barrier options under the Heston model. This is reasonable because the payoffs of Asian options depend on the average price between time $0$ and $T$, making them more robust against small deviations from the ideal paths in the simulations. In contrast, the payoffs of barrier options crucially depend on whether the price reaches a certain threshold during the process, and this event is more sensitive to the aforementioned deviations. Therefore, a more accurate simulation of the Ito process (which requires a larger $N$) is needed to achieve the same accuracy for barrier options as for Asian options.

\section{Option pricing under the Heston model on quantum computers} 
\label{sec:option_pricing_heston_model_quantum_algorithms}
In this section, we describe our quantum algorithms for pricing Asian and barrier options under the Heston model, and estimate how many resources, i.e., T-count, T-depth and number of qubits, they require to achieve certain accuracy in their outputs on certain input instances. These algorithms are based on combining quantum amplitude estimation ~\cite{Brassard_2002,grinko2021iterative} and strong/weak Euler schemes for SDEs, respectively. Our resource estimation indicates that the quantum algorithms based on weak Euler scheme are far more efficient than the ones based on strong Euler scheme.

\subsection{Basic strategy}
Given a specification of the Heston model, including the parameters $r$, $\kappa$, $\theta$, $\xi$, $\rho$ and initial condition $S_0$, $\nu_0$, we first choose a numerical scheme for simulating its dynamics. Here we use either strong Euler or weak Euler methods (see Eqs.~\eqref{eq:strong_euler_for_heston} and~\eqref{eq:weak_euler_for_heston}). (As shown in Section \ref{sec:classical_simulation_results}, order 2.0 weak Taylor method does not seem to have an advantage over strong and weak Euler methods, despite being significantly more complicated. So we will not develop quantum algorithms for option pricing based on this numerical scheme.)

Let $Y^1_0=0$ and $Y^2_0=\nu_0$, and let $Y^k_n$ be defined recursively as Eq.~\eqref{eq:strong_euler_for_heston} or Eq.~\eqref{eq:weak_euler_for_heston} for $k=1,2$ and $n=1,\dots,N$. Then  $Y^1_n$ and $Y^2_n$ approximate the log return and asset variance at time $nh$ in the ideal process, respectively, i.e., 
\begin{align}
Y^1_n \approx R_{nh}=\ln{S_{nh}/S_0}, ~&~Y^2_n \approx \nu_{nh},    
\end{align}
provided that $h$ is sufficiently small. 

The input to our problem also includes the specification of an Asian or barrier option. Suppose it has payoff function $f$ and expiration time $T$. Without loss of generality, we rescale $f$ so that it takes values in $[0, 1]$. This facilitates the construction of quantum circuits for option pricing, because we want to encode the payoffs of the option for possible paths into quantum amplitudes and they cannot exceed $1$.

Let $\vec S=(S_0, S_h, \dots, S_T)$ and $\vec R=(R_0, R_h, \dots, R_T)$ be the paths of asset price and log return in the ideal process, respectively (note that $S_0$ is fixed and $R_0=0$). Then the no-arbitrage price of the option is $e^{-rT}\E{f(\vec S)}$. In fact, it is more convenient to work with log returns than prices. So we re-define the payoff in terms of $\vec R$, i.e., there exists a function $\tilde{f}$ such that
\begin{align}
\tilde{f}(\vec R)=f(\vec S).
\end{align}
Next, let $\vec Y^k=(Y_0^k, Y_1^k, \dots, Y_N^k)$ be the path of $Y^k_n$ in the simulated process, for $k=1,2$. Then we get
\begin{align}
e^{-rT}\E{\tilde{f}(\vec R)} \approx e^{-rT}\E{\tilde{f}(\vec Y^1)}.    
\end{align}

Our goal is to estimate $\E{\tilde{f}(\vec Y^1)}$ for the chosen numerical scheme. Note that the random path in the simulated process can be parameterized by a random variable $\omega$ drawn from a probability distribution $p(\omega)$ on a set $\Omega$. For example, in strong Euler method, we define $\omega=(\Delta W_n^j/\sqrt{h}:~0\le n\le N-1, 1\le j \le 2)$, and $p$ is the $2N$-dimensional standard normal distribution on $\Omega=\R^{2N}$. Meanwhile, in weak Euler method, we define
$\omega=(\Delta \hat{W}_n^j/\sqrt{h}:~0\le n\le N-1, 1\le j \le 2)$, and $p$ is the uniform distribution on $\Omega=\lrcb{1, -1}^{2N}$. Then each $Y^k_n$ is a deterministic function of $\omega$. Thus, our task is reduced to estimating 
\begin{align}
\E{\tilde{f}(\vec Y^1(\omega))} = \sum_{\omega \in \Omega} p(\omega) \tilde{f}(\vec Y^1(\omega)).
\end{align}
Here we abuse the notation for strong Euler method (which should have  integration instead of summation) without causing confusion.

Classically, one often estimates $\E{\tilde{f}(\vec Y^1(\omega))}$ by Monte Carlo methods. That is, one simulates the stochastic process, and generates $M$ random paths, and computes the mean of the payoffs of the option for these paths. This leads to an unbiased estimator of the target quantity whose standard deviation scales as $O(1/\sqrt{M})$, which means that we need to collect $O(1/\epsilon^2)$ samples to reach precision $\epsilon$ in our estimate.

Quantumly, we can utilize the technique of amplitude estimation~\cite{Brassard_2002} to achieve quadratic speedup in the estimation of the same quantity. Specifically, suppose $\A$ is a unitary operation such that
\begin{align}
    \A\ket{0}\ket{0^m} = \sqrt{a} \ket{0} \ket{\psi_0} + \sqrt{1-a} \ket{1} \ket{\psi_1},
    \label{eq:circuit_a}
\end{align}
where $a=\E{\tilde{f}(\vec Y^1(\omega))}$ is the target quantity, and $\ket{\psi_0}$ and $\ket{\psi_1}$ are some normalized $m$-qubit states. Quantum amplitude estimation (QAE) can estimate $a$ within additive error $\epsilon$ with probability $1-O(1)$ by making $O(1/\epsilon)$ uses of the unitary operation
\begin{align}
    Q=\A R_0 \A^{-1} S_0,
\end{align}
where $R_0=I_{m+1}-2\ket{0^{m+1}}\bra{0^{m+1}}$ \footnote{Following  convention, we use $I_n$ to denote the $n$-qubit identity operator.} and 
$S_0 = I_{m+1}-2 \ket{0}\bra{0}\otimes I_m$ are reflection operators.
QAE consists of running quantum phase estimation (QPE) on the unitary operator $Q$ with initial state $\A\ket{0^{m+1}}$ and inferring $a$ from the measurement outcomes of this circuit. 

A drawback of standard Quantum Amplitude Estimation (QAE) is that it requires performing a quantum Fourier transform on multiple ancilla qubits, which is expensive to implement. To overcome this limitation, several variants of QAE have been proposed \cite{aaronson2020quantum, suzuki2020amplitude, wang2021minimizing, grinko2021iterative, GiurgicaTiron2022lowdepthalgorithms, Plekhanov2022variationalquantum} that do not involve the use of ancilla qubits or quantum Fourier transform. In particular, the iterative quantum amplitude estimation (IQAE) of~\cite{grinko2021iterative} has shown excellent empirical performance. Therefore, we will utilize this algorithm for our task. This algorithm can estimate $a$ within additive error $\epsilon$ with probability $1-\delta$ by making
\begin{align}
    N_{\rm oracle} \le \frac{1.4}{\epsilon} \ln{\frac{2}{\delta} \operatorname{log}_2\lrb{\frac{\pi}{4\epsilon}}}
    \label{eq:n_oracle}
\end{align}
uses of $Q$, for any $\epsilon>0$ and $\delta \in (0,1)$.

Now our task is reduced to building a quantum circuit $\A$ that satisfies the condition Eq.~\eqref{eq:circuit_a}. This can be achieved by concatenating three unitary operations $U_1$, $U_2$ and $U_3$, where $U_1$ simulates the Heston model and creates a proper superposition of the log-return paths, $U_2$ calculates the payoffs of the option for these paths simultaneously, and $U_3$ encodes the payoffs into quantum amplitudes. Formally, we define
\begin{align}
    \A = U_3 U_2 U_1 
\end{align}
where
\begin{align}
    &U_1 \ket{00\dots 0}_{A}\ket{00\dots 0}_{F}=\sum_{\omega \in \Omega} \sqrt{p(\omega)}\ket{\vec Y^1(\omega)}_{A} \ket{\Phi(\omega)}_F, \label{eq:def_u1}&\\
    &U_2 \ket{\vec Y^1(\omega)}_{A}\ket{0}_C\ket{00\dots 0}_G=\ket{\vec Y^1(\omega)}_{A}\ket{\tilde{f}(\vec Y^1(\omega))}_C \ket{\Psi(\omega)}_G, & ~\forall \omega \in \Omega,
    \label{eq:def_u2}\\
    &U_3 \ket{z}_C \ket{0}_D \ket{00\dots 0}_H=
    \ket{z}_C \lrsb{\sqrt{z}\ket{0}_D
    +\sqrt{1-z}\ket{1}_D} \ket{\Xi(z)}_H,  & ~\forall z \in [0, 1].
    \label{eq:def_u3}
\end{align}
Here $\ket{\Phi(\omega)}$ and $\ket{\Psi(\omega)}$ are some normalized states depending on $\omega$, and $\ket{\Xi(z)}$ is a normalized state depending on $z$. One can verify that 
\begin{align}
\norm{(I_{ACFGH}\otimes\bra{0}_D)\A \ket{00\dots 0}_{ACDFGH}}^2=\E{\tilde{f}(\vec Y^1(\omega))},
\end{align}
as desired.

In the next three subsections, we will construct the circuits for $U_1$, $U_2$ and $U_3$, respectively. Each circuit is composed of the basic modules introduced in Appendices \ref{apd:fixed_point_arithmetics}, \ref{apd:opt_circuit_usin} and \ref{apd:prep_gaussian_states}, including
various fixed-point quantum arithmetic operations, a block-encoding $U_{\rm sin}$ of the sine function, a unitary operation $U_{\rm gauss}$ for preparing Gaussian states (for strong Euler method only), as well as Clifford gates. The costs of these modules are also given in these appendices. Here we assume Clifford + T gate set, and measure the cost of an operation in terms of the T-count, T-depth and number of ancilla qubits in the circuit for implementing this operation. Every real number will be represented by $n$ qubits in the circuits (see Appendix \ref{apd:fixed_point_arithmetics} for more details). To avoid distraction from our main messages, we will ignore the errors in quantum arithmetic operations, $U_{\rm sin}$ and $U_{\rm gauss}$ for now, and postpone the error analysis to Section \ref{sec:error_analysis}.

\subsection{Implementing the operator $U_1$}
\label{sec:implement_U1}
To facilitate the implementation of $U_1$, we re-write the update rules of strong Euler method as
\begin{align}
Y^1_{j+1} &= Y^1_j - Y^2_j \cdot \textcolor{blue}{\frac{h}{2}}
+ \textcolor{orange}{\alpha_j} \cdot \sqrt{Y_j^2} \cdot \textcolor{blue}{ \rho \sqrt{h}}
+ \textcolor{orange}{\beta_j} \cdot \sqrt{Y_j^2} \cdot \textcolor{blue}{ \sqrt{h(1-\rho^2})}
+ \textcolor{blue}{r h}, \label{eq:strong_euler_update_rule1}\\
Y^2_{j+1} &= Y^2_j  - Y^2_j \cdot \textcolor{blue}{\kappa h} 
+ \textcolor{orange}{\alpha_j} \cdot \sqrt{Y^2_j} \cdot \textcolor{blue}{ \xi \sqrt{h}}
+ \textcolor{blue}{\kappa \theta h},\label{eq:strong_euler_update_rule2}
\end{align}
where $\alpha_j, \beta_j \sim N(0, 1)$, and re-write the update rules of weak Euler method as
\begin{align}
Y^1_{j+1} &= Y^1_j - Y^2_j \cdot \textcolor{blue}{\frac{h}{2}}
+ \textcolor{orange}{(-1)^{\alpha_j}} \cdot \sqrt{Y^2_j} \cdot \textcolor{blue}{ \rho \sqrt{h}}
+ \textcolor{orange}{(-1)^{\beta_j}} \cdot \sqrt{Y^2_j} \cdot \textcolor{blue}{ \sqrt{h(1-\rho^2})}
+ \textcolor{blue}{r h}, \label{eq:weak_euler_update_rule1}\\
Y^2_{j+1} &= Y^2_j  - Y^2_j \cdot \textcolor{blue}{\kappa h} 
+ \textcolor{orange}{(-1)^{\alpha_j}} \cdot \sqrt{Y^2_j} \cdot \textcolor{blue}{ \xi \sqrt{h}}
+ \textcolor{blue}{\kappa \theta h},\label{eq:weak_euler_update_rule2}
\end{align}
where $\alpha_j, \beta_j=0$ or $1$ with probability $1/2$. Note that the blue terms in the above equations are constants and can be pre-computed classically.

\paragraph{$U_1$ for strong Euler scheme.}
We implement $U_1$ as the product of $N+2$ unitary operations:
\begin{align}
    U_1 = V_{N-1}V_{N-2}\dots V_1 V_0 Q_2 Q_1.
\end{align}
The goal of $Q_1$ and $Q_2$ is to prepare the initial state and load the necessary Gaussian distributions for $\alpha_j$'s and $\beta_j$'s, respectively. Then $V_0$, $V_1$, $\dots$, $V_{N-1}$ performs the $N$ iterations of updates in strong Euler scheme.

Specifically, $Q_1$ prepares the initial state $\ket{Y^1_0}_{A_1} \ket{Y^2_0}_{B_1}$:
\begin{align}
    Q_1\ket{0}_{A_1} \ket{0}_{B_1} = \ket{Y^1_0}_{A_1} \ket{Y^2_0}_{B_1} = \ket{0}_{A_1} \ket{\nu_0}_{B_1} 
\end{align}
This operation can be implemented with $X$ gates only.

Meanwhile, $Q_2$ creates a quantum state that encodes the distribution of $(\vec \alpha, \vec \beta)$, where $\vec \alpha=(\alpha_0, \alpha_1, \dots, \alpha_{N-1})$ and
$\vec \beta=(\beta_0, \beta_1, \dots, \beta_{N-1})$. Ideally, each $\alpha_i$ and $\beta_i$ should have standard normal distribution. But on a real computer, they can only take a finite number of possible values. Thus, we have to approximate the original continuous distribution with a discrete distribution. To this end, we introduce a unitary operation $U_{\rm gauss}$ such that 
\begin{align}
U_{\rm gauss}\ket{0} \approx \frac{1}{Z}\sum_{i=0}^{M-1} e^{-\frac{x_i^2}{4}} \ket{x_i},    
\end{align}
where $x_i=(-1+\frac{2i}{M})\eta$ for $0\le i\le M-1$, $Z=\sum_{i=0}^{M-1} e^{-x_i^2/2}$, and $M$, $\eta$ are appropriately chosen parameters. The implementation of $U_{\rm gauss}$ is given in Appendix \ref{apd:prep_gaussian_states}. Then we define $Q_2=U_{\rm gauss}^{\otimes 2N}$ and perform it on $2N$ $n$-qubit registers $E_0, E_1, \dots, E_{2N-2}, E_{2N-1}$, obtaining
\begin{align}
   Q_2\ket{00\dots 0}_{E_0E_1\dots E_{2N-2}E_{2N-1}} 
   &=
   \sum_{\vec \alpha, \vec \beta} \sqrt{\tilde{p}(\vec \alpha, \vec \beta)} \ket{\alpha_0}_{E_0} \ket{\beta_0}_{E_1} \dots \ket{\alpha_{N-1}}_{E_{2N-2}} \ket{\beta_{N-1}}_{E_{2N-1}},
\end{align}
where $\tilde{p}$ is a discrete approximation of the $2N$-dimensional standard normal distribution.

Next, for $j=0,1,\dots,N-1$, we define $V_j$ as a unitary operation satisfying:
\begin{align}
&V_j \ket{Y^1_j}_{A_j} \ket{Y^2_j}_{B_j} \ket{\alpha_j}_{E_{2j}}\ket{\beta_j}_{E_{2j+1}} \ket{0}_{A_{j+1}}\ket{0}_{B_{j+1}}
\nonumber \\
&=\ket{Y^1_j}_{A_j}\ket{Y^2_j}_{B_j}
\ket{\alpha_j}_{E_{2j}} \ket{\beta_j}_{E_{2j+1}} \ket{Y^1_{j+1}}_{A_{j+1}}\ket{Y^2_{j+1}}_{B_{j+1}}.
\label{eq:def_vn}
\end{align}
That is, it computes $Y^1_{j+1}$ and $Y^2_{j+1}$ from $Y^1_j$, $Y^2_j$, $\alpha_j$ and $\beta_j$. We implement $V_j$ by concatenating 8 unitary operations:
\begin{align}
    V_j = R_j^{\dagger} S_j^{\dagger} T_j^{\dagger} H_j T_j S_j R_j G_j.
\end{align}
These operations work as follows:
\begin{enumerate}
\item $G_j$ copies $Y_j^1$ and $Y_j^2$ to registers $A_{j+1}$ and $B_{j+1}$, respectively:
\begin{align}
    G_j \ket{Y_j^1}_{A_j}\ket{Y_j^2}_{B_j}\ket{0}_{A_{j+1}}\ket{0}_{B_{j+1}}
    =   \ket{Y_j^1}_{A_j}\ket{Y_j^2}_{B_j}\ket{Y_j^1}_{A_{j+1}}\ket{Y_j^2}_{B_{j+1}}
\end{align}
This step requires only CNOT gates.

\item $R_j$ computes the square root of $Y_j^2$ and saves it in an ancilla register $J$:
\begin{align}
    R_j \ket{Y_j^2}_{B_j} \ket{0}_{J}
    =
    \ket{Y_j^2}_{B_n} \ket{\sqrt{Y_j^2}}_{J}.
\end{align}
This step requires a square root operation.

\item Let $\vec \Lambda_j=(\alpha_j \sqrt{Y_j^2}, \beta_j \sqrt{Y_j^2})$. Then $S_j$ computes these terms and saves them in an ancilla register $L$:
\begin{align}
    S_j \ket{\sqrt{Y_j^2}}_{J} \ket{\alpha_j}_{E_{2j}}\ket{\beta_j}_{E_{2j+1}} \ket{\vec 0}_{L}
    =
    \ket{\sqrt{Y_j^2}}_{J} \ket{\alpha_j}_{E_{2j}} \ket{\beta_j}_{E_{2j+1}} \ket{\vec \Lambda_j}_{L}
\end{align}
This step requires 2 multiplication operations.

\item Let $\vec \Gamma_j = (-Y_j^2 \cdot {\frac{h}{2}}, \alpha_j\sqrt{Y_j^2} \cdot { \rho \sqrt{h}}, \beta_j\sqrt{Y_j^2} \cdot { \sqrt{h(1-\rho^2})}, -Y^2_j \cdot {\kappa h}, \alpha_j\sqrt{Y_j^2} \cdot { \xi \sqrt{h}})$. Then $T_j$ computes these terms and saves them in an ancilla register $M$:
\begin{align}
    T_j \ket{Y_j^2}_{B_j} \ket{\vec \Lambda_j}_{L} \ket{\vec 0}_{M}
    =
    \ket{Y_j^2}_{B_j} \ket{\vec \Lambda_j}_{L} \ket{\vec \Gamma_j}_{M}
\end{align}
This step requires 5 multiplication-with-a-constant operations.

\item $H_j$ adds the terms in $\vec \Gamma_j$ and the two constant terms $rh$, $\kappa \theta h$ to the values in registers $A_{j+1}$ and $B_{j+1}$, transforming them into $Y^1_{j+1}$ and $Y^2_{j+1}$ eventually: 
\begin{align}
    H_j \ket{Y_j^1}_{A_{j+1}}\ket{Y_j^2}_{B_{j+1}}  \ket{\vec \Gamma_j}_{M}
    = \ket{Y_{j+1}^1}_{A_{j+1}}\ket{Y_{j+1}^2}_{B_{j+1}} \ket{\vec \Gamma_j}_{M}.
\end{align}
This step can be accomplished with 5 addition and 2 addition-with-a-constant operations.

\item $T_j^\dagger$ uncomputes $\ket{\vec \Gamma_j}$ and restores register $M$ back to the zero state. The step has the same cost as that of $T_j$.

\item $S_j^\dagger$ uncomputes $\ket{\vec \Lambda_j}$ and restores register $L$ back to the zero state.
The step has the same cost as that of $S_j$.

\item $R_j^\dagger$ uncomputes $\ket{\sqrt{Y_j^2}}$ and restores register $J$ back to the zero state. The step has the same cost as that of $R_j$. 

\end{enumerate}
Overall, we have implemented $V_j$ by using 5 addition, 2 addition-with-a-constant, 4 multiplication, 10 multiplication-with-a-constant, 2 square root operations, and Clifford gates, with the help of ancilla registers $J$, $M$, $L$ (and other ancilla qubits necessary for performing the arithmetic operations).

\paragraph{$U_1$ for weak Euler scheme.}
The implementation of $U_1$ for weak Euler scheme is similar to the one for strong Euler scheme, except that a few steps are different (in fact, simpler) now.

We still decompose $U_1$ into $N+2$ unitary operations:
\begin{align}
    U_1 = V_{N-1}V_{N-2}\dots V_1 V_0 Q_2 Q_1.
\end{align}
$Q_1$ is the same as before, but now $Q_2$ loads the uniform distribution on $\lrcb{0,1}^{2N}$. So we simply have $Q_2=H^{\otimes 2N}$ (each register $E_j$ contains a single qubit now).

For $j=0,1,\dots,N-1$, the unitary operator $V_j$ still performs the transformation:
\begin{align}
&V_j \ket{Y^1_j}_{A_j} \ket{Y^2_j}_{B_j}  \ket{\alpha_j}_{E_{2j}}\ket{\beta_j}_{E_{2j+1}} \ket{0}_{A_{j+1}}\ket{0}_{B_{j+1}}
\nonumber\\
&=\ket{Y^1_j}_{A_j}\ket{Y^2_j}_{B_j}  \ket{\alpha_j}_{E_{2j}} \ket{\beta_j}_{E_{2j+1}} \ket{Y^1_{j+1}}_{A_{j+1}}\ket{Y^2_{j+1}}_{B_{j+1}}
\end{align}
Now we decompose $V_j$ into 6 unitary operations:
\begin{align}
    V_j = R_j^\dagger T_j^\dagger H_j T_j R_j G_j.
\end{align}
These operations work as follows:
\begin{itemize}
\item $G_j$ copies $Y_j^1$ and $Y_j^2$ to registers $A_{j+1}$ and $B_{j+1}$, respectively:
\begin{align}
    G_j \ket{Y_j^1}_{A_j}\ket{Y_j^2}_{B_j}\ket{0}_{A_{j+1}}\ket{0}_{B_{j+1}}
    =   \ket{Y_j^1}_{A_j}\ket{Y_j^2}_{B_j}\ket{Y_j^1}_{A_{j+1}}\ket{Y_j^2}_{B_{j+1}}
\end{align}
This step requires only CNOT gates.

\item $R_j$ computes the square root of $Y_{n}^2$ and saves it in an ancilla register $J$:
\begin{align}
    R_j \ket{Y_{n}^2}_{B_j} \ket{0}_{J}
    =
    \ket{Y_{n}^2}_{B_j} \ket{\sqrt{Y_j^2}}_{J}.
\end{align}
This step requires a square root operation.

\item Let $\vec \Gamma_j = (-Y_j^2 \cdot {\frac{h}{2}}, \sqrt{Y_j^2} \cdot { \rho \sqrt{h}}, \sqrt{Y_j^2} \cdot { \sqrt{h(1-\rho^2})}, -Y^2_j \cdot {\kappa h}, \sqrt{Y_j^2} \cdot { \xi \sqrt{h}})$. Then $T_j$ computes these terms and saves them in an ancilla register $M$:
\begin{align}
    T_j \ket{Y_j^2}_{B_j} \ket{\sqrt{Y_j^2}}_{J} \ket{\vec 0}_{M}
    =
    \ket{Y_j^2}_{B_j} \ket{\sqrt{Y_j^2}}_{J} \ket{\vec \Gamma_j}_{M}
\end{align}
This step requires 5 multiplication-with-a-constant operations.

\item Depending on the values of $\alpha_j$ and $\beta_j$ (which are either 0 or 1), $H_j$ adds/subtracts the terms in $\Gamma_j$ to/from the values in registers $A_{j+1}$ and $B_{j+1}$. It also adds the two constant terms $rh$, $\kappa \theta h$ to those quantities. Eventually, we obtain $Y^1_{j+1}$ and $Y^2_{j+1}$ in registers $A_{j+1}$ and $B_{j+1}$, respectively: 
\begin{align}
    &H_j \ket{Y_j^1}_{A_{j+1}}\ket{Y_j^2}_{B_{j+1}} \ket{\alpha_j}_{E_{2j}}\ket{\beta_j}_{E_{2j+1}} \ket{\vec \Gamma_j}_{M} \nonumber \\
    &= \ket{Y_{j+1}^1}_{A_{j+1}}\ket{Y_{j+1}^2}_{B_{j+1}}\ket{\alpha_j}_{E_{2j}}\ket{\beta_j}_{E_{2j+1}} \ket{\vec \Gamma_j}_{M}.
\end{align}
This step can be accomplished with 5 addition and 2 addition-with-a-constant operations and Clifford gates.

\item $T_j^\dagger$ uncomputes $\ket{\vec \Gamma_j}$ and restores register $M$ back to the zero state. The step has the same cost as that of $T_j$. 

\item $R_j^\dagger$ uncomputes $\ket{\sqrt{Y_j^2}}$ and restores register $J$ back to the zero state. The step has the same cost as that of $R_j$. 
\end{itemize}
Overall, we have implemented $V_j$ by using 5 addition, 2 addition-with-a-constant, 10 multiplication-with-a-constant, 2 square root operations,  and Clifford gates, with the assistance of ancilla registers $J$ and $M$ (and other ancilla qubits necessary for performing the arithmetic operations).

\paragraph{Remark:} In principle, the quantum circuit for each $V_j$ does not need $R_j^\dagger$, $T_j^\dagger$ or $S_j^\dagger$ (if present) in the end. We put them there to release the ancilla registers $J$, $M$ or $L$ (if present) so that they can be reused in future iterations. If we eliminate some or all of these operations, then we would need a new batch of qubits in each iteration. Consequently, the time efficiency of the algorithm will improve while  its space efficiency will worsen.

\subsection{Implementing the operator $U_2$}
\label{sec:implement_u2}
The implementation of $U_2$ depends on the specific option under question. So we will discuss the cases of Asian and barrier options separately.

\noindent \paragraph{Asian option.}
Here we consider an Asian call. Asian puts can be handled similarly.

Suppose the Asian call option has strike price $K$ and expiration time $T$. Then its payoff function is 
\begin{align}
    f_0(\vec S) =\lrb{\frac{1}{N}\sum_{j=1}^N S_{jh}-K}^+.
\end{align}
Let $Z$ be (an upper bound on) the maximum payoff of the option \footnote{In practice, we estimate $Z$ by simulating the stochastic process on a classical computer and generating multiple random paths and returning the maximum payoff among these paths. Suppose $Z_M$ is the maximum payoff among $M$ random paths. Then with high confidence, we conclude that the probability of a random path having payoff larger than $Z_M$ is at most $O(1/M)$. By picking sufficiently large $M$, we can ensure that this probability is extremely low. Therefore, we could safely ignore the paths whose payoffs are larger than $Z_M$ in our calculation.}. Then the normalized payoff function is
\begin{align}   
    f(\vec S) =\frac{1}{Z}\lrb{\frac{1}{N}\sum_{j=1}^N S_{jh}-K}^+.
\end{align}
In terms of the log returns $\vec R$, the normalized payoff function becomes
\begin{align}
\tilde{f}(\vec R)=\frac{1}{Z}\lrb{\frac{S_0}{N}\sum_{j=1}^N e^{R_{jh}} - K}^+=\lrb{c \sum_{j=1}^N e^{R_{jh}} - k}^+,
\end{align}
where $c=S_0/(NZ)$ and $k=K/Z$ do not depend on the path.

For convenience, we define $X(\omega)=c \sum_{j=1}^N e^{Y_j^1(\omega)} - k$. Then $\tilde{f}(\vec Y^1(\omega))=X^+(\omega)$. Our goal is to construct a unitary operation $U_2$ such that 
\begin{align}
    U_2\ket{\vec Y^1(\omega)}_A \ket{0}_C \ket{00\dots 0}_G =\ket{\vec Y^1(\omega)}_A \ket{X^+(\omega)}_C \ket{\Psi(\omega)}_G, ~&~\forall \omega \in \Omega,
    \label{eq:u2_asian}
\end{align}
where $\ket{\Psi(\omega)}$ is a normalized state depending on $\omega$. This can be accomplished by concatenating three unitary operations: 
\begin{align}
    U_2=U_{2,3}U_{2,2}U_{2,1},
\end{align}
where
\begin{align}
    U_{2,1}\ket{Y_1^1, Y_2^1,\dots,Y_N^1}_A \ket{0,0,\dots,0}_{G_1}&=\ket{Y_1^1, Y_2^1,\dots,Y_N^1}_A \ket{e^{Y_1^1},e^{Y_2^1},\dots,e^{Y_N^1}}_{G_1}, \\
        U_{2,2}\ket{e^{Y_1^1},e^{Y_2^1},\dots,e^{Y_N^1}}_{G_1} \ket{0}_{G_2} &=
        \ket{e^{Y_1^1},e^{Y_2^1},\dots,e^{Y_N^1}}_{G_1} \ket{c \sum_{j=1}^N e^{Y_j^1} - k}_{G_2}, \\
    U_{2,3} \ket{X}_{G_2} \ket{0}_C
    &=\ket{X}_{G_2}  \ket{X^+}_C.
\end{align}
That is, $U_{2,1}$ computes the exponentials of the $Y_j^1$'s, $U_{2,2}$ computes $X$ from these exponentials, and $U_{2,3}$ computes $X^+$ from $X$. The registers $G_1$ and $G_2$ contain $Nn$ and $n$ qubits, respectively, and they form the subsystem $G$ in Eq.~\eqref{eq:u2_asian}.

Next, we analyze the costs of the three steps. $U_{2,1}$ requires $N$ exponential operations. Meanwhile, $U_{2,2}$ can be implemented with $N-1$ addition, one multiplication-with-a-constant and one subtraction-with-a-constant operations, and Clifford gates. Finally, $U_{2,3}$ can be implemented with $n$ Toffoli gates and Clifford gates.

\noindent \paragraph{Barrier option.} Here we consider an up-and-out put. The other types of barrier options can be handled similarly.

Suppose the up-and-out put option has strike price $K$, barrier $B$ and 
expiration time $T$. Then its payoff function is
\begin{align}
f_0(\vec S)=1_{\max\limits_{1\le j \le N}S_{jh}\le B} \cdot \lrb{K-S_T}^+.   
\end{align}
Let $Z$ be (an upper bound on) the maximum payoff of the option. Then the normalized payoff function is
\begin{align}
f(\vec S)= 1_{\max\limits_{1\le j \le N}S_{jh}\le B} \cdot 
\frac{\lrb{K-S_T}^+}{Z}.   
\end{align}
In terms of the log returns $\vec R$, the normalized payoff function becomes
\begin{align}
\tilde{f}(\vec R)= 1_{\max\limits_{1\le j \le N}R_{jh} \le \ln{B/S_0}} \cdot \frac{\lrb{K-S_0 e^{R_T}}^+}{Z}
=1_{\max\limits_{1\le j \le N}R_{jh} \le \ln{B/S_0}} \cdot \lrb{k-ce^{R_T}}^+
\end{align}
where $c=S_0/Z$ and $k=K/Z$ do not depend on the path. 

For convenience, we define $\eta_j(\omega)=1$ if $Y^1_j(\omega)\le \ln{B/S_0}$ and $0$ otherwise, for $j=1,2,\dots,N$, and define 
$\eta(\omega)=\eta_1(\omega)\eta_2(\omega)\dots \eta_N(\omega)$. In addition, let $X(\omega)=k-ce^{Y^1_N(\omega)}$. Then we have 
$\tilde{f}(\vec Y^1(\omega))= \eta(\omega) X^+(\omega)$.

Our goal is to construct a unitary operation $U_2$ such that 
\begin{align}
    U_2\ket{\vec Y^1(\omega)}_A \ket{0}_C \ket{00\dots 0}_G =\ket{\vec Y^1(\omega)}_A \ket{\eta(\omega)X^+(\omega)}_C \ket{\Psi(\omega)}_G, ~&~\forall \omega \in \Omega,
    \label{eq:u2_barrier}
\end{align}
where $\ket{\Psi(\omega)}$ is a normalized state depending on $\omega$. This can be accomplished by concatenating five unitary operations: 
\begin{align}
    U_2=U_{2,5}U_{2,4}U_{2,3}U_{2,2}U_{2,1}
\end{align}
where
\begin{align}
    U_{2,1} \ket{Y_1^1, Y_2^1,\dots,Y_N^1}_A \ket{0,0,\dots,0}_{G_1}&=\ket{Y_1^1, Y_2^1,\dots,Y_N^1}_A \ket{\eta_1, \eta_2,\dots, \eta_N}_{G_1}, \\
    U_{2,2} \ket{\eta_1, \eta_2, \dots, \eta_N}_{G_1} \ket{0}_{G_2}&= \ket{\eta_1, \eta_2, \dots, \eta_N}_{G_1} \ket{\eta_1 \eta_2 \dots \eta_N}_{G_2}, \\ 
    U_{2,3} \ket{Y_N^1}_{A_N} \ket{0}_{G_3}&=\ket{Y_N^1}_{A_N} \ket{k-c e^{Y_N^1}}_{G_3}, \\
    U_{2,4} \ket{X}_{G_3} \ket{0}_{G_4} &= \ket{X}_{G_3} \ket{X^+}_{G_4}, \\
    U_{2,5} \ket{\eta}_{G_2} \ket{X^+}_{G_4} \ket{0}_C &= \ket{\eta}_{G_2} \ket{X^+}_{G_4} \ket{\eta X^+}_C.
\end{align}
That is, $U_{2,1}$ computes the $\eta_j$'s from the $Y_j^1$'s, $U_{2,2}$ computes $\eta$ from the $\eta_j$'s, $U_{2,3}$ computes $X$ from $Y_N^1$, $U_{2,4}$ computes $X^+$ from $X$, and $U_{2,5}$ computes $\eta X^+$ from $\eta$ and $X^+$. The registers $G_1$, $G_2$, $G_3$ and $G_4$ contain $N$, $1$,  $n$ and $n$ qubits, respectively, and they form the subsystem $G$ in Eq.~\eqref{eq:u2_barrier}.

The costs of the five steps are as follows. $U_{2,1}$ requires $N$ comparison (with a constant) operations. $U_{2,2}$ can be implemented as an $(N+1)$-qubit Toffoli gate. $U_{2,3}$ requires one exponential, one multiplication-with-a-constant, and one addition-with-a-constant operations. $U_{2,4}$ can be implemented with $n$ Toffoli gates and Clifford gates. Finally, $U_{2,5}$ can be implemented with $n$ Toffoli gates.

\subsection{Implementing the operator $U_3$}
\label{sec:implement_u3}
The operator $U_3$ appears frequently in the literature (e.g. \cite{Rebentrost2018quantum,Chakrabarti2021thresholdquantum, hhl}) and we adopt a common method to realize it. Specifically, we decompose $U_3$ into two unitary operations:
\begin{align}
    U_3 = U_{3,2} U_{3,1},
\end{align}
where
\begin{align}
    U_{3,1}\ket{z}_C \ket{0}_H &=\ket{z}_C\ket{\arcsin{\sqrt{z}}}_H, \\
    U_{3,2}\ket{\arcsin{\sqrt{z}}}_H \ket{0}_D &=\ket{\arcsin{\sqrt{z}}}_H \lrsb{\sqrt{z}\ket{0}_D +\sqrt{1-z}\ket{1}_D},
\end{align}
for all $z \in [0, 1]$. Here the register $H$ contains $n$ qubits. We implement $U_{3,1}$ as an arcsin-of-square-root operation defined in Appendix \ref{apd:fixed_point_arithmetics}. Meanwhile, we implement $U_{3,2}$ as a $U_{\rm sin}$ operation defined in Appendix \ref{apd:opt_circuit_usin}.

\subsection{Resource analysis} 
\label{sec:resource_analysis}
So far, we have constructed the circuit ${\cal A}$ which satisfies Eq.~\eqref{eq:circuit_a} by composing the modules in  Appendices \ref{apd:fixed_point_arithmetics},
\ref{apd:opt_circuit_usin} and \ref{apd:prep_gaussian_states} as well as Clifford gates. Suppose it contains $m+1$ qubits.
Our algorithm makes $N_{\rm oracle}$ uses of the unitary operation $Q={\cal A}R_0 {\cal A}^{-1} S_0$, where $R_0=I_{m+1}-2\ket{0^{m+1}}\bra{0^{m+1}}$ can be implemented with an  $(m+1)$-qubit Toffoli gate and Clifford gates, and $S_0=-Z\otimes I_m$ costs no T gate. So the total number of T gates in our algorithm is 
\begin{align}
N_{\rm oracle}\tcount{Q}=N_{\rm oracle}\lrsb{2\tcount{\cal A}+\tcount{\rm Toffoli_{m+1}}}.    
\end{align}

In Appendices \ref{apd:fixed_point_arithmetics},
\ref{apd:opt_circuit_usin} and \ref{apd:prep_gaussian_states}, we introduce the fixed-point quantum arithmetic operations, a block-encoding $U_{\rm sin}$ of the sine function, and a unitary operation $U_{\rm gauss}$ for preparing Gaussian states, respectively, and analyze the costs of implementing these operations. From now on, we will adopt the notation in these appendices.

The number of T gates in ${\cal A}$ is
\begin{align}
\tcount{\cal A}=\tcount{U_1}+\tcount{U_2}+\tcount{U_3}    
\end{align}
where $U_1$, $U_2$ and $U_3$ satisfy Eqs.~\eqref{eq:def_u1}, \eqref{eq:def_u2} and \eqref{eq:def_u3} respectively.

The number of T gates in $U_1$ depends on which numercial scheme we use. For strong Euler method, it is
\begin{align}
\tcount{U_1}=&\tcount{Q_1}+\tcount{Q_2}+\sum_{j=0}^{N-1}\tcount{V_j}\\
=&N [2\tcount{U_{\rm gauss}(n, \eta, \epsilon_{\rm prep}, \epsilon_{\rm gauss})} + 5\tcount{{\rm ADD}_n} \nonumber \\
&+2\tcount{{\rm ADD}\_{\rm CONST}_n} 
+4 \tcount{{\rm MUL}_{n,p}} \nonumber \\
&+10 \tcount{{\rm MUL}\_{\rm CONST}_{n,p}}
+2 \tcount{{\rm SQRT}_n}].
\end{align}
By contrast, for weak Euler method, it is
\begin{align}
\tcount{U_1}=&\tcount{Q_1}+\tcount{Q_2}+\sum_{j=0}^{N-1}\tcount{V_j}\\
=& N [5\tcount{{\rm ADD}_n} +
2\tcount{{\rm ADD}\_{\rm CONST}_n} \nonumber \\
&+10 \tcount{{\rm MUL}\_{\rm CONST}_{n,p}}
+2 \tcount{{\rm SQRT}_n}],
\end{align}
which is much smaller, as will be shown in Section \ref{sec:case_studies}.

The number of T gates in $U_2$ depends on the type of the option under consideration. For example, for an Asian call, it is
\begin{align}
\tcount{U_2}=&\tcount{U_{2,1}}+\tcount{U_{2,2}}+\tcount{U_{2,3}}\\
=&N\tcount{{\rm EXP}_{n,p,\epsilon_{exp}}}
+(N-1)\tcount{{\rm ADD}_n} \nonumber\\
&+\tcount{{\rm MUL}\_{\rm CONST}_{n,p}}
+\tcount{{\rm SUB}\_{\rm CONST}_{n}}
+n\tcount{{\rm Toffoli}_3}.
\end{align}
Meanwhile, for an up-and-out put, it is
\begin{align}
\tcount{U_2}=&\tcount{U_{2,1}}+\tcount{U_{2,2}}+\tcount{U_{2,3}} 
+\tcount{U_{2,4}}+\tcount{U_{2,5}} \\
=& N \tcount{{\rm COMP}\_{\rm CONST}_n}
+ \tcount{{\rm Toffoli}_{N+1}}
+ \tcount{{\rm EXP}_{n,p,\epsilon_{exp}}} \nonumber \\
&+ \tcount{{\rm MUL}\_{\rm CONST}_{n,p}}
+ \tcount{{\rm ADD}\_{\rm CONST}_{n}}
+ 2n \tcount{{\rm Toffoli}_3}.
\end{align}

Finally, the number of T gates in $U_3$ is
\begin{align}
\tcount{U_3}=&\tcount{U_{3,1}}+\tcount{U_{3,2}}\\
=& \tcount{{\rm ARCSIN}\_{\rm SQRT}_{n, p, \epsilon_{\rm arcsin}}}
+\tcount{U_{\rm sin}(n,\epsilon_{\rm sin})}.
\end{align}

To obtain the T-depth of our algorithm, we replace ${\rm T}\_{\rm count}$ with ${\rm T}\_{\rm depth}$ in the above equations. Note that this yields a conservative estimate of the necessary T-depth, because we could in principle improve it by running some modules, e.g., $Q_1$ and $Q_2$, in parallel. But this kind of optimization is unlikely to significantly reduce the T-depth of the circuit, given the highly-sequential nature of our algorithm. Thus, we will not pursue this direction here.

To determine the number of qubits in the circuit $\cal A$, we scan through the operations in it, check the number of qubits necessary to store the permanent and temporary information and perform the current operation at each moment, and take the maximum among them. The number of qubits in $Q$ is larger than that of $\cal A$ by $1$.

\subsection{Error analysis}
\label{sec:error_analysis}

There are multiple sources of errors in our algorithm:

\begin{enumerate}
\item Amplitude estimation can only produce an estimate of the true value with certain accuracy and confidence. 

\item The results of fixed-point arithmetic operations contain errors due to the finite-precision representation of real numbers and/or the piecewise-polynomial approximation of exponential and arcsin functions in their implementation (see Appendix \ref{apd:fixed_point_arithmetics} for more details).

\item We use the circuit in Appendix \ref{apd:opt_circuit_usin} to implement $U_{\rm sin}$. This circuit contains $R_x/R_y/R_z$ rotations which can only be synthesized approximately with Clifford and T gates. In this work, we utilize the method of~\cite{bocharov2015efficient} to synthesize $R_x/R_y/R_z$ rotations. 

\item For a similar reason, $U_{\rm gauss}$ cannot be implemented exactly either.

\item In our algorithm, each $\alpha_j$ and $\beta_j$ can only take values from a finite set of numbers. This poses a problem for strong Euler method, as it changes the probability distribution of the random path. Consequently, the expectation of the payoff in the simulation is different from the ideal one. 

\item Finally, even if $U_{\rm gauss}$ is implemented perfectly, it only approximately prepares a quantum state that encodes a discrete version of standard normal distribution (see Appendix \ref{apd:prep_gaussian_states} for more details).
\end{enumerate}
Note that the last three issues are specific to strong Euler method. They do not exist for weak Euler method.

Next, we analyze each of the above errors individually:

\begin{itemize}
\item The first error can be bounded analytically, and Eq.~\eqref{eq:n_oracle} gives an upper bound on the number of oracle queries needed to achieve the desired accuracy and confidence in amplitude estimation.

\item The third error can be bounded as follows. Suppose each $U_{\rm sin}$ is implemented with $\epsilon_{\rm sin}$ precision. Then since there are $2 N_{\rm oracle}$ such operations in our circuit, the output state of the circuit is at most $2 N_{\rm oracle} \epsilon_{\rm sin}$-away (in trace distance) from the ideal state, which means that the deviation of the final result due to this factor is at most $2 N_{\rm oracle} \epsilon_{\rm sin}$ too. 

\item Similarly, if strong Euler method is employed and each $U_{\rm gauss}$ is implemented with precision $\epsilon_{\rm gauss}$, then the fourth error is at most $4 N N_{\rm oracle} \epsilon_{\rm gauss}$.

\item The fifth error can be also bounded easily. In the algorithm based on strong Euler method, we need $4 N N_{\rm oracle}$ copies of the state encoding a discrete version of standard normal distribution. Suppose each of them is prepared with $\epsilon_{\rm prep}$ accuracy in trace distance. Then this leads to an error of at most $4 N N_{\rm oracle} \epsilon_{\rm prep}$ in the final result. 

\item In principle, we could derive a theoretical upper bound on the error incurred by imperfect fixed-point arithmetic operations, as done in~\cite{Chakrabarti2021thresholdquantum}. However, we find that such a bound is quite loose, as it always assumes the worst scenario, i.e., each arithmetic operation contains the largest possible error, and the error of a sequence of arithmetic operations is the sum of these individual errors, which almost never happens. To better bound this error in practical situations, we simulate the stochastic process using strong/weak Euler method \emph{under fixed-point arithmetic conditions} on a classical computer, generate multiple random paths, and compute the average distance between the true and calculated payoffs of the option for these paths. This procedure turns out to be highly efficient \footnote{Here we aim to estimate the difference between the payoffs of the option for two random paths that are always close to each other. The variance of this random variable is quite small. Thus, it takes few samples to reach high accuracy in this estimation. }.

\item Moreover, if the algorithm is based on strong Euler method, then we choose 
each $\alpha_j$ and $\beta_j$ from a discrete version of standard normal distribution in the above simulation. Consequently, the estimated error will be the combined effect of flawed fixed-point arithmetic operations and discrete approximation of Gaussian distributions.

\end{itemize}

To summarize, if the algorithm is based on strong Euler method, then the error in our final estimate of $\E{\tilde{f}(\vec Y^1(\omega))}$ is at most 
\begin{align}
    \epsilon_{\rm estimate} + \epsilon_{\rm arithm/disc} + 2 N_{\rm oracle} \epsilon_{\rm sin} + 4 N N_{\rm oracle} \lrb{\epsilon_{\rm prep}+\epsilon_{\rm gauss}}
\end{align}
 where
\begin{itemize}
    \item $\epsilon_{\rm estimate}$ is the precision of amplitude estimation,
    \item $\epsilon_{\rm arithm/disc}$ is the error due to imperfect fixed-point arithmetic operations and discrete approximation of Gaussian distributions, 
    \item $\epsilon_{\rm sin}$ is the precision of implementing each $U_{\rm sin}$,
    \item $\epsilon_{\rm prep}$ be the precision of preparing each Gaussian state,
    \item $\epsilon_{\rm gauss}$ be the precision of implementing each $U_{\rm gauss}$,
\end{itemize}
whereas if the algorithm is based on weak Euler method, then the final error is at most
\begin{align}
    \epsilon_{\rm estimate} + \epsilon_{\rm arithm} + 2 N_{\rm oracle} \epsilon_{\rm sin} 
\end{align}
where
\begin{itemize}
    \item $\epsilon_{\rm estimate}$ is the precision of amplitude estimation,
    \item $\epsilon_{\rm arithm}$ is the error due to imperfect fixed-point arithmetic operations,    
    \item $\epsilon_{\rm sin}$ is the precision of implementing each $U_{\rm sin}$.
\end{itemize}

\subsection{Case studies} 
\label{sec:case_studies}
Next, we apply our algorithms to four instances of option pricing under the Heston model to test their performance in practice. Specifically, we consider four settings of the Heston model which are described in Table \ref{tab:case_study_heston_specs}, and estimate the value of an Asian or barrier option under each setting. The specifications of the options are listed in Table \ref{tab:case_study_option_specs}.

\begin{table}[htbp!]
    \centering
\begin{tabular}{ |c|c|c|c|c|c|c|c| } 
 \hline
Setting & $r$ & $\rho$ & $\kappa$ & $\theta$ & $\xi$ & $S_0$ & $\nu_0$ 
\\
 \hline 
No. 1 & $0.03$ &  $-0.1$ & $2$ & $0.12$ 
& $0.3$ & $100$ & $0.1$ \\ 
 \hline
No. 2 & $0.05$ &  $-0.1$ & $2$ & $0.04$ 
& $0.2$ & $100$ & $0.05$ \\ 
 \hline
No. 3 & $0.03$ &  $0$ & $2$ & $0.03$ 
& $0.2$ & $100$ & $0.03$ \\ 
 \hline
No. 4 & $0.05$ &  $-0.1$ & $2$ & $0.09$ 
& $0.2$ & $100$ & $0.06$ \\ 
 \hline
\end{tabular}
\caption{The settings of the Heston model studied in Section \ref{sec:case_studies}. Each setting includes the model parameters $r$, $\rho$, $\kappa$, $\theta$, $\xi$ and initial condition $S_0$ and $\nu_0$.}
\label{tab:case_study_heston_specs}
\end{table}

\begin{table}[htbp!]
    \centering
\begin{tabular}{ |c|c|c|c|c|c|c| } 
 \hline
Instance & Setting & Option type & $K$ & $B$ & $T$ & $Z$
\\
 \hline 
No. 1 & No. 1 & Asian~call & $90$ & N/A & $1.0$ & $200$\\ 
 \hline
No. 2 & No. 2 & Asian~put & $110$ & N/A & $1.0$ & $100$\\ 
 \hline
No. 3 & No. 3 & Up-and-out call & $90$ & $170$ & $1.0$ & $80$\\ 
 \hline
No. 4 & No. 4 & Down-and-in put & $110$ & $80$ & $1.0$ & $100$\\ 
 \hline
\end{tabular}
\caption{The problem instances studied in Section \ref{sec:case_studies}. The specification of each instance includes the setting of the Heston model and the type, strike $K$, barrier $B$ (if needed) and expiration time $T$ of the option. Here we also present an upper bound $Z$ on the maximum payoff of the option which is used to normalize the payoff function.}
\label{tab:case_study_option_specs}
\end{table}

For each problem instance, we develop the quantum algorithms based on strong and weak Euler schemes, and tune their parameters, e.g., $\epsilon_{\rm sin}$, $\epsilon_{\rm gauss}$, $\epsilon_{\rm prep}$, to make sure that these two algorithms achieve (roughly) the same accuracy and confidence in their results. The specific settings of the parameters are given in Table \ref{tab:case_study_paramter_setting}. Note that the weak-Euler-based algorithm uses fewer qubits to represent each real number and implements each $U_{\rm sin}$ with lower precision than the strong-Euler-based algorithm, yet it still produces as accurate result as the latter.

\begin{table}[htbp!]
    \centering
\begin{tabular}{ |c|c|c|c|c|c|c|c|c|c| } 
 \hline
Instance & \thead{Numerical \\ Scheme} & $N$ & $n$ & $p$ & 
$\epsilon_{\rm sin}$ 
& $\epsilon_{\rm gauss}$ & $\epsilon_{\rm prep}$ & $\epsilon_{\rm estimate}$ 
\\
 \hline 
No. 1 & Strong Euler & $256$ & $29$ & $11$ 
& $10^{-9}$ & $10^{-12}$ & $10^{-12}$ & $10^{-3}$  \\
No. 1 & Weak Euler & $256$ & $27$ & $11$  
& $10^{-8}$ & N/A & N/A & $10^{-3}$ \\ 
 \hline
No. 2 & Strong Euler & $256$ & $29$ & $10$ 
& $10^{-9}$ & $5\times 10^{-12}$ & $5 \times 10^{-12}$ & $10^{-3}$  \\
No. 2 & Weak Euler & $256$ & $27$ & $10$  
& $10^{-8}$ & N/A & N/A & $10^{-3}$ \\ 
 \hline
No. 3 & Strong Euler & $1024$ & $32$ & $10$ 
& $10^{-9}$ & $5\times 10^{-13}$ & $5 \times 10^{-13}$ & $10^{-3}$  \\
No. 3 & Weak Euler & $1024$ & $29$ & $10$ 
& $5 \times 10^{-9}$ & N/A & N/A & $10^{-3}$ \\ 
 \hline
No. 4 & Strong Euler & $1024$ & $30$ & $10$
& $10^{-9}$ & $10^{-12}$ & $10^{-12}$ & $10^{-3}$  \\
No. 4 & Weak Euler & $1024$ & $29$ & $10$ 
& $5\times 10^{-9}$ & N/A & N/A & $10^{-3}$ \\ 
 \hline

\end{tabular}
\caption{The parameter settings in our experiments, including the underlying numerical scheme, the number $N$ of time steps, the parameters $n$ and $p$ about the fixed-point representation of real numbers, 
the precision $\epsilon_{\rm sin}$ of implementing each $U_{\rm sin}$, the precision $\epsilon_{\rm gauss}$ of implementing each $U_{\rm gauss}$, the precision $\epsilon_{\rm prep}$ of preparing each Gaussian state, and the accuracy $\epsilon_{\rm estimate}$ of amplitude estimation. Moreover, we set $\eta=6.0$ in all Gaussian state preparation procedures, and set $\epsilon_{\rm exp}=\epsilon_{\rm arcsin}=10^{-6}$, and set the failure probability of amplitude estimation to $\delta=0.1$ in all experiments.}
\label{tab:case_study_paramter_setting}
\end{table}

Table \ref{tab:case_study_resource_costs} summarizes the resources required by the two quantum algorithms to reach the same accuracy and confidence on each problem instance. 

\begin{table}[htbp!]
    \centering
\begin{tabular}{ |c|c|c|c|c|c|c| } 
 \hline
 \thead{Instance} &
\thead{Numerical\\Scheme} & N &\thead{Error} & \thead{T-count} & \thead{T-depth} & \thead{Number of \\ Logical Qubits} 
\\
 \hline 
No. 1 & Strong Euler &  256 &$1.3\times 10^{-3}$ & $4.1\times 10^{13}$ & $2.9\times 10^{13}$ 
& $3.8\times 10^4$\\ 
No. 1 & Weak Euler &  256 & $1.3\times 10^{-3}$ & $2.4\times 10^{11} $ & $1.2\times 10^{11}$ 
& $2.2\times 10^4$\\ 
 \hline
No. 2 & Strong Euler &  256 & $1.2\times 10^{-3}$ & $3.9\times 10^{13}$ & $2.7\times 10^{13}$ 
& $3.8\times 10^4$\\ 
No. 2 & Weak Euler &  256 & $1.2\times 10^{-3}$ & $2.3\times 10^{11}$ & $ 1.1\times 10^{11}$ 
& $2.2\times 10^4$\\ 
 \hline
No. 3 & Strong Euler & 1024 & $1.3\times 10^{-3}$ & $1.7\times 10^{14} $ & $1.2\times 10^{14}$ 
& $1.3\times 10^5$\\ 
No. 3 & Weak Euler & 1024 & $1.3\times 10^{-3}$ & $4.4\times 10^{11}$ & $ 2.2\times 10^{11}$ 
& $6.4\times 10^4$\\ 
 \hline
No. 4 & Strong Euler &  1024 & $1.4\times 10^{-3}$ & $1.7\times 10^{14}$ & $1.2\times 10^{14}$ 
& $1.3\times 10^5$\\ 
No. 4 & Weak Euler &  1024 & $1.4\times 10^{-3}$ & $4.4\times 10^{11}$ & $2.2\times 10^{11}$ 
& $6.4\times 10^4$\\ 
 \hline
\end{tabular}
\caption{The resources required by strong- and weak-Euler-based algorithms to achieve the same accuracy and confidence on each problem instance, including the T-counts, T-depths and numbers of logical qubits in the circuits. Note that the weak-Euler-based algorithm is far more efficient than the strong-Euler-based algorithm on every instance. Here the target quantity is $\E{\tilde{f}(\vec Y^1(\omega)}$ for the corresponding numerical scheme with $N$ time steps. }
\label{tab:case_study_resource_costs}
\end{table}

One can see that the algorithm based on weak Euler scheme has much smaller T-count and T-depth than the one based on strong Euler scheme, and uses fewer qubits as well. In fact, the strong-Euler-based algorithm demands hundreds of times more T gates than the weak-Euler-based algorithm on all instances. This is mainly due to the expensive procedure of preparing the Gaussian states needed by the former. Specifically, Table \ref{tab:case_study_tcount_each_operation} shows the T-counts of the operations $U_1$, $U_2$, $U_3$ and $Q$ as well as the number of applications of $Q$ in each algorithm on each instance. Note that the strong-Euler-based algorithm has much more costly $U_1$ than the weak-Euler-based algorithm. We find that in the this algorithm, $Q_2=U_{\rm gauss}^{\otimes 2N}$ dominates the consumption of T gates within $U_1$, i.e., $\tcount{Q_2} \ge 0.99 \tcount{U_1}$. This is mostly because $U_{\rm gauss}$ contains a large number, i.e., $>10^5$, of $R_y/R_z$ rotations, each of which needs to be synthesized with Clifford and T gates with high precision. On the other hand, the weak-Euler-based algorithm does not need to prepare Gaussian states and hence simulates the stochastic process with much lower costs. 

\begin{table}[htbp!]
\centering
\begin{tabular}{ |c|c|c|c|c|c|c| } 
 \hline
 \thead{Instance} &
\thead{Numerical\\Scheme} & \thead{$\tcount{U_1}$} & \thead{$\tcount{U_2}$} & \thead{$\tcount{U_3}$} & \thead{$\tcount{Q}$} &\thead{$N_{\rm oracle}$} 
\\
 \hline 
No. 1 & Strong Euler &  $2.8\times 10^{9}$ & $1.0\times 10^7$ & $ 7.0\times 10^4$ 
& $5.6\times 10^{9} $ & $7363$\\ 
No. 1 & Weak Euler &  $6.4\times 10^6$ & $9.3\times 10^6$ & $6.0\times 10^4$ 
& $3.2\times 10^7$ & $7363$\\
 \hline
No. 2 & Strong Euler &  $2.6\times 10^{9}$ & $1.0\times 10^7$ & $ 6.8\times 10^4$ 
& $5.3 \times 10^{9}$ & $7363$\\
No. 2 & Weak Euler &  $6.4\times 10^6$ & $9.2\times 10^6$ & $6.0\times 10^4$ 
& $3.2\times 10^7$ & $7363$\\
 \hline
No. 3 & Strong Euler &  $1.2\times 10^{10}$ & $3.0\times 10^5$ & $ 8.0\times 10^4$ 
& $2.4\times 10^{10}$ & $7363$\\
No. 3 & Weak Euler &  $2.9\times 10^7$ & $2.7\times 10^5$ & $6.8\times 10^4$ 
& $6.0\times 10^7$ & $7363$\\
 \hline
No. 4 & Strong Euler &  $1.1\times 10^{10}$ & $2.8\times 10^5$ & $ 7.2\times 10^4$ 
& $2.3\times 10^{10} $ & $7363$\\
No. 4 & Weak Euler &  $2.9\times 10^7$ & $2.7\times 10^5$ & $6.8\times 10^4$ 
& $6.0\times 10^7$ & $7363$\\
 \hline
\end{tabular}
\caption{The costs of implementing $U_1$, $U_2$, $U_3$, $Q$ and the number of applications of $Q$ in each algorithm on each instance.}
\label{tab:case_study_tcount_each_operation}
\end{table}

We emphasize that the above resource estimates hold under the experimental settings delineated in Tables \ref{tab:case_study_heston_specs}, \ref{tab:case_study_option_specs} and \ref{tab:case_study_paramter_setting} only. Nevertheless, the conclusion that the algorithm based on weak Euler scheme is far superior to the one based on strong Euler scheme remains valid in general.

\section{Conclusion and outlook}
\label{sec:conclusion}
To summarize, we have developed quantum algorithms for pricing Asian and barrier options under the Heston model, and estimated their costs and errors on example instances under typical market conditions. These algorithms are based on combining classical numerical methods for stochastic differential equations and quantum amplitude estimation technique. In particular, we empirically showed that, despite its simplicity, weak Euler method achieves the same level of accuracy as strong Euler method for option pricing under the Heston model. Furthermore, by eliminating the expensive procedure of preparing Gaussian states, the quantum algorithm based on weak Euler scheme achieves drastically better efficiency than the one based on strong Euler scheme. Our results shed light on the possibility of using quantum computers to accelerate option pricing under stochastic volatility in the future.

Even though our quantum algorithms achieve only a quadratic speedup over classical Monte Carlo methods, they avoid several pitfalls that can make many quantum algorithms fast in theory but potentially slow in practice. First, our algorithms do not require quantum random access memory (QRAM), which could be difficult to implement. Instead, we receive the problem specification in classical form and then build and execute the quantum circuits. Second, our algorithms directly produce the estimate of the target quantity, unlike other approaches—such as those based on quantum linear system solvers \cite{hhl}—which must first generate a quantum state encoding the solution and then repeatedly measure it to obtain the final result, leading to additional overhead. Third, our numerical results suggest that the costs of our algorithms are relatively insensitive to the problem specification. Their performance does not rely on assumptions like the well-conditioning of a data matrix, which could be difficult to satisfy in practice. Additionally, our algorithms use computational resources similar to or fewer than those required for many quantum chemistry simulations~\cite{PRXQuantum.2.030305,PRXQuantum.4.040303,PhysRevResearch.3.033055,Berry2019qubitizationof}. For these reasons, we believe that option pricing under stochastic volatility could be among the first useful applications of quantum computers.

To our knowledge, this is the first work on fault-tolerant quantum algorithms with gate-by-gate level instructions and resource estimates for pricing exotic options under a stochastic volatility model. Previous works  on derivative pricing have been restricted to the case that the asset price follows a geometric Brownian motion with constant drift and volatility. Due to different problem settings, it is hard to fairly compare our results and previous results. Among the previous works, ~\cite{Chakrabarti2021thresholdquantum} is most similar to ours and considers the pricing of autocallable and Target Accrual Redemption Forward (TARF) derivatives in the GBM case. In their problems, they have $20$ time steps, while we have over $200$ or $1000$ time steps here. If one compares the average number of T gates per time step, then our cost is similar to theirs, even though we consider a more complicated model which requires more arithmetic operations to simulate. Moreover, ~\cite{Chakrabarti2021thresholdquantum} uses strong Euler method to simulate their stochastic process and thus needs to prepare Gaussian states (which is done heuristically using a variational algorithm in their work). We believe their results can quite possibly be improved by utilizing weak Euler method instead. 

Our resource estimates can be used to derive the requirement for quantum hardware to demonstrate practical advantage for option pricing under stochastic volatility. Assuming a target of $\sim 10^3$ seconds for pricing a barrier option, the quantum processor would need to execute each layer of T gates at a rate of $\sim 10$MHz, i.e., logical clockspeed, where each layer consists of $\sim 10$ T gates. While~\cite{Chakrabarti2021thresholdquantum} also concludes that $\sim 10$MHz per layer of T gate is needed for their algorithms to provide practical advantage, their algorithms have $\sim 10^3$ T gates per layer, which, compared to $\sim 10$ T gates per layer, is more difficult to apply in parallel at a high rate, especially in early fault tolerant devices with limited number of magic state factories. In this sense, our hardware requirement is less stringent than theirs. We stress that in practice, the simulation duration for option pricing under stochastic volatility could be longer than $10^3$ seconds if a more accurate pricing, i.e., smaller $\epsilon$, is desired, in which case the hardware threshold is further relaxed because the quantum computational costs grow more slowly than the classical costs.

There are multiple research directions that deserve further exploration:
\begin{itemize}
    \item So far, we have employed strong/weak Euler method for SDEs to simulate the dynamics of the Heston model. As a consequence, our algorithm has time cost proportional to the evolution time $T$. Is it possible to utilize more sophisticated strategy, e.g., spectral methods~\cite{bouland2023quantum}, to develop quantum algorithms for the same problem with only $\poly{\log{T}}$ runtime?
    \item Our algorithm consists of a large number of arithmetic operations which become a main bottleneck of the algorithm. It would be beneficial to investigate whether alternative techniques similar to those in~\cite{Stamatopoulos2024derivativepricing} are applicable to stochastic volatility models.
    \item Besides Monte Carlo simulation, another approach to pricing options is to reduce the problem to a PDE to be solved numerically~\cite{shreve2004stochastic}. It would be interesting to analyze the cost of the quantum version of this approach and compare it with ours.
    \item Can we develop efficient quantum algorithms for pricing American options under stochastic volatility? This might require quantization of the binomial tree method~\cite{shreve2005stochastic} or generalization of the results in~\cite{Doriguello2022quantum}.
    \item Recently, there has been progress in the development of low-depth amplitude estimation algorithms~\cite{wang2021minimizing, koh2022foundations, GiurgicaTiron2022lowdepthalgorithms} which are more suitable for early fault-tolerant quantum computers~\cite{zhang2022computingground, wang2022statepreparation, wang2022classicallyboosted, wang2023quantum, wang2023faster, katabarwa2023early}. It would be interesting to apply these algorithms to option pricing under stochastic volatility models and see how the cost would change in the depth-limited setting. This would help us understand the potential of achieving quantum advantage for this problem in the near future.
\end{itemize}

\appendix

\section{Fixed-point quantum arithmetic operations}
\label{apd:fixed_point_arithmetics}
In this appendix, we describe the elementary quantum arithmetic operations used in our algorithm, and analyze the costs of implementing them.

We use two's complement to represent real numbers in our circuits. That is, every number is represented by an $n$-bit string in which the most significant bit indicates the sign of the number, the next $p-1$ bits are used to denote the integral part, and the other $n-p$ bits are used to denote the fractional part. Here $n$ and $p$ are fixed throughout the computation. 

Formally, if a real number $x$ has format $x_{n-1} x_{n-2}\dots x_{n-p}. x_{n-p-1} \dots x_0$, where $x_i \in \zo$ for each $i$, then it has the value
\begin{align}
x = -x_{n-1} 2^{p-1} + x_{n-2} 2^{p-2} + \dots + x_{n-p} 2^0 + x_{n-p-1}   2^{-1} + \dots + x_0  2^{-n+p}.
\label{eq:def_two_complement}
\end{align} 
Note that if $x_{n-1}=1$, then $x$ is negative; otherwise, $x$ is positive or $0$. This number is represented by the state $\ket{x_{n-1},x_{n-2},\dots,x_0}$ on a quantum computer.

Let $S_{n,p}=\lrcb{z 2^{-n+p}:~z \in [-2^{n-1}, 2^{n-1})\cap \Z}$ be the set of real numbers that can be written as in Eq.~\eqref{eq:def_two_complement}. Note that $S_{n,p} \subset [-2^{p-1}, 2^{p-1})$. Moreover, for any $x \in [-2^{p-1}, 2^{p-1})$, let $\overline{x}=2^{p-n}\lfloor{2^{n-p}x} \rfloor$ be the largest number in $S_{n,p}$ that does not exceed $x$. This number is at most $2^{p-n}$-away from $x$.

We choose sufficiently large $p$ so that overflow never happens in our algorithm. That is, e.g., if we need to compute $x+y$ somewhere, then $|x+y|<2^{p-1}$, and similarly for other arithmetic operations. This is a necessary condition to ensure that the fixed-point representation remains valid for the entire computation.

\paragraph{Multiqubit Toffoli gates.}
The $n$-qubit Toffoli gate is defined as follows:
\begin{align}
{\rm Toffoli}_n:    \ket{x_1,x_2,\dots,x_{n-1}}\ket{x_n} \to     \ket{x_1,x_2,\dots,x_{n-1}}\ket{x_n \oplus x_1 x_2 \dots x_{n-1}}, 
\end{align}
for all $x_1, x_2,\dots, x_{n} \in \zon$. It plays a crucial role in realizing logical AND and OR operations as well as the reflection operator $R_0=I-2\ket{00\dots 0}\bra{00\dots 0}$ in amplitude estimation.

We utilize two different methods to implement this gate depending on the scenario:
\begin{itemize}
    \item The method of~\cite{Jones2013low} achieves $4(n-2)$ T-count and $n-2$ T-depth, but requires $n-1$ ancilla qubits. We employ it to implement most of the multiqubit Toffoli gates in our algorithm, as it is more time-efficient. 
    \item For $n\ge 5$, the method of~\cite{Amy2021phase} needs a single ancilla qubit, but has $16n-60$ T-count and at most $16n-60$ T-depth. We use it to implement the multiqubit Toffoli gates related to $R_0$, as it is more space-efficient (and we do not run many $R_0$'s in our algorithm).

\end{itemize}

\paragraph{Addition/Subtraction.}
The quantum adder and subtractor are defined as follows:
\begin{align}
  {\rm ADD}_n:  \ket{x}\ket{y} \to \ket{x}\ket{{x+y}},\\
  {\rm SUB}_n:  \ket{x}\ket{y} \to \ket{x}\ket{{x-y}},  
\end{align}
for $x,y \in S_{n,p}$ such that $|x \pm y| < 2^{p-1}$. We utilize the method of~\cite{Gidney2018halvingcostof} to implement these operations which has $4n-4$ T-count and $2n-2$ T-depth and requires $n-1$ ancilla qubits.

Sometimes we only need to add/subtract a known constant to/from the number in a register. So it is useful to introduce the following variant of quantum adder and subtractor:
\begin{align}
  {\rm ADD}\_{\rm CONST}_n(c):  \ket{x} \to \ket{{x+c}},\\
  {\rm SUB}\_{\rm CONST}_n(c):  \ket{x} \to \ket{{x-c}},
\end{align}
for $x \in S_{n,p}$ such that $|x \pm c| < 2^{p-1}$, for given $c \in S_{n,p}$. These operations can be implemented by a variant of the method of~\cite{Gidney2018halvingcostof} which achieves $4n-8$ T-count and $2n-4$ T-depth and demands $2n-2$ ancilla qubits. 

In addition, our algorithm also needs controlled-adders and controlled-subtractors which are defined as follows:
\begin{align}
  {\rm c-ADD}_n:  \ket{a}\ket{{x}}\ket{{y}} \to 
  \begin{cases}
  \ket{a}\ket{{x}}\ket{{x+y}}, & {\rm if}~a=1, \\
  \ket{a}\ket{{x}}\ket{{y}}, & {\rm otherwise},
  \end{cases}\\
  {\rm c-SUB}_n:  \ket{a}\ket{{x}}\ket{{y}} \to 
  \begin{cases}
  \ket{a}\ket{{x}}\ket{{x-y}}, & {\rm if}~a=1, \\
  \ket{a}\ket{{x}}\ket{{y}}, & {\rm otherwise},
  \end{cases}
\end{align}
for $a \in \zo$ and $x,y \in S_{n,p}$ such that $|x \pm y| < 2^{p-1}$ if $a=1$. One can use the method of~\cite{Gidney2018halvingcostof} to implement these operations in $8n-4$ T-count and $4n-2$ T-depth, with the help of $2n-1$ ancilla qubits.

\paragraph{Comparison.} We often need to determine whether the number in a register exceeds a threshold or not. To this end, we introduce the following quantum comparator:
\begin{align}
    {\rm COMP}\_{\rm CONST}_n(c): \ket{x} \ket{0} \to 
  \begin{cases}
  \ket{x}\ket{1}, & {\rm if}~x \ge c, \\
  \ket{x}\ket{0}, & {\rm otherwise},
  \end{cases}    
\end{align}
for all $x \in S_{n,p}$, for given $c \in S_{n,p}$. This operation can be realized by computing $x-c$, copying its sign bit, and uncomputing $x-c$, i.e., 
\begin{align}
\ket{{x}}\ket{0}\ket{0}
&\to 
\ket{{x}}\ket{x}\ket{0} \nonumber\\
&\to
\ket{{x}}\ket{x-c}\ket{0} \nonumber\\
&\to 
\ket{{x}}\ket{x-c}\ket{\sgn{x-c}} \nonumber\\
&\to
\ket{{x}}\ket{{x}}\ket{\sgn{x-c}} \nonumber\\
&\to
\ket{{x}}\ket{{0}}\ket{\sgn{x-c}},    
\end{align}
where the first, third and last steps require only CNOT gates, the second 
and fourth steps require ${\rm SUB}\_{\rm CONST}_n(c)$ and its inverse. Thus, we can implement ${\rm COMP}\_{\rm CONST}_n(c)$ in $8n-16$ T-count and $4n-8$ T-depth, with the assistance of $3n-2$ ancilla qubits.

\paragraph{Multiplication.}
The quantum multiplier is defined as follows:
\begin{align}
  {\rm MUL}_{n,p}:  \ket{{x}}\ket{{y}} \ket{{0}} \to \ket{{x}}\ket{{y}}\ket{\overline{xy}},
\end{align}
for $x,y \in S_{n,p}$ such that $|x y| < 2^{p-1}$. Using standard shift-and-add strategy, one can be implement this operation as a sequence of controlled-adders. The number of T gates in the resulting circuit is
\begin{align}
    {\rm T}_{\rm count}({\rm MUL}_{n,p}) &= 
    \sum_{i=0}^{p-1}     {\rm T}_{\rm count}({\rm c-ADD}_{n-i})  
    +  \sum_{i=1}^{n-p}     {\rm T}_{\rm count}({\rm c-ADD}_{n-i})   \\
        &=\sum_{i=0}^{p-1} [8(n-i)-4] +\sum_{i=1}^{n-p} [8(n-i)-4] \\
        &=4n^2-8n+8pn-8p^2+8p.
\end{align}
Furthermore, this circuit has T-depth $2n^2-4n+4pn-4p^2+4p$, and requires $2n-1$ ancilla qubits.

Sometimes we only need to multiply the number in a register by a known constant. Therefore, it is useful to introduce the following variant of quantum multiplier:
\begin{align}
  {\rm MUL}\_{\rm CONST}_{n,p}(c):  \ket{{x}}  \ket{{0}} \to \ket{{x}} \ket{\overline{cx}},
\end{align}
for $x \in S_{n,p}$ such that $|cx| < 2^{p-1}$, for given $c \in S_{n,p}$. This operation can be implemented as a sequence of adders. The number of T gates in the resulting circuit is
\begin{align}
    {\rm T}_{\rm count}({\rm MUL}\_{\rm CONST}_{n,p}) &= 
    \sum_{i=0}^{p-1}     {\rm T}_{\rm count}({\rm ADD}_{n-i})  
    +  \sum_{i=1}^{n-p}     {\rm T}_{\rm count}({\rm ADD}_{n-i})   \\
        &=\sum_{i=0}^{p-1} [4(n-i)-4] +\sum_{i=1}^{n-p} [4(n-i)-4] \\
        &=2n^2-6n+4pn-4p^2+4p.
\end{align}
Moreover, this circuit has T-depth $n^2-3n+2pn-2p^2+2p$, and uses $n-1$ ancilla qubits.

\paragraph{Square Root.}
The quantum version of square root is defined as:
\begin{align}
    {\rm SQRT}_{n}: \ket{{x}}\ket{{0}} \to \ket{{x}}\ket{\overline{\sqrt{x}}},
\end{align}
for all $x \in S_{n,p} \cap \R^{\ge 0}$. \cite{munoz2018t} gives a strategy for implementing a similar operation which maps $\ket{{x}}\ket{{0}}$ to $\ket{Z_x}\ket{\overline{\sqrt{x}}}$, where $Z_x$ depends on $x$. We change it to a clean version by the do-copy-undo trick, and also replace its adders by the ones of~\cite{Gidney2018halvingcostof} which cost fewer T gates. The resulting circuit has 
\begin{align}
    {\rm T}_{\rm count}({\rm SQRT}_{n}) &= 2\sum_{i=1}^{\lceil n/2\rceil} {\rm T}_{\rm count}({\rm ADD}_{2i}) + 2 {\rm T}_{\rm count}({\rm c-ADD}_{2 \lceil n/2 \rceil}) \\
        &=2\sum_{i=1}^{\lceil n/2 \rceil} (8i-4) + 2(16\lceil n/2 \rceil -4) \\
        &= 8 \lceil n/2 \rceil ^2+32 \lceil n/2 \rceil -8
\end{align}
T gates. Furthermore, it has T-depth $4 \lceil n/2 \rceil ^2+16 \lceil n/2 \rceil-4$ and uses $\lceil 3.5n \rceil $ ancilla qubits.

\paragraph{Piecewise Polynomial.} A function $g$ on a domain $[a,b)$ is called an $(M, d)$-piecewise polynomial if we can partition $[a, b)$ into $M$ subintervals $[a_0, a_1)$, $[a_1, a_2)$, $\dots$, $[a_{M-1}, a_M)$ (where $a=a_0<a_1<\dots<a_M=b$) such that $g$ equals a degree-$d$ polynomial $p_i$ on the $i$-th subinterval, i.e., $g(x)=p_i(x)$ for all $x \in [a_{i-1}, a_{i})$, for $i=1,2,\dots,M$. Piecewise polynomials are interesting because many natural functions, e.g., $\operatorname{tanh}(x)$, $\operatorname{sin}(x)$, $\operatorname{exp}(-x^2)$, can be well-approximated by them with small numbers of subintervals or low polynomial degrees. 

Let $pp$ be an $(M, d)$-piecewise polynomial. We define its quantum version as:
\begin{align}
    {\rm PPoly}_{n,p,M,d}: \ket{{x}}\ket{{0}} \to
    \ket{{x}}\ket{\overline{pp(x)}}
\end{align}
for $x \in S_{n,p}$ such that $|pp(x)| < 2^{p-1}$. \cite{haner2018optimizing}  develops a method to implement a similar operation which maps $\ket{{x}}\ket{{0}}\ket{0}$ to $\ket{{x}}\ket{\overline{pp(x)}} \ket{\Gamma_x}$, where $\Gamma_x$ depends on $x$. Here we change it to a clean version by the do-copy-undo trick, and also replace its adders and multiqubit Toffoli gates with the more efficient ones of~\cite{Gidney2018halvingcostof} and~\cite{Jones2013low} respectively. This leads to a circuit that contains
\begin{align}
        {\rm T}_{\rm count}( {\rm PPoly}_{n,p,M,d}) &= 
    4M {\rm T}_{\rm count}({\rm COMP}\_{\rm CONST}_n)    
    +  2d [{\rm T}_{\rm count}({\rm MUL}_{n,p}) + {\rm T}_{\rm count}({\rm ADD}_n)] \nonumber\\
    &+4dM {\rm T}_{\rm count}({\rm Toffoli}_{\lceil \logb{M}\rceil+1}) \\
        &= 4M(8n-16)+2d[(4n^2-8n+8pn-8p^2+8p) + (4n-4)] \nonumber\\
        &+16dM(\lceil \logb{M}\rceil-1)\\
        &= 8d(n^2-n+2pn-2p^2+2p-1) + 32M(n-2) \nonumber \\
        &+ 16dM (\lceil\logb{M}\rceil-1)
\end{align}
T gates. Moreover, it has T-depth $4d \max(n^2-2n+2pn-2p^2+2p, M (\lceil\logb{M}\rceil-1)) + 16M(n-2)+4d(n-1)$ and requires $(d+4)n + 2\lceil \logb{M} \rceil$ ancilla qubits.

\paragraph{Exponential.} In our algorithms, we need to compute the exponentials of log returns which are then used to compute the payoff of an option. We adopt the strategy of~\cite{haner2018optimizing} to approximately implement such exponentiation, obtaining the unitary operation
\begin{align}
    {\rm EXP}_{n,p,\epsilon}: \ket{x}\ket{0} \to
    \ket{x}\ket{\overline{{\rm pp_{exp}}(x)}}, 
\end{align}
for $x \in S_{n,p} \cap D$ such that $|e^x| < 2^{p-1}$. Here $\epsilon>0$ is the target accuracy, and ${\rm pp_{exp}}$ is an $(M_\epsilon, d_\epsilon)$-piecewise polynomial that is $\epsilon$-close to $\exp{x}$ on the relevant domain $D$. In our case, the domain is $D=[R_{\rm min}, R_{\rm max}]$, where $R_{\rm min}$ and $R_{\rm max}$ are the minimum and maximum possible log returns at any time in the stochastic process, respectively. The cost of implementing ${\rm EXP}_{n,p,\epsilon}$ is identical to that of ${\rm PPoly}_{n,p,M_\epsilon,d_\epsilon}$.

\paragraph{Arcsin of Square Root.} The implementation of $U_3$ in Section \ref{sec:implement_u3} requires to compute the arcsin of square root of a given non-negative number, i.e., $\arcsin{\sqrt{x}}$ for $x\ge 0$. We employ the strategy of~\cite{Chakrabarti2021thresholdquantum} (which generalizes the one of~\cite{haner2018optimizing}) to approximately implement this arithmetic operation, obtaining the unitary operation
\begin{align}
    {\rm ARCSIN}\_{\rm SQRT}_{n,p,\epsilon}: \ket{x}\ket{0} \to
    \ket{x}\ket{\overline{g(x)}}.
\end{align}
for all $x \in S_{n,p}\cap [0,1]$, where 
\begin{align}
g(x)=\begin{cases}
    {\rm pp_{arcsin}}(\overline{\sqrt{x}}), &{\rm~if~} x<1/4, \\
    \frac{\pi}{2}-{\rm pp_{arcsin}}(\overline{\sqrt{1-x}}), &{\rm~otherwise},    
\end{cases} 
\end{align}
 in which ${\rm pp_{arcsin}}$ is an $(M_\epsilon, d_\epsilon)$-piecewise polynomial that is $\epsilon$-close to $\arcsin{x}$ on the domain $[-1/2, 1/2]$. Note that
\begin{align}
    \arcsin{\sqrt{x}} = \frac{\pi}{2} - \arcsin{\sqrt{1-x}},
\end{align}
for all $x \in [0, 1]$. Thus, if we ignore the round-off error in square rooting, then $g(x)$ is always $\epsilon$-close to $\arcsin{\sqrt{x}}$ regardless of the value of $x$.

Specifically, given an initial state $\ket{x}\ket{0}$, we run the following steps to obtain the target state $\ket{x}\ket{\overline{g(x)}}$:
\begin{enumerate}
    \item Append a $1$-qubit register $W$ and set its state to $\ket{z}$, where $z=1$ if $x<1/4$, and $0$ otherwise, by using ${\rm COMP}\_{\rm CONST}_n(1/4)$. 
    \item Append an $n$-qubit register $X$ and set its state to $\ket{y}$, where $y=x$ if $z=1$, and $1-x$ otherwise, by using ${\rm c-SUB}_n$ and Clifford gates.
    \item Append an $n$-qubit register $Y$ and set its state to $\ket{w}$, where $w=\overline{\sqrt{y}}$, by using ${\rm SQRT}_n$.
    \item Append an $n$-qubit register $Z$ and set its state to $\ket{v}$, where $v=\overline{{\rm pp_{arcsin}}(w)}$, by using ${\rm PPoly}_{n,p,M_\epsilon,d_\epsilon}$.
    \item Set the state of register $Z$ to $\ket{v}$ if $z=1$, and $\ket{\overline{\pi/2-v}}$ otherwise, by using ${\rm c-SUB}_n$ and Clifford gates.
    \item Copy the value of register $Z$ to the target register by using CNOT gates.
    \item Perform the inverse of the first five steps to restore the ancilla registers $W$, $X$, $Y$ and $Z$ to the zero state.
\end{enumerate}
The resulting circuit has T-count
\begin{align}
{\rm T}_{\rm count}({\rm ARCSIN}\_{\rm SQRT}_{n,p,\epsilon}) =& 
2{\rm T}_{\rm count}({\rm PPoly}_{n,p,M_\epsilon,d_\epsilon})
+2{\rm T}_{\rm count}({\rm COMP}\_{\rm CONST}_n) \nonumber\\
&+2{\rm T}_{\rm count}({\rm SQRT}_n)+4{\rm T}_{\rm count}({\rm c-SUB}_n) \\
=& 16d_\epsilon(n^2-n+2pn-2p^2+2p-1) + 64M_\epsilon(n-2) \nonumber  \\
& + 32d_\epsilon M_\epsilon (\lceil\logb{M_\epsilon}\rceil-1) +16 \lceil n/2 \rceil ^2+48n\nonumber \\
&+64 \lceil n/2\rceil-64
\end{align}
Similarly, one can get that its T-depth is $8d_\epsilon \max(n^2-2n+2pn-2p^2+2p, M_\epsilon (\lceil\logb{M_\epsilon}\rceil-1))
+ 32M_\epsilon(n-2) +8d_\epsilon(n-1) +8 \lceil n/2 \rceil^2+24n
+32 \lceil n/2\rceil-32$. In addition, the circuit requires $(d_\epsilon+7)n + 2\lceil \logb{M_\epsilon} \rceil + 1$ ancilla qubits.

We summarize the costs of the above quantum arithmetic operations in 
Table 8.

\begin{adjustbox}{addcode={\begin{minipage}{\width}}{\caption{
The costs of implementing the fixed-point quantum arithmetic operations used in this work.}\end{minipage}}, rotate=90, center, float=table}
\begin{tabular}{ |c|c|c|c|c|c| } 
\hline
\thead{Operation} & \thead{T-count} & \thead{T-depth} & \thead{Number of \\ ancilla qubits} 
& \thead{Techniques} \\
 \hline 
\thead{${\rm Toffoli}_n$} &  $4n-8$ & $n-2$ & $n-1$ 
& \cite{Jones2013low}\\ 
\thead{${\rm Toffoli}_n$} &  $16n-60$ & $16n-60$ & $1$ 
& \cite{Amy2021phase}\\ 
\thead{${\rm ADD}_n$} & $4n-4$ & $2n-2$ & $n-1$ 
& \cite{Gidney2018halvingcostof}\\
\thead{${\rm SUB}_n$} & $4n-4$ & $2n-2$ & $n-1$ 
& \cite{Gidney2018halvingcostof}\\
\thead{${\rm c-ADD}_n$} & $8n-4$ & $4n-2$ & $2n-1$  
& \cite{Gidney2018halvingcostof}\\
\thead{${\rm c-SUB}_n$} & $8n-4$ & $4n-2$ & $2n-1$  
& \cite{Gidney2018halvingcostof}\\
\thead{${\rm ADD}\_{\rm CONST}_n$} & $4n-8$ & $2n-4$ & $2n-2$ 
& \cite{Gidney2018halvingcostof}\\
\thead{${\rm SUB}\_{\rm CONST}_n$} & $4n-8$ & $2n-4$ & $2n-2$ 
& \cite{Gidney2018halvingcostof}\\
\thead{${\rm COMP}\_{\rm CONST}_n$} & $8n-16$ & $4n-8$ & $3n-2$ 
& \cite{Gidney2018halvingcostof}\\
\thead{${\rm MUL}_{n,p}$} & $4n^2-8n+8pn-8p^2+8p$ & $2n^2-4n+4pn-4p^2+4p$
& $2n-1$ 
& \cite{Gidney2018halvingcostof}\\
\thead{${\rm MUL}\_{\rm CONST}_{n,p}$} & $2n^2-6n+4pn-4p^2+4p$ & $n^2-3n+2pn-2p^2+2p$ & $n-1$ 
& \cite{Gidney2018halvingcostof}\\
\thead{${\rm SQRT}_{n}$} &  $8\lceil n/2 \rceil ^2+32 \lceil n/2 \rceil-8$ & $4 \lceil n/2 \rceil ^2 + 16 \lceil n/2 \rceil-4$ & $\lceil 3.5n \rceil$ 
& \cite{Gidney2018halvingcostof}, \cite{munoz2018t}\\
\thead{${\rm PPOLY}_{n,p,M,d}$} & \makecell{$8d(n^2-n+2pn-2p^2+2p-1)$ \\ $+ 32M(n-2) + 16dM (\lceil\logb{M}\rceil-1)$} &  \makecell{$
4d \max(n^2-2n+2pn-2p^2+2p, $\\ $ M (\lceil\logb{M}\rceil-1))$ \\ 
$ + 16M(n-2)+4d(n-1)$
}
&\makecell{$(d+4)n $\\$+ 2\lceil \logb{M} \rceil$} 
&\makecell{\cite{Jones2013low}, \cite{Gidney2018halvingcostof},\\ \cite{haner2018optimizing}}\\
\thead{${\rm EXP}_{n,p,\epsilon}$} & \makecell{$8d_\epsilon(n^2-n+2pn-2p^2+2p-1)$ \\ $+ 32M_\epsilon(n-2) + 16d_\epsilon M_\epsilon (\lceil\logb{M_\epsilon}\rceil-1)$} &  \makecell{$4d_\epsilon \max(n^2-2n+2pn-2p^2+2p, $\\ $ M_\epsilon (\lceil\logb{M_\epsilon}\rceil-1))$ \\ 
$ + 16M_\epsilon(n-2)+4d_\epsilon(n-1)$}
& \makecell{$(d_\epsilon+4)n $\\$+ 2\lceil \logb{M_\epsilon} \rceil$} 
&
\makecell{\cite{Jones2013low}, \cite{Gidney2018halvingcostof},\\ \cite{haner2018optimizing}}\\
\thead{${\rm ARCSIN}\_{\rm SQRT}_{n,p,\epsilon}$} & \makecell{
$16d_\epsilon(n^2-n+2pn-2p^2+2p-1)$ \\
$+ 64M_\epsilon(n-2)+32d_\epsilon M_\epsilon (\lceil\logb{M_\epsilon}\rceil-1)$ \\ $+16 \lceil n/2 \rceil ^2+48n+64\lceil n/2 \rceil-64$
} &  \makecell{
$8d_\epsilon \max(n^2-2n+2pn-2p^2+2p, $ \\ 
$M_\epsilon (\lceil\logb{M_\epsilon}\rceil-1))$ \\
$+ 32M_\epsilon(n-2) +8d_\epsilon(n-1)$ \\
$+8 \lceil n/2 \rceil^2+24n
+32 \lceil n/2\rceil -32$
}
& \makecell{$(d_\epsilon+7)n$\\ $+ 2\lceil \logb{M_\epsilon} \rceil + 1$} 
& \makecell{\cite{Chakrabarti2021thresholdquantum}, \cite{Jones2013low}, \\ \cite{Gidney2018halvingcostof}, \cite{haner2018optimizing}}\\
 \hline
\end{tabular}
\label{table:costs_of_arithmetic_operations}
\end{adjustbox}

\section{Resource-optimized block-encoding of the sine function}
\label{apd:opt_circuit_usin}

In this appendix, we present a low-cost block-encoding of the sine function. This construction is useful for the implementation of $U_{3,2}$ in Section \ref{sec:implement_u3} and $U_{\rm gauss}$ in Appendix \ref{apd:prep_gaussian_states}.

Let $n \in \Z^+$, $N=2^n$, and $a, b \in \R$ satisfy $-1 \le a < b \le 1$. We aim to implement an $(n+1)$-qubit unitary operation $U_{\rm sin}$ such that
\begin{align}
    U_{\rm sin} \ket{i} \ket{0} = \ket{i} \lrsb{\mysin{y(i)}\ket{0} + \mycos{y(i)}\ket{1}}, ~\forall i=0,1,\dots,N-1,
    \label{eq:def_u_for_sin}
\end{align}
where $y(i)=a + (b-a) i/N$ for each $i$. Note that $U_{\rm sin}$ satisfies
\begin{align}
\lrb{I_n \otimes \bra{0}} U_{\rm sin} \lrb{I_n \otimes \ket{0}}
=\sum_{i=0}^{N-1} \mysin{y(i)} \ket{i}\bra{i}.
\end{align}    
In other words, $U_{\rm sin}$ is a block-encoding of the (shifted) sine function.

We assume the Clifford + T gate set, and aim to optimize the T-count and depth, while neglecting the costs of Clifford gates. This is so because the quantum computational costs are dominated by T gates~\cite{bravyi2005universal}. In the circuits that implement the state-preparation method in~\cite{mcardle2022quantum}, many of the T gates are used in the fault-tolerant implementation of rotation gates~\cite{bocharov2015efficient}, i.e., $R_z(\theta)$. Therefore, we aim to reduce the rotation gate-count and depth. 

We start with a basic circuit for $U_{\rm sin}$ which contains $n$ controlled-$R_y$ rotations, one uncontrolled $R_y$ rotation and an $X$ gate: 
\begin{equation}\label{circ:Usin}
\scalebox{0.75}{ 
    \Qcircuit @C=.7em @R=.5em @!R{
    \lstick{\ket{z}} & \gate{R_y(2^{1-n}\Delta)}& \gate{R_y(2^{2-n}\Delta)} &\qw&\cdots& & \gate{R_y(\Delta)}& \gate{R_y(2a)}&\gate{X} &\qw \\
    \lstick{\ket{i_1}} & \ctrl{-1}& \qw &\qw&\cdots& &\qw& \qw &\qw &\qw\\
    \lstick{\ket{i_2}} & \qw& \ctrl{-2} &\qw&\cdots& &\qw& \qw &\qw &\qw &\mbox{,}\\
    \lstick{\vdots} &  &  & & & \lstick{\vdots} & & & &\\
    \lstick{\ket{i_n}} & \qw& \qw &\qw&\cdots& &\ctrl{-4}& \qw &\qw&\qw
    }
}
\end{equation}
where $\Delta=b-a$, and $i_j$ is the $j$-th bit of $i$, i.e., $i=\sum_{j=1}^n 2^{j-1}i_j$. One can verify that this circuit implements a unitary operation that fulfills the condition \eqref{eq:def_u_for_sin}.
 
Conventionally, the controlled-$R_y(\theta)$ gates are compiled using~\cite{Nielsen_Chuang_2000}:
\begin{equation}
\label{circ:cry_synthesis}
\scalebox{0.75}{
    \Qcircuit @C=.7em @R=.5em @!R{
    & \gate{R_y(\theta)} & \qw \\
    & \ctrl{-1} & \qw
    }
    =
    \Qcircuit @C=1em @R=.7em @!R{
    &\gate{S^\dag}&\gate{H}& \targ & \gate{R_z(-\theta/2)} &\targ & \gate{R_z(\theta/2)}&\gate{H}&\gate{S}&\qw \\
    &\qw&\qw& \ctrl{-1} & \qw &\ctrl{-1} & \qw &\qw& \qw &\qw
    }}.
\end{equation}
This leads to an overall $U_{\rm sin}$ circuit with a $R_z$-count and $R_z$-depth of $2n$. Note that the final uncontrolled $R_y$ operation is transformed by the $S$ and $H$ gates into an $R_z$ rotation, and thus, can be merged with the previous $R_z$ rotation. 

Here we compile the controlled-rotation gates, with the assistance of ancilla qubits, into a single layer of uncontrolled rotation gates (up to Clifford gates) instead. Specifically, we repeatedly apply the circuit identities:
\begin{equation}
\scalebox{0.75}{
    \Qcircuit @C=.7em @R=.5em @!R{
    \lstick{\ket{0}}& \qw &\qw & \lstick{\ket{0}}& \lstick{\ket{0}} &\qw & \qw & \targ& \targ & \gate{R_z(-\theta/2)} & \targ & \targ& \qw & \qw & \qw & & \lstick{\ket{0}}\\
    & \gate{R_y(\theta)} & \qw &
\push{\rule{.3em}{0em}=\rule{.3em}{0em}} & & \gate{S^\dag} & \gate{H} & \qw &\ctrl{-1} & \gate{R_z(\theta/2)} & \ctrl{-1} & \qw& \gate{H} & \gate{S} & \qw & & \mbox{,}\\
    & \ctrl{-1} & \qw & & &\qw & \qw & \ctrl{-2} & \qw & \qw & \qw & \ctrl{-2} & \qw & \qw & \qw
    }}
\end{equation}
taken from~\cite{Wang2021resourceoptimized}, and
\begin{equation}
\scalebox{0.75}{
    \Qcircuit @C=.7em @R=.5em @!R{
    & \qw & \qw & \qw & \targ & \targ & \gate{R_z(\theta_3)} & \qw & & &\targ &\targ &\gate{R_z(\theta_3)} & \qw & \qw & \qw \\
    & \gate{R_z(\theta_1)} & \targ & \targ & \qw & \qw & \qw & \qw & & &\qw & \qw &\gate{R_z(\theta_1)} &\targ & \targ & \qw \\
    & \gate{R_z(\theta_2)} & \ctrl{-1} & \qw &\ctrl{-2} & \qw & \gate{R_z(\theta_4)} & \qw &
\push{\rule{.3em}{0em}=\rule{.3em}{0em}} &  & \ctrl{-2} &\qw &\gate{R_z(\theta_2+\theta_4)} & \ctrl{-1} & \qw & \qw \\
    & \qw & \qw & \ctrl{-2} &\qw & \qw & \qw & \qw & & &\qw & \qw & \qw & \qw & \ctrl{-2} & \qw\\
    & \qw & \qw & \qw & \qw & \ctrl{-4} & \qw & \qw & & &\qw &\ctrl{-4} & \qw&\qw & \qw& \qw
    }}
\end{equation}
to~\eqref{circ:Usin} and obtain
\begin{equation}\label{circ:Usin_opt}
\scalebox{0.75}{
\Qcircuit @C=.7em @R=.7em @!R{  
  \lstick{\ket{z}} & \gate{S^\dag} & \gate{H} & \qw &\ctrl{1} &\gate{R_z\left( \lrb{1-2^{-n}} \Delta +2 a \right )} & \ctrl{1} &\qw & \gate{H} & \gate{S} &\gate{X}&\qw\\
  \lstick{\ket{0}} & \targ & \qw  & \qw & \targ & \gate{R_z\left(-2^{-n}\Delta\right)} & \targ&\qw & \qw & \targ & \qw &\qw\\
  \lstick{\ket{0}} & \qw & \targ  & \qw & \targ \qwx &\gate{R_z\left(-2^{1-n}\Delta\right)} & \targ \qwx  &\qw & \targ & \qw& \qw&\qw\\
  \lstick{\vdots} &  &  &  & \qwx & \lstick{\vdots} & \qwx &  &  & &\\
  \lstick{\ket{0}} & \qw & \qw & \targ & \targ\qwx & \gate{R_z\left(- 2^{-1}\Delta\right)} & \targ \qwx &\targ &\qw&\qw&\qw&\qw&\mbox{,}\\
  \lstick{\ket{i_1}} & \ctrl{-4} & \qw & \qw &\qw &\qw&\qw &\qw & \qw &\ctrl{-4}&\qw&\qw\\
  \lstick{\ket{i_2}} & \qw & \ctrl{-4} & \qw&\qw&\qw&\qw&\qw &\ctrl{-4} &\qw&\qw&\qw\\
  \lstick{\vdots} &  &  & & & \lstick{\vdots} & & & & & &\\
  \lstick{\ket{i_n}} & \qw & \qw & \ctrl{-4}&\qw&\qw&\qw &\ctrl{-4}&\qw &\qw &\qw&\qw \gategroup{2}{6}{5}{6}{.7em}{--}
}
}
\end{equation}
where the $R_z$-count and $R_z$-depth have been reduced to $n+1$ and 1, respectively, at the cost of $n$ reusable ancilla qubits. Furthermore, we note that this circuit optimization is applicable to any series of consecutive controlled-rotation gates about the same axis, which are applied to the same target qubit and controlled by different qubits, e.g.,~\eqref{circ:Usin}, regardless of the rotation angles.

Next, we show an optimized implementation of a layer of $n$ $R_z$ gates of the form $\otimes_{i=0}^{n-1} R_z(2^i \theta)$, which we use to apply the gates in the dashed box in~\eqref{circ:Usin_opt}. We begin with the phase catalysis circuit from~\cite{Gidney2019efficientmagicstate}:
\begin{equation}\label{circ:gidney}
\scalebox{0.75}{
\Qcircuit @C=.7em @R=.5em @!R{
   \lstick{\ket{\psi_0}} & \qw & \ctrl{1} & \qw & \ctrl{3} & \qw & \ctrl{3} &\qw & \qw & \ctrl{2}&\qw &\qw & \rstick{ R_z(\theta) \ket{\psi_0} } \\
    \lstick{\ket{\psi_1}} & \qw & \targ & \ctrl{2} & \qw & \qw & \qw &\ctrl{2} &\ctrl{1} &\targ &\qw&\qw&\rstick{ R_z(\theta) \ket{\psi_1}}\\
    \lstick{ R_z(\theta) \ket{+} }& \targ & \targ\qwx & \ctrl{1} & \qw & \qw & \qw &\ctrl{1} &\targ & \targ \qwx &\targ&\qw&\rstick{ R_z(\theta) \ket{+} }\\
    & & & {} & \targ & \gate{R_z\left( 2\theta \right)} & \targ & \qw & & &\\
}}
\end{equation}
where we have borrowed the notation for a relative-phase Toffoli gate, also known as temporary logical-AND, and its measurement-feedforward uncomputation, i.e.,
\begin{equation}
\scalebox{0.75}{
\Qcircuit @C=0.7em @R=.5em {
& \qw & \ctrl{2} & \qw & & & & \ctrl{2} & \qw & \targ & \gate{T^\dagger} & \targ & \qw & \qw & \qw &  & & & \qw & \ctrl{2} & \qw & & &\qw &\qw & \ctrl{1}& \qw \\
& \qw & \ctrl{1} & \qw & \push{\rule{.3em}{0em}=\rule{.3em}{0em}} & &  & \qw & \ctrl{1}& \targ \qwx & \gate{T^\dagger} & \targ \qwx & \qw & \qw & \qw & & \mbox{ , } & & \qw & \ctrl{1} & \qw & \push{\rule{.3em}{0em}=\rule{.3em}{0em}} & &\qw & \qw & \gate{Z} & \qw \\
& & & \qw &  & & \lstick{\ket{T}} &\targ & \targ & \ctrl{-2} & \gate{T}& \ctrl{-2} & \gate{H} & \gate{S}& \qw & &  & & \qw&\qw &  & & &\gate{H} & \meter & \cctrl{-1} & 
}}
\end{equation}
from~\cite{Gidney2018halvingcostof}. Using~\eqref{circ:gidney}, two $R_z(\theta)$ can be applied at the cost of a $R_z(2 \theta)$ and a $R_z(\theta) \ket{+}$ catalyst state. We now show that the desired layer of $\R_z$ gates can be effected by applying~\eqref{circ:gidney} recursively. In particular, instead of applying the $R_z(2 \theta)$ in~\eqref{circ:gidney} directly, we apply it using another~\eqref{circ:gidney} circuit where we let $\theta$ is replaced by $2\theta$; in the second application of~\eqref{circ:gidney}, the $R_z(4\theta)$ is effected using yet another~\eqref{circ:gidney}, and so on and so forth. In order to apply $\otimes_{i=0}^{n-1} R_z(2^i \theta)$,~\eqref{circ:gidney} needs to be applied $n$ times recursively, as shown below
\begin{equation}\label{circ:gidney_recursive}
\scalebox{0.65}{
\Qcircuit @C=.7em @R=.5em @!R{
   \lstick{\ket{\psi_{0}}} & \qw &\ctrl{2} &\qw & \ctrl{3} &\qw &\qw&\qw &\qw &\qw &\qw &\qw &\qw  &\qw  &\qw  &\qw  &\qw &\qw &\qw  & \qw& \qw& \qw &\qw &\qw &\qw & \qw& \qw&\qw & \qw& \ctrl{3} &\qw &\qw &\ctrl{1} &\qw&\qw&\rstick{ R_z( \theta) \ket{\psi_{0}} } \\
    \lstick{\ket{\psi_{1}}}& \qw &\targ &\ctrl{2}&\qw &\qw &\qw &\qw & \qw&\qw &\qw  &\qw & \qw &\qw  &\qw  &\qw  &\qw & \qw& \qw &\qw &\qw &\qw  &\qw &\qw & \qw&\qw &\qw &\qw &\qw & \qw & \ctrl{2}&\ctrl{1} &\targ &\qw &\qw& \rstick{ R_z(\theta) \ket{\psi_{1}} }\\
    \lstick{ R_z(\theta) \ket{+} }& \targ &\targ \qwx &\ctrl{1}&\qw &\qw &\qw &\qw & \qw& \qw&\qw  &\qw &\qw  & \qw & \qw &\qw  &\qw &\qw &\qw  & \qw&\qw & \qw &\qw &\qw &\qw &\qw &\qw & \qw& \qw& \qw & \ctrl{1}&\targ &\targ\qwx &\targ &\qw & \rstick{ R_z(\theta) \ket{+} }\\
    & & & \qwx &\targ \qwx&\ctrl{1}&\qw&\ctrl{1} & \qw&  & \cdots & &  &  &  &  & & &  & & &  & & \cdots& & \ctrl{1}&\qw & \qw &\ctrl{1}& \targ \qwx&\qw  & & &  &\\
    & & & & & &  &  & & &  & &  &  &  &  & & &  & & &  & & & & & & & &  & & &  &\\
    & & & & & & & \ddots& & &  & &  &  &  &  & & &  & & & & & & &\adots &  & & & & & &  & & &  &\\
    & & & & & &  &  & & &  & &  &  &  &  & & &  & & &  & & & & & & & &  & & &  &\\
    & & & & & & & \qwx& \targ \qwx&\ctrl{2} & \qw &  \ctrl{3}&  \qw&  \qw&\qw  &\qw  &\qw &\qw &\qw  &\qw & \ctrl{3}& \qw & \qw& \ctrl{2}& \targ\qwx& \qw\qwx& & & &  & & &  &\\
    & & & & & & & \lstick{\ket{\psi_{n-1}}}& \qw&\targ\qwx & \ctrl{2}& \qw  & \qw & \qw & \qw &  \qw& \qw&\qw &\qw  &\qw &\qw &\ctrl{2} & \ctrl{1} & \targ \qwx &\qw &\qw & \rstick{ R_z(2^{n-2}\theta) \ket{\psi_{n-1}}} & & &  & & &  & \mbox{.}\\
    & & & & & & & \lstick{ R_z(2^{n-2}\theta) \ket{+} }& \targ & \targ\qwx & \ctrl{1}& \qw & \qw & \qw & \qw & \qw & \qw&\qw &\qw  & \qw&\qw &\ctrl{1} & \targ & \targ \qwx &\targ &\qw & \rstick{ R_z(2^{n-2}\theta) \ket{+}} & & &  & & &  &\\
    & & & & & & & & & & & \targ& \ctrl{1} & \qw & \ctrl{3} & \qw & \ctrl{3} &\qw & \qw & \ctrl{2}& \targ& \qw & & & & & & & & & &  &\\
    & & & & & & & & & & \lstick{\ket{\psi_n}} & \qw & \targ & \ctrl{2} & \qw & \qw & \qw &\ctrl{2} &\ctrl{1} &\targ &\qw&\qw&\rstick{ R_z(2^{n-1}\theta) \ket{\psi_n}} & & & & & & & & &  &\\
    & & & & & & & & & & \lstick{ R_z(2^{n-1}\theta) \ket{+} }& \targ & \targ\qwx & \ctrl{1} & \qw & \qw & \qw &\ctrl{1} &\targ & \targ \qwx &\targ&\qw&\rstick{ R_z(2^{n-1}\theta) \ket{+} } & & & & & & & & &  &\\
    & & & & & & & & & & & & & {} & \targ & \gate{R_z\left( 2^{n}\theta \right)} & \targ & \qw & & & & & & & & & & &  &\\
}}
\end{equation}
We apply this recursive phase catalysis circuit with $\theta = -2^{-n} \Delta$, and remove $\ket{\psi_0}$ and the CNOTs controlled by it to implement the desired $R_z$ layer, i.e., $\otimes_{i=1}^{n} R_z(-2^{-i} \Delta)$, in~\eqref{circ:Usin_opt}.
It is worth pointing out that this circuit is directly applicable to quantum simulation algorithms~\cite{nam2019low,Shaw2020quantumalgorithms,campbell2021early,kan2022lattice,kan2022simulating} where $R_z$ layers of this exact form are widely used to perform phase kickbacks.

Now we analyze the savings in computational costs due to our optimization. Here, we synthesize each $R_z$ rotation using repeat-until-success (RUS) circuits in~\cite{bocharov2015efficient}. Then, each $R_z$ rotation costs $1.15 \logb{1/\epsilon}$ T gates, where $\epsilon$ is the synthesis error per rotation gate. 

Suppose we aim to implement a $U_{\rm sin}$ with precision $\epsilon_{\rm sin}$. Then, each $R_z$ rotation in the unoptimized circuit, i.e., \eqref{circ:Usin} compiled using \eqref{circ:cry_synthesis}, can incur at most $\epsilon_{\rm sin}/(2n)$ error, and thus costs $1.15 \logb{2n/\epsilon_{\rm sin}}$ T gates. Then, the entire circuit costs $3.3 n \logb{2n/\epsilon_{\rm sin}}$ T gates. The T-depth is the same as the T-count. Moreover, the circuit requires 1 ancilla qubit (which is necessary for the synthesis of all $R_z$ rotations).

In contrast, our optimized circuit, assisted by a catalyst state, i.e., $\otimes_{i=0}^{n-1} R_z(2^i \theta)$ $\ket{+}^{\otimes n}$, requires only $n$ relative-phase Toffoli gates and two $R_z$ gates. Since the catalyst state only needs to be synthesized once, after which it can be reused, and $U_{\sin}$ is typically invoked many times, i.e., $\gg n$, in an algorithm, for the sake of comparing with the unoptimized circuit, we neglect the one-time synthesis cost of $\ket{\psi_g}$, which we will address shortly. Then, each $R_z$ has an error budget of $\epsilon_{\sin}/2$, and thus costs $1.15 \logb{2/\epsilon_{\rm sin}}$ T gates. In total, the circuit costs $4n + 3.3\logb{2/\epsilon_{\rm sin}}$ T gates. The T-depth is then $n + 1.15 \logb{2/\epsilon_{\rm sin}} + 1$. The number of ancilla qubits required is $2n+2$ (two for synthesizing the $R_z$ gates simultaneously and 
the remaining $3n$
are shown in~\eqref{circ:gidney_recursive}). We stress that the ancilla qubit count is insignificant compared to that required by other subroutines, e.g., arithmetic circuits, of the overall pricing algorithm. In summary, our optimization reduces the T-count and T-depth of $U_{\sin}$ from $O(n\logb{n/\epsilon_{\rm sin}})$ to $O(n+ \logb{1/\epsilon_{\sin}})$, while retaining small constant factors and effectively not affecting the overall qubit count.


Let $U_{\rm sin}(n, \epsilon_{\rm sin})$ denote the final circuit consisting of Clifford and T gates. Our results can be summarized as follows. With the conventional method, we have
\begin{align}
    \tcount{U_{\rm sin}(n, \epsilon_{\rm sin})}&=3.3n\logb{2n/\epsilon_{\rm sin}}, \\
    \tdepth{U_{\rm sin}(n, \epsilon_{\rm sin})}&=3.3 n\logb{2n/\epsilon_{\sin}},  \\
    \numancilla{U_{\rm sin}(n, \epsilon_{\rm sin})}&=1,
\end{align}
whereas with our optimization, we get
\begin{align}
    \tcount{U_{\rm sin}(n, \epsilon_{\rm sin})}&=4n + 3.3\logb{2/\epsilon_{\rm sin}}, \\
    \tdepth{U_{\rm sin}(n, \epsilon_{\rm sin})}&=n + 1.15 \logb{2/\epsilon_{\rm sin}} + 1,  \\
    \numancilla{U_{\rm sin}(n, \epsilon_{\rm sin})}&=
    3n+2.
\end{align}

Returning to the catalyst state synthesis, we note that the state can be synthesized with $n$ $R_z$ rotations at a cost of $1.15n \logb{n/\epsilon_{\rm cat}}$, where $\epsilon_{\rm cat}$ is the error budget for this synthesis. Note that this only needs to be done once, as the catalyst state can be reused. In our algorithm, we choose sufficiently small $\epsilon_{\rm cat}$ so that the error in the final result caused by the imperfection of catalyst state preparation is negligible. For example, for the instances considered in Section \ref{sec:case_studies}, 
we can pick $\epsilon_{\rm cat}=10^{-50}$ to ensure that this error is at most $10^{-30}$. Then the synthesis of the catalyst state consumes at most $10^5$ T gates, whereas the other components of the algorithm take more than $10^{10}$ T gates. In other words, the cost of synthesizing the catalyst state is insignificant compared to the total cost of the algorithm.

\section{Quantum circuits for preparing the Gaussian states}
\label{apd:prep_gaussian_states}
Preparing a quantum state whose amplitudes are described by a given function is a fundamental problem in quantum information science and has been extensively studied in the past decades (e.g. \cite{grover2000synthesis, grover2002creating, holmes2020efficient, GarciaRipoll2021quantuminspired, gharibian2015tensor, rattew2022preparing, Bausch2022fastblackboxquantum, wang2022inverse}). In this appendix, we utilize the method of \cite{mcardle2022quantum} to prepare quantum states that encode Gaussian distributions, as these states are necessary for the algorithm based on strong Euler scheme. This method has the merits that it does not rely on amplitude oracles (which are often implemented with expensive coherent arithmetic), requires few ancilla qubits, and has rigorous performance guarantee,

Let us first review the main result of~\cite{mcardle2022quantum}. Suppose we want to prepare an $n$-qubit state whose amplitudes are described by a function $f:[a,b]\to \C$, i.e.,
\begin{align}
    \ket{\psi_f} = \frac{1}{{\cal N}_f}\sum_{i=0}^{N-1}f(x(i))\ket{i},
\end{align}
where $N=2^n$, $x(i)=a+(b-a)i/N$ for each $i$, and ${\cal N}_f = \sqrt{\sum_{i=0}^{N-1} |f(x(i))|^2}$. Let 
\begin{align}
&\norm{f}_2^{[\infty]}=\sqrt{\int_a^b \abs{f(x)}^2 dx},    \\
&\norm{f}_2^{[N]}=\sqrt{\frac{b-a}{N}\sum_{i=0}^{N-1} \abs{f(x(i))}^2}, 
\end{align}
be the $L^2$ norm of $f$ and its discrete approximation. Then let
\begin{align}
&F_f^{[\infty]}=\frac{\norm{f}_2^{[\infty]}}{\sqrt{(b-a)|f|_{\rm max}^2}}, \\
&F_f^{[N]}=\frac{\norm{f}_2^{[N]}}{\sqrt{(b-a)|f|_{\rm max}^2}}. 
\end{align}
be the $L^2$-norm filling-fraction of $f$ and its discrete approximation. Theorem 1 of~\cite{mcardle2022quantum} states that if there exists a polynomial $p$ of degree $d_\delta$ such that 
\begin{align}
\abs{p(\sin(j/N))-\frac{f(x(j))}{|f|_{\rm max}}}\le \delta =\epsilon \cdot \min\lrb{F_f^{[N]}, F_{\tilde{f}}^{[N]}},    
\end{align}
for all $j \in [0, N]$, where $\tilde{f}(x(j))=p(\sin(j/N))$, then we can prepare a quantum state $\ket{\psi_{\tilde{f}}}$ that is $\epsilon$-close in trace distance to $\ket{\psi_f}$ using $O\left(n d_\delta / F_{\tilde{f}}^{[N]}\right)$ gates. 

Now we apply this method to prepare an $n$-qubit state whose amplitudes are described by a Gaussian function. Precisely, we have $f(x)=e^{-x^2/4}$, $a=-\eta$ and $b=\eta$ for appropriately chosen $\eta$ in our case, and hence our target state is
\begin{align}
    \ket{\psi} = \frac{1}{Z} \sum_{i=0}^{N-1} e^{-\frac{x(i)^2}{4}} \ket{i},
    \label{eq:ideal_gaussian_state}
\end{align}
where $x(i)=(2i-N)\eta/N$ for $i=0,1,\dots,N-1$, and $Z=\sum_{i=0}^{N-1} e^{-x(i)^2/2}$. We follow the procedure of~\cite{mcardle2022quantum} to prepare this state within $\epsilon_{\rm prep}$ precision in trace distance. This leads to a circuit consisting of multi-qubit Toffoli gates, controlled and uncontrolled $R_x/R_y/R_z$ rotations, and Clifford gates:
\begin{enumerate}
    \item We combine circuits~\eqref{circ:Usin_opt} and~\eqref{circ:gidney_recursive} (see Appendix \ref{apd:opt_circuit_usin} for details) to implement an $(n+1)$-qubit unitary operation $U_{\sin}$ (shown in~\eqref{circ:Usin}) such that 
    \begin{align}
    (\bra{0} \otimes I_n) U_{\sin} (\ket{0} \otimes I_n) = \sum_{i=0}^{N-1} \sin(x(i)/\eta) \ket{i}\bra{i}.
    \end{align}
    The entire circuit requires $3n+1$ ancilla qubits (relative to the $n$-qubit input, and $n$ ancilla qubits are used to store a reusable catalyst state), $4n$ T gates (contributed by $n$ relative-phase Toffoli gates), two $R_z$ rotations and Clifford gates. Note that the $4n$ T gates can be arranged in $n+1$ layers, and the two $R_z$ rotations can be executed in parallel with the assistance of 2 extra ancilla qubits. See Appendix~\ref{apd:opt_circuit_usin} for a discussion on the one-time synthesis cost of the catalyst, which we will neglect here.
    \item Let $g(z)=\exp{-\frac{\eta^2}{4} \operatorname{arcsin}^2(z)}$ for $z \in [-\sin(1), \sin(1)]$. Then for any $x \in [-\eta, \eta]$, we have 
    $f(x)=g(\sin(x/\eta))$. Corollary 1 in Appendix D of~\cite{mcardle2022quantum} shows that for any $\beta, \delta>0$,
    there exists a polynomial $p$ of degree at most 
    \begin{align}
    d_\delta = \frac{\frac{\pi^2}{8} \beta + \ln{1/\delta}}{1-\sin(1)}-1
    \end{align}
    such that 
    \begin{align}
    \abs{\exp{-\frac{\beta}{2} \operatorname{arcsin}^2(z)}-p(z)} \le \delta,        
    \end{align}
    for all $z \in [-\sin(1), \sin(1)]$. 
    Here we set $\beta=\eta^2/2$ and $\delta=\epsilon_{\rm prep} \cdot \min\lrb{F_f^{[N]}, F_{\tilde{f}}^{[N]}}$ 
    Note that the proof of Lemma 5 in Appendix E of~\cite{mcardle2022quantum} shows that $F_{\hat{f}}^{[N]} \ge \frac{1}{5 \sqrt[\leftroot{-3}\uproot{3}4]{\beta}}$ provided that $N \ge \sqrt{\beta}$ and $|f(x(i))-\hat{f}(x(i))| \le 1/4$ for each $i$. This implies that 
    $F_{f}^{[N]}, F_{\hat{f}}^{[N]} \ge \sqrt[\leftroot{-3}\uproot{3}4]{ 2}/\lrb{5 \sqrt{\eta}}$ and $\delta \ge \epsilon_{\rm prep} \sqrt[\leftroot{-3}\uproot{3}4]{ 2}/\lrb{5 \sqrt{\eta}}$ in our case.\\
    Now let $\tilde{f}(x)=p(\sin(x/\eta))$. Then we have $\abs{f(x)-\tilde{f}(x)}\le \delta$ for all $x \in [-\eta, \eta]$. Moreover, we can use quantum eigenvalue transformation (QET) to obtain an $(n+3)$-qubit unitary operation $U_{\tilde{f}}$ such that
    \begin{align}
        (\bra{0}^{\otimes 3} \otimes I_n) U_{\tilde{f}} (\ket{0}^{\otimes 3} \otimes I_n) = \frac{1}{2} \sum_{i=0}^{N-1} \frac{\tilde{f}(x(i))}{|\tilde{f}|_{\rm max}}\ket{i}\bra{i}.
    \end{align}
    This QET circuit requires $2$ ancilla qubits (relative to $U_{\sin}$), $d_\delta$ applications of $U_{\sin}$/$U_{\sin}^\dag$, $2d_\delta$ controlled-$R_z$ rotations, and Clifford gates.
    \item Applying $U_{\tilde{f}} \lrb{I_3 \otimes H^{\otimes n}}$ on $\ket{0}^{\otimes n+3}$ yields a state from which $\ket{\psi_{\tilde{f}}}$ can be obtained probabilistically: 
    \begin{align}
        (\bra{0}^{\otimes 3} \otimes I_n) U_{\tilde{f}} \lrb{I_3 \otimes H^{\otimes n}} \ket{0}^{\otimes n+3} = \frac{F_{\tilde{f}}^{[N]}}{2} \ket{\psi_{\tilde{f}}}.
    \end{align}
    One can use exact amplitude amplification (EAA)~\cite{mcardle2022quantum} to boost this probability to one. This requires one ancilla qubit (relative to $U_{\tilde{f}}$), $2k+1$ applications of $U_{\tilde{f}}$/$U_{\tilde{f}}^\dag$, 
    $2k+1$ $R_y$ rotations, $k$ 4-qubit Toffoli gates,
    $k$ $(n+4)$-qubit Toffoli gates, 
    and Clifford gates, where 
    \begin{align}
        k=\left\lceil \frac{\pi}{4\operatorname{arcsin}\left({F_{\tilde{f}}^{[N]}}/{2}\right)} -\frac{1}{2} \right\rceil.
    \end{align}
\end{enumerate}

Overall, the above circuit acts on $4n+4$ qubits and contains 
\begin{itemize}
    \item $4n d_\delta (2k+1)$ T gates,
    \item $2 d_\delta (2k+1)$ $R_z$ rotations,
    \item $2d_\delta (2k+1)$ controlled-$R_z$ rotations,
    \item $2k+1$ $R_y$ rotations,
    \item $k$ $(n+4)$-qubit Toffoli gates,
    \item $k$ $4$-qubit Toffoli gates, 
\end{itemize}
and Clifford gates. Note that among the $4n+4$ qubits, $n$ of them are used to store a reusable catalyst state. Moreover, the $4n d_\delta (2k+1)$ T gates can be arranged in $d_\delta (n+1)(2k+1)$ layers, and the $2 d_\delta (2k+1)$ $R_z$ rotations can be executed as $d_\delta (2k+1)$ layers with the help of 2 additional ancilla qubits.

Next, we use the method of~\cite{Jones2013low} to decompose the multi-qubit Toffoli gates into Clifford and T gates. Meanwhile, we also decompose each controlled-$R_z$ rotation into two $R_z$ rotations and Clifford gates as implied by \eqref{circ:cry_synthesis}. Consequently, the total number of $R_y/R_z$ rotations in the circuit becomes 
\begin{align}
M_\delta = (6 d_\delta + 1)(2k+1).
\end{align}
Suppose we want to implement the whole circuit with $\epsilon_{\rm gauss}$ precision. Then each $R_y/R_z$ rotation can incur at most ${\epsilon_{\rm gauss}}/{M_\delta}$ error. We use the method of~\cite{bocharov2015efficient} to synthesize each $R_y/R_z$ rotation with this precision, which requires $1.15 \logb{{M_\delta}/{\epsilon_{\rm gauss}}}$ T gates and Clifford gates with the help of an additional ancilla qubit. 

Let $U_{\rm gauss}(n, \eta, \epsilon_{\rm prep}, \epsilon_{\rm gauss})$ denote the final circuit consisting of Clifford and T gates. We conclude that
\begin{align}
    \tcount{U_{\rm gauss}(n, \eta, \epsilon_{\rm prep}, \epsilon_{\rm gauss})}
    =&
    4n d_\delta (2k+1)
    +
    k \tcount{{\rm Toffoli}_{n+4}}
    +k \tcount{{\rm Toffoli}_{4}}\nonumber\\
    &
    +M_\delta \cdot 1.15 \logb{{M_\delta}/{\epsilon_{\rm gauss}}} \nonumber \\
    =&
    4n d_\delta (2k+1)+
    4k (n+4) \nonumber \\
    &+1.15 (6 d_\delta + 1)(2k+1) \logb{{M_\delta}/{\epsilon_{\rm gauss}}}, \\
    \tdepth{U_{\rm gauss}(n, \eta, \epsilon_{\rm prep}, \epsilon_{\rm gauss})}
    =&
    d_\delta (n+1)(2k+1) +
    k \tdepth{{\rm Toffoli}_{n+4}}
    +k \tdepth{{\rm Toffoli}_{4}} \nonumber\\
    &+ 
    (5d_\delta+1) (2k+1)   
    \cdot 1.15 \logb{{M_\delta}/{\epsilon_{\rm gauss}}} \nonumber \\
    =&
    d_\delta (n+1)(2k+1) +
    k (n+4) \nonumber \\
    &+1.15 (5d_\delta+1) (2k+1) \logb{{M_\delta}/{\epsilon_{\rm gauss}}}, \\
    \numancilla{U_{\rm gauss}(n, \eta, \epsilon_{\rm prep}, \epsilon_{\rm gauss})}
    =& 4+n +\max\lrb{\numancilla{{\rm Toffoli}_{n+4}}, 2n+2} \nonumber\\
    =& 3n+6.
\end{align}
Note that the $M_\delta$ $R_y/R_z$ rotations are arranged in $(5d_\delta+1) (2k+1)$ layers instead of $(6d_\delta+1) (2k+1)$ layers because, as mentioned above, we execute each pair of $R_z$ rotations in the circuit for $U_{\rm sin}$ simultaneously with the assistance of 2 extra ancilla qubits. In addition, about the ancilla qubits, we need $n$ of them to store a reusable catalyst state, $1$ to construct $U_{\rm sin}$, $2$ for QET, and $1$ for EAA. These $n+4$ ancilla qubits cannot be used to facilitate the implementation of $R_y/R_z$ rotations and multiqubit Toffoli gates. For those operations, we need additional $2n+2$ ancilla qubits. Thus, the total number of ancilla qubits is $3n+6$ which is insignificant compared to that required by other components of the overall pricing algorithm.

\bibliographystyle{quantum}
\bibliography{refs}

\begin{thebibliography}{10}

\bibitem{herman2023quantum}
Dylan Herman, Cody Googin, Xiaoyuan Liu, Yue Sun, Alexey Galda, Ilya Safro, Marco Pistoia, and Yuri Alexeev.
\newblock ``Quantum computing for finance''.
\newblock \href{https://dx.doi.org/10.1038/s42254-023-00603-1}{Nature Reviews Physics {\bf 5}, 450--465}~(2023).

\bibitem{Rebentrost2018quantum}
Patrick Rebentrost, Brajesh Gupt, and Thomas~R. Bromley.
\newblock ``Quantum computational finance: Monte carlo pricing of financial derivatives''.
\newblock \href{https://dx.doi.org/10.1103/PhysRevA.98.022321}{Phys. Rev. A {\bf 98}, 022321}~(2018).

\bibitem{Stamatopoulos2020option}
Nikitas Stamatopoulos, Daniel~J. Egger, Yue Sun, Christa Zoufal, Raban Iten, Ning Shen, and Stefan Woerner.
\newblock ``Option {P}ricing using {Q}uantum {C}omputers''.
\newblock \href{https://dx.doi.org/10.22331/q-2020-07-06-291}{{Quantum} {\bf 4}, 291}~(2020).

\bibitem{Stamatopoulos2022towardsquantum}
Nikitas Stamatopoulos, Guglielmo Mazzola, Stefan Woerner, and William~J. Zeng.
\newblock ``Towards {Q}uantum {A}dvantage in {F}inancial {M}arket {R}isk using {Q}uantum {G}radient {A}lgorithms''.
\newblock \href{https://dx.doi.org/10.22331/q-2022-07-20-770}{{Quantum} {\bf 6}, 770}~(2022).

\bibitem{An2021quantumaccelerated}
Dong An, Noah Linden, Jin-Peng Liu, Ashley Montanaro, Changpeng Shao, and Jiasu Wang.
\newblock ``Quantum-accelerated multilevel {M}onte {C}arlo methods for stochastic differential equations in mathematical finance''.
\newblock \href{https://dx.doi.org/10.22331/q-2021-06-24-481}{{Quantum} {\bf 5}, 481}~(2021).

\bibitem{Chakrabarti2021thresholdquantum}
Shouvanik Chakrabarti, Rajiv Krishnakumar, Guglielmo Mazzola, Nikitas Stamatopoulos, Stefan Woerner, and William~J. Zeng.
\newblock ``A {T}hreshold for {Q}uantum {A}dvantage in {D}erivative {P}ricing''.
\newblock \href{https://dx.doi.org/10.22331/q-2021-06-01-463}{{Quantum} {\bf 5}, 463}~(2021).

\bibitem{alcazar2022quantum}
Javier Alcazar, Andrea Cadarso, Amara Katabarwa, Marta Mauri, Borja Peropadre, Guoming Wang, and Yudong Cao.
\newblock ``Quantum algorithm for credit valuation adjustments''.
\newblock \href{https://dx.doi.org/10.1088/1367-2630/ac5003}{New Journal of Physics {\bf 24}, 023036}~(2022).

\bibitem{bouland2023quantum}
Adam Bouland, Aditi Dandapani, and Anupam Prakash.
\newblock ``A quantum spectral method for simulating stochastic processes, with applications to monte carlo''~(2023).
\newblock  \href{http://arxiv.org/abs/2303.06719}{arXiv:2303.06719}.

\bibitem{Stamatopoulos2024derivativepricing}
Nikitas Stamatopoulos and William~J. Zeng.
\newblock ``Derivative {P}ricing using {Q}uantum {S}ignal {P}rocessing''.
\newblock \href{https://dx.doi.org/10.22331/q-2024-04-30-1322}{{Quantum} {\bf 8}, 1322}~(2024).

\bibitem{shreve2004stochastic}
Steven Shreve.
\newblock ``{Stochastic Calculus for Finance II: Continuous-Time Models}''.
\newblock Volume~11.
\newblock Springer New York, NY. ~(2004).

\bibitem{shreve2005stochastic}
Steven Shreve.
\newblock ``{Stochastic Calculus for Finance I: The Binomial Asset Pricing Model}''.
\newblock \href{https://dx.doi.org/10.1007/978-0-387-22527-2}{Springer New York, NY}. ~(2005).

\bibitem{Kloeden_Platen_1992}
Peter~E. Kloeden and Eckhard Platen.
\newblock ``{Numerical Solution of Stochastic Differential Equations}''.
\newblock \href{https://dx.doi.org/10.1007/978-3-662-12616-5}{Springer Berlin, Heidelberg}. ~(1992).

\bibitem{Brassard_2002}
Gilles Brassard, Peter H{\o}yer, Michele Mosca, and Alain Tapp.
\newblock ``Quantum amplitude amplification and estimation''.
\newblock \href{https://dx.doi.org/10.1090/conm/305/05215}{Quantum Computation and Information {\bf 305}, 53--74}~(2002).

\bibitem{black1973pricing}
Fischer Black and Myron Scholes.
\newblock ``The pricing of options and corporate liabilities''.
\newblock \href{https://dx.doi.org/10.1086/260062}{Journal of Political Economy {\bf 81}, 637--654}~(1973).

\bibitem{merton1973theory}
Robert~C Merton.
\newblock ``Theory of rational option pricing''.
\newblock \href{https://dx.doi.org/10.2307/3003143}{The Bell Journal of economics and management science {\bf 4}, 141--183}~(1973).

\bibitem{heston1993closed}
Steven~L Heston.
\newblock ``A closed-form solution for options with stochastic volatility with applications to bond and currency options''.
\newblock \href{https://dx.doi.org/10.1093/rfs/6.2.327}{The Review of Financial Studies {\bf 6}, 327--343}~(1993).

\bibitem{beliaeva2010simple}
Natalia~A Beliaeva et~al.
\newblock ``A simple approach to pricing american options under the heston stochastic volatility model''.
\newblock \href{https://dx.doi.org/10.3905/jod.2010.17.4.025}{The Journal of Derivatives {\bf 17}, 25--43}~(2010).

\bibitem{alos2012decomposition}
Elisa Al{\`o}s.
\newblock ``A decomposition formula for option prices in the heston model and applications to option pricing approximation''.
\newblock \href{https://dx.doi.org/10.1007/s00780-012-0177-0}{Finance and Stochastics {\bf 16}, 403--422}~(2012).

\bibitem{CHIARELLA20122034}
Carl Chiarella, Boda Kang, and Gunter~H. Meyer.
\newblock ``The evaluation of barrier option prices under stochastic volatility''.
\newblock \href{https://dx.doi.org/10.1016/j.camwa.2012.03.103}{Computers \& Mathematics with Applications {\bf 64}, 2034--2048}~(2012).

\bibitem{he2018closed}
Xin-Jiang He and Song-Ping Zhu.
\newblock ``A closed-form pricing formula for european options under the heston model with stochastic interest rate''.
\newblock \href{https://dx.doi.org/10.1016/j.cam.2017.12.011}{Journal of Computational and Applied Mathematics {\bf 335}, 323--333}~(2018).

\bibitem{grinko2021iterative}
Dmitry Grinko, Julien Gacon, Christa Zoufal, and Stefan Woerner.
\newblock ``Iterative quantum amplitude estimation''.
\newblock \href{https://dx.doi.org/10.1038/s41534-021-00379-1}{npj Quantum Information {\bf 7}, 52}~(2021).

\bibitem{Nielsen_Chuang_2000}
Michael~A. Nielsen and Isaac~L. Chuang.
\newblock ``{Quantum Computation and Quantum Information}''.
\newblock \href{https://dx.doi.org/10.1017/CBO9780511976667}{Cambridge University Press}. Cambridge, U.K.~(2000).

\bibitem{bravyi2005universal}
Sergey Bravyi and Alexei Kitaev.
\newblock ``Universal quantum computation with ideal clifford gates and noisy ancillas''.
\newblock \href{https://dx.doi.org/10.1103/PhysRevA.71.022316}{Phys. Rev. A {\bf 71}, 022316}~(2005).

\bibitem{mcardle2022quantum}
Sam McArdle, András Gilyén, and Mario Berta.
\newblock ``Quantum state preparation without coherent arithmetic''~(2022).
\newblock  \href{http://arxiv.org/abs/2210.14892}{arXiv:2210.14892}.

\bibitem{higham2001analgorithmic}
Desmond~J. Higham.
\newblock ``An algorithmic introduction to numerical simulation of stochastic differential equations''.
\newblock \href{https://dx.doi.org/10.1137/S0036144500378302}{SIAM Review {\bf 43}, 525--546}~(2001).

\bibitem{andersen2001extended}
Leif~BG Andersen and Rupert Brotherton-Ratcliffe.
\newblock ``Extended libor market models with stochastic volatility''.
\newblock \href{https://dx.doi.org/10.2139/ssrn.294853}{Available at SSRN 294853}~(2001).

\bibitem{Glasserman2003montecarlo}
Paul Glasserman.
\newblock ``{Monte Carlo Methods in Financial Engineering}''.
\newblock \href{https://dx.doi.org/10.1007/978-0-387-21617-1}{Springer New York, NY}. ~(2003).

\bibitem{milstein2004stochastic}
Grigori~N Milstein and Michael~V Tretyakov.
\newblock ``Stochastic numerics for mathematical physics''.
\newblock \href{https://dx.doi.org/10.1007/978-3-662-10063-9}{Springer Berlin, Heidelberg}. ~(2004).

\bibitem{giles2008multilevel}
Michael~B Giles.
\newblock ``Multilevel monte carlo path simulation''.
\newblock \href{https://dx.doi.org/10.1287/opre.1070.0496}{Operations Research {\bf 56}, 607--617}~(2008).

\bibitem{lord2010comparison}
Roger Lord, Remmert Koekkoek, and Dick~Van Dijk.
\newblock ``A comparison of biased simulation schemes for stochastic volatility models''.
\newblock \href{https://dx.doi.org/10.1080/14697680802392496}{Quantitative Finance {\bf 10}, 177--194}~(2010).

\bibitem{aaronson2020quantum}
Scott Aaronson and Patrick Rall.
\newblock ``Quantum approximate counting, simplified''.
\newblock In Symposium on Simplicity in Algorithms.
\newblock \href{https://dx.doi.org/10.1137/1.9781611976014.5}{Pages 24--32}.
\newblock SIAM~(2020).

\bibitem{suzuki2020amplitude}
Yohichi Suzuki, Shumpei Uno, Rudy Raymond, Tomoki Tanaka, Tamiya Onodera, and Naoki Yamamoto.
\newblock ``Amplitude estimation without phase estimation''.
\newblock \href{https://dx.doi.org/10.1007/s11128-019-2565-2}{Quantum Information Processing {\bf 19}, 1--17}~(2020).

\bibitem{wang2021minimizing}
Guoming Wang, Dax~Enshan Koh, Peter~D. Johnson, and Yudong Cao.
\newblock ``Minimizing estimation runtime on noisy quantum computers''.
\newblock \href{https://dx.doi.org/10.1103/PRXQuantum.2.010346}{PRX Quantum {\bf 2}, 010346}~(2021).

\bibitem{GiurgicaTiron2022lowdepthalgorithms}
Tudor Giurgica-Tiron, Iordanis Kerenidis, Farrokh Labib, Anupam Prakash, and William Zeng.
\newblock ``Low depth algorithms for quantum amplitude estimation''.
\newblock \href{https://dx.doi.org/10.22331/q-2022-06-27-745}{{Quantum} {\bf 6}, 745}~(2022).

\bibitem{Plekhanov2022variationalquantum}
Kirill Plekhanov, Matthias Rosenkranz, Mattia Fiorentini, and Michael Lubasch.
\newblock ``Variational quantum amplitude estimation''.
\newblock \href{https://dx.doi.org/10.22331/q-2022-03-17-670}{{Quantum} {\bf 6}, 670}~(2022).

\bibitem{hhl}
Aram~W. Harrow, Avinatan Hassidim, and Seth Lloyd.
\newblock ``Quantum algorithm for linear systems of equations''.
\newblock \href{https://dx.doi.org/10.1103/PhysRevLett.103.150502}{Phys. Rev. Lett. {\bf 103}, 150502}~(2009).

\bibitem{bocharov2015efficient}
Alex Bocharov, Martin Roetteler, and Krysta~M. Svore.
\newblock ``Efficient synthesis of universal repeat-until-success quantum circuits''.
\newblock \href{https://dx.doi.org/10.1103/PhysRevLett.114.080502}{Phys. Rev. Lett. {\bf 114}, 080502}~(2015).

\bibitem{PRXQuantum.2.030305}
Joonho Lee, Dominic~W. Berry, Craig Gidney, William~J. Huggins, Jarrod~R. McClean, Nathan Wiebe, and Ryan Babbush.
\newblock ``Even more efficient quantum computations of chemistry through tensor hypercontraction''.
\newblock \href{https://dx.doi.org/10.1103/PRXQuantum.2.030305}{PRX Quantum {\bf 2}, 030305}~(2021).

\bibitem{PRXQuantum.4.040303}
Nicholas~C. Rubin, Dominic~W. Berry, Fionn~D. Malone, Alec~F. White, Tanuj Khattar, A.~Eugene DePrince, Sabrina Sicolo, Michael K\"uehn, Michael Kaicher, Joonho Lee, and Ryan Babbush.
\newblock ``Fault-tolerant quantum simulation of materials using bloch orbitals''.
\newblock \href{https://dx.doi.org/10.1103/PRXQuantum.4.040303}{PRX Quantum {\bf 4}, 040303}~(2023).

\bibitem{PhysRevResearch.3.033055}
Vera von Burg, Guang~Hao Low, Thomas H\"aner, Damian~S. Steiger, Markus Reiher, Martin Roetteler, and Matthias Troyer.
\newblock ``Quantum computing enhanced computational catalysis''.
\newblock \href{https://dx.doi.org/10.1103/PhysRevResearch.3.033055}{Phys. Rev. Res. {\bf 3}, 033055}~(2021).

\bibitem{Berry2019qubitizationof}
Dominic~W. Berry, Craig Gidney, Mario Motta, Jarrod~R. McClean, and Ryan Babbush.
\newblock ``Qubitization of {A}rbitrary {B}asis {Q}uantum {C}hemistry {L}everaging {S}parsity and {L}ow {R}ank {F}actorization''.
\newblock \href{https://dx.doi.org/10.22331/q-2019-12-02-208}{{Quantum} {\bf 3}, 208}~(2019).

\bibitem{Doriguello2022quantum}
Jo\~{a}o~F. Doriguello, Alessandro Luongo, Jinge Bao, Patrick Rebentrost, and Miklos Santha.
\newblock ``{Quantum Algorithm for Stochastic Optimal Stopping Problems with Applications in Finance}''.
\newblock In Fran\c{c}ois Le~Gall and Tomoyuki Morimae, editors, 17th Conference on the Theory of Quantum Computation, Communication and Cryptography (TQC 2022).
\newblock \href{https://dx.doi.org/10.4230/LIPIcs.TQC.2022.2}{Volume 232 of Leibniz International Proceedings in Informatics (LIPIcs), pages 2:1--2:24}.
\newblock Dagstuhl, Germany~(2022). Schloss Dagstuhl -- Leibniz-Zentrum f{\"u}r Informatik.

\bibitem{koh2022foundations}
Dax~Enshan Koh, Guoming Wang, Peter~D. Johnson, and Yudong Cao.
\newblock ``{Foundations for Bayesian inference with engineered likelihood functions for robust amplitude estimation}''.
\newblock \href{https://dx.doi.org/10.1063/5.0042433}{Journal of Mathematical Physics {\bf 63}, 052202}~(2022).

\bibitem{zhang2022computingground}
Ruizhe Zhang, Guoming Wang, and Peter Johnson.
\newblock ``Computing {G}round {S}tate {P}roperties with {E}arly {F}ault-{T}olerant {Q}uantum {C}omputers''.
\newblock \href{https://dx.doi.org/10.22331/q-2022-07-11-761}{{Quantum} {\bf 6}, 761}~(2022).

\bibitem{wang2022statepreparation}
Guoming Wang, Sukin Sim, and Peter~D. Johnson.
\newblock ``State {P}reparation {B}oosters for {E}arly {F}ault-{T}olerant {Q}uantum {C}omputation''.
\newblock \href{https://dx.doi.org/10.22331/q-2022-10-06-829}{{Quantum} {\bf 6}, 829}~(2022).

\bibitem{wang2022classicallyboosted}
Guoming Wang.
\newblock ``Classically-boosted quantum optimization algorithm''~(2022).
\newblock  \href{http://arxiv.org/abs/2203.13936}{arXiv:2203.13936}.

\bibitem{wang2023quantum}
Guoming Wang, Daniel~Stilck Fran{\c{c}}a, Ruizhe Zhang, Shuchen Zhu, and Peter~D. Johnson.
\newblock ``Quantum algorithm for ground state energy estimation using circuit depth with exponentially improved dependence on precision''.
\newblock \href{https://dx.doi.org/10.22331/q-2023-11-06-1167}{{Quantum} {\bf 7}, 1167}~(2023).

\bibitem{wang2023faster}
Guoming Wang, Daniel~Stilck França, Gumaro Rendon, and Peter~D. Johnson.
\newblock ``Faster ground state energy estimation on early fault-tolerant quantum computers via rejection sampling''~(2023).
\newblock  \href{http://arxiv.org/abs/2304.09827}{arXiv:2304.09827}.

\bibitem{katabarwa2023early}
Amara Katabarwa, Katerina Gratsea, Athena Caesura, and Peter~D. Johnson.
\newblock ``Early fault-tolerant quantum computing''.
\newblock \href{https://dx.doi.org/10.1103/PRXQuantum.5.020101}{PRX Quantum {\bf 5}, 020101}~(2024).

\bibitem{Jones2013low}
Cody Jones.
\newblock ``Low-overhead constructions for the fault-tolerant toffoli gate''.
\newblock \href{https://dx.doi.org/10.1103/PhysRevA.87.022328}{Phys. Rev. A {\bf 87}, 022328}~(2013).

\bibitem{Amy2021phase}
Matthew Amy and Neil~J. Ross.
\newblock ``Phase-state duality in reversible circuit design''.
\newblock \href{https://dx.doi.org/10.1103/PhysRevA.104.052602}{Phys. Rev. A {\bf 104}, 052602}~(2021).

\bibitem{Gidney2018halvingcostof}
Craig Gidney.
\newblock ``Halving the cost of quantum addition''.
\newblock \href{https://dx.doi.org/10.22331/q-2018-06-18-74}{{Quantum} {\bf 2}, 74}~(2018).

\bibitem{munoz2018t}
Edgard Mu{\~n}oz-Coreas and Himanshu Thapliyal.
\newblock ``T-count and qubit optimized quantum circuit design of the non-restoring square root algorithm''.
\newblock \href{https://dx.doi.org/10.1145/3264816}{ACM Journal on Emerging Technologies in Computing Systems (JETC) {\bf 14}, 1--15}~(2018).

\bibitem{haner2018optimizing}
Thomas H{\"a}ner, Martin Roetteler, and Krysta~M Svore.
\newblock ``Optimizing quantum circuits for arithmetic''~(2018).
\newblock  \href{http://arxiv.org/abs/1805.12445}{arXiv:1805.12445}.

\bibitem{Wang2021resourceoptimized}
Qingfeng Wang, Ming Li, Christopher Monroe, and Yunseong Nam.
\newblock ``Resource-{O}ptimized {F}ermionic {L}ocal-{H}amiltonian {S}imulation on a {Q}uantum {C}omputer for {Q}uantum {C}hemistry''.
\newblock \href{https://dx.doi.org/10.22331/q-2021-07-26-509}{{Quantum} {\bf 5}, 509}~(2021).

\bibitem{Gidney2019efficientmagicstate}
Craig Gidney and Austin~G. Fowler.
\newblock ``Efficient magic state factories with a catalyzed {$|CCZ\rangle$} to {$2|T\rangle$} transformation''.
\newblock \href{https://dx.doi.org/10.22331/q-2019-04-30-135}{{Quantum} {\bf 3}, 135}~(2019).

\bibitem{nam2019low}
Yunseong Nam and Dmitri Maslov.
\newblock ``Low-cost quantum circuits for classically intractable instances of the hamiltonian dynamics simulation problem''.
\newblock \href{https://dx.doi.org/10.1038/s41534-019-0152-0}{npj Quantum Information {\bf 5}, 44}~(2019).

\bibitem{Shaw2020quantumalgorithms}
Alexander~F. Shaw, Pavel Lougovski, Jesse~R. Stryker, and Nathan Wiebe.
\newblock ``Quantum {A}lgorithms for {S}imulating the {L}attice {S}chwinger {M}odel''.
\newblock \href{https://dx.doi.org/10.22331/q-2020-08-10-306}{{Quantum} {\bf 4}, 306}~(2020).

\bibitem{campbell2021early}
Earl~T Campbell.
\newblock ``Early fault-tolerant simulations of the hubbard model''.
\newblock \href{https://dx.doi.org/10.1088/2058-9565/ac3110}{Quantum Science and Technology {\bf 7}, 015007}~(2021).

\bibitem{kan2022lattice}
Angus Kan and Yunseong Nam.
\newblock ``Lattice quantum chromodynamics and electrodynamics on a universal quantum computer''~(2022).
\newblock  \href{http://arxiv.org/abs/2107.12769}{arXiv:2107.12769}.

\bibitem{kan2022simulating}
Angus Kan and Yunseong Nam.
\newblock ``Simulating lattice quantum electrodynamics on a quantum computer''.
\newblock \href{https://dx.doi.org/10.1088/2058-9565/aca0b8}{Quantum Science and Technology {\bf 8}, 015008}~(2022).

\bibitem{grover2000synthesis}
Lov~K. Grover.
\newblock ``Synthesis of quantum superpositions by quantum computation''.
\newblock \href{https://dx.doi.org/10.1103/PhysRevLett.85.1334}{Phys. Rev. Lett. {\bf 85}, 1334--1337}~(2000).

\bibitem{grover2002creating}
Lov Grover and Terry Rudolph.
\newblock ``Creating superpositions that correspond to efficiently integrable probability distributions''~(2002).
\newblock  \href{http://arxiv.org/abs/quant-ph/0208112}{arXiv:quant-ph/0208112}.

\bibitem{holmes2020efficient}
Adam Holmes and A.~Y. Matsuura.
\newblock ``Efficient quantum circuits for accurate state preparation of smooth, differentiable functions''~(2020).
\newblock  \href{http://arxiv.org/abs/2005.04351}{arXiv:2005.04351}.

\bibitem{GarciaRipoll2021quantuminspired}
Juan~Jos{\'{e}} Garc{\'{i}}a-Ripoll.
\newblock ``Quantum-inspired algorithms for multivariate analysis: from interpolation to partial differential equations''.
\newblock \href{https://dx.doi.org/10.22331/q-2021-04-15-431}{{Quantum} {\bf 5}, 431}~(2021).

\bibitem{gharibian2015tensor}
Sevag Gharibian, Zeph Landau, Seung~Woo Shin, and Guoming Wang.
\newblock ``Tensor network non-zero testing''.
\newblock \href{https://dx.doi.org/10.26421/QIC15.9-10-7}{Quantum Info. Comput. {\bf 15}, 885–889}~(2015).

\bibitem{rattew2022preparing}
Arthur~G. Rattew and Bálint Koczor.
\newblock ``Preparing arbitrary continuous functions in quantum registers with logarithmic complexity''~(2022).
\newblock  \href{http://arxiv.org/abs/2205.00519}{arXiv:2205.00519}.

\bibitem{Bausch2022fastblackboxquantum}
Johannes Bausch.
\newblock ``Fast {B}lack-{B}ox {Q}uantum {S}tate {P}reparation''.
\newblock \href{https://dx.doi.org/10.22331/q-2022-08-04-773}{{Quantum} {\bf 6}, 773}~(2022).

\bibitem{wang2022inverse}
Shengbin Wang, Zhimin Wang, Runhong He, Shangshang Shi, Guolong Cui, Ruimin Shang, Jiayun Li, Yanan Li, Wendong Li, Zhiqiang Wei, and Yongjian Gu.
\newblock ``Inverse-coefficient black-box quantum state preparation''.
\newblock \href{https://dx.doi.org/10.1088/1367-2630/ac93a8}{New Journal of Physics {\bf 24}, 103004}~(2022).

\end{thebibliography}

\end{document}